\documentclass[a4paper]{aa}
\usepackage{txfonts,amssymb,natbib,lscape}
\usepackage[dvips]{graphicx}
\bibpunct{(}{)}{;}{a}{}{,}
\title{Absorption line indices in the $UV$.\\Empirical and theoretical
stellar population models}
\author{C. Maraston\inst{1}\and L. Nieves Colmen\'arez\inst{2}\and R.Bender\inst{2,3}\and D.Thomas\inst{1}}
\institute{University of Portsmouth, Dennis Sciama Building, Burnaby Road, Portsmouth, PO1 3QL, UK \and  Max-Planck-Institute f\"ur extraterrestrische Physik, Giessenbachstr. 1, D-85741 Garching Germany \and
  Universit\"ats-Sternwarte M\"unchen, Scheinerstr. 1, 81679 M\"unchen,
  Germany}
\offprints{Claudia Maraston., \email{claudia.maraston@port.ac.uk}}
\authorrunning{Maraston, Nieves, Bender \& Thomas}
\titlerunning{Absorption line indices in the $UV$}
\date{Received 2006 December 18; accepted 2008 September 12} 
\abstract{}{Stellar absorption lines in the optical (e.g. the Lick
system) have been extensively studied and constitute an important stellar
population diagnostic for galaxies in the local universe and up to 
moderate redshifts. Proceeding towards higher look-back times, galaxies
are younger and the ultraviolet becomes the relevant spectral region
where the dominant stellar populations shine. A comprehensive study of
ultraviolet absorption lines of stellar population models is however still lacking. With this
in mind, we study absorption line indices in the far and mid-ultraviolet in order to determine
age and metallicity indicators for $UV$-bright stellar populations in the local universe 
as well as at high redshift.}  {We
explore empirical and theoretical spectral libraries and use
evolutionary population synthesis to compute synthetic line indices of
stellar population models. From the empirical side, we exploit the IUE-low resolution
library of stellar spectra and system of absorption lines, from which
we derive analytical functions (fitting functions) describing the
strength of stellar line indices as a function of gravity, temperature
and metallicity. The fitting functions are entered into an
evolutionary population synthesis code in order to compute the
integrated line indices of stellar populations models. The same line
indices are also directly evaluated on theoretical spectral energy
distributions of stellar population models based on Kurucz high-resolution
synthetic spectra, 
In order to select indices that can be used as age and/or metallicity indicators for distant galaxies and globular clusters, we compare the models to data of
template globular clusters from the Magellanic Clouds with independently
known ages and metallicities.}{We provide synthetic line indices in the wavelength range $\sim 1200~\AA$ 
to $\sim 3000~\AA$ for stellar populations of various ages and
metallicities.This adds several new indices to the already well-studied CIV and SiIV absorptions.
Based on the comparison with globular cluster data, we select a set of 11 indices blueward of the 2000 $\AA$ rest-frame that allows to recover well the ages and the metallicities of the clusters. These indices are ideal to study ages and metallicities of young galaxies at high redshift. We also provide the synthetic high-resolution stellar population SEDs.}
{}

\begin{document}

  \maketitle 
	\section{Introduction}
	\label{sec:intro}

  \defcitealias{fan_iv}{F92}
  \defcitealias{claudia2005}{M05}

  The far and mid-ultraviolet region of the electromagnetic spectrum
  ($\lambda \sim1200\!-\!3200~\AA$) traces the hot
  component of galaxy stellar populations, that in young galaxies is
  made up of luminous O and B-type stars. These leave a characteristic
  imprint in the integrated spectra in the form of numerous absorption
  features, related to key chemical species like silicon, carbon,
  iron, magnesium, etc. Old  stellar populations can also produce a
  hot stellar component if sufficient mass-loss occurs, as shown by
  well-known phenomena such as the extreme blue horizontal branch in
  globular clusters \citep[e.g.][]{deboer} and the $UV$-upturn of
  elliptical galaxies \citep[e.g.][]{dorman, burstein, gr_ren}.  In this
 first paper we focus on young and massive stars as producers of $UV$ light.

  Thanks to substantial improvement in observational capabilities, it
  is now feasible to obtain the rest-frame ultraviolet spectrum of
  galaxies up to very high redshift \citep[see e.g.,] [] {yee96, steetal96,
  pettini, mehlert01, demello04, cimattietal04, mcaretal04, daddietal05, popetal08}. 
  From the modelling side, extensive work is being invested in understanding the intrinsic $UV$ spectrum
  of young stellar populations as a function of their basic parameters -
  age, metallicity, star formation history and Initial Mass Function -
  (\citealt[hereafter SB99]{robert1993, leitherer91, leitherer95,
  leitherer99}; \citealt{demello}). In these works, the $UV$
  spectra of stellar populations are obtained by using observed
  stellar spectra of O,B stars in the Milky Way and Magellanic Clouds,
  or synthetic spectra derived from model atmospheres.
  \citet{rix2004} have used
  high-resolution $UV$ theoretical stellar spectra \citep{adi}, in
  order to extend the modelling to sub-solar and
  super-solar metallicities. This work focuses on the \ion{Si}{iv}{}
  and \ion{C}{iv}{} lines that are the most prominent absorption
  features in the $UV$. 
  A well-known
  limitation imposed by the use of empirical libraries is the narrow
  metallicity range spanned by the real stars - usually not too
  different from solar. On the other hand, a clear
  advantage of using real stars is that the empirical spectra should
  contain the effect of stellar winds that affect the photospheric
  lines of massive stars and are complicated to model
  \citep[e.g.][]{kudritzki87}.
  
  In this work we take a complementary approach to existing models.
  We follow a twofold strategy and compute both empirically-based models
  as well as theoretical ones. 
  
  From the empirical side we exploit the full potential 
  of the \citet[][\citetalias{fan_iv} hereafter]{fan_iv} empirical library of \emph{IUE} spectra and index
  system for population synthesis models. 
  
  We construct
  analytical polynomial fits (fitting functions, FF) that trace the empirical indices as
  functions of stellar parameters. FF can be easily incorporated into
  an evolutionary synthesis code in order to predict the integrated
  indices of stellar populations. Another advantage is that fluctuations in the spectra of stars with similar atmospheric parameters are averaged out. Absorption line indices have the
  advantage of being insensitive to reddening, which is a serious issue
  in the $UV$ of young stellar populations. This is particularly
  important for high-redshift studies where little is known about dust reddening.
  To our knowledge, this is the first work in which the 
  \citetalias{fan_iv} library of stellar groups and index system is
  thoroughly examined for evolutionary population synthesis studies,
  although individual indices in the system have been considered in
  previous works \citep[see e.g.:][]{fan_ii, ponder, heapetal98, lotz00}.
 
  In parallel, we compute fully theoretical indices by incorporating a high-resolution version of the Kurucz  
  library of stellar spectra \citep{rodetal05} in the Maraston (2005) evolutionary population synthesis code.
  
  We compare the models with
  observations of young star clusters in the Magellanic Clouds. Star
  clusters with independently known ages and metallicities are in principle the
  ideal templates for stellar population models and this comparison should  
  give indications on which indices are reliable tracers of ages and metallicities and
  can be applied with confidence to study distant galaxes. Care has to be taken 
  since spectral lines are also sensitive to specific element ratios of individual elements, which
  may be the case for some of the indices we study here.
 
  This paper is structured as follows. In Sect.~\ref{sec:eps} we
  briefly recall the evolutionary population synthesis code we adopt and
  its ingredients. In Sect.~\ref{sec:lib} we summarise the relevant
  features of the IUE index system and empirical spectral
  library. Sect.~\ref{sec:ffunc} explains the construction of the
  fitting functions, their behaviour with changes in stellar atmospheric
  parameters and how we include metallicity effects. In
  Sect.~\ref{sec:ssp} we describe the construction of the stellar
  population models, both empirical and theoretical and in Sect.~6 we test the models 
  with star cluster data. In Sect.~\ref{sec:concl} we provide a summary and conclusions.

  \section{Evolutionary population synthesis ingredients}
  \label{sec:eps}

  The evolutionary population synthesis (EPS) technique allows the
  computation of the spectro-photometric properties of stellar
  populations using stellar evolutionary tracks \citep[][among
  others]{buzzoni, bc93, worthey_models, vazdekis96, fioc,
  claudia_mod, leitherer99, bc03, claudia2005, sch07}. The main target of EPS
  models are the stellar populations that cannot be resolved into
  individual stars. EPS models provide the theoretical framework for
  interpreting such systems, in particular for deriving ages and
  chemical abundances.

\begin{figure*}[!ht]
    \centering
    \includegraphics[width=.36\hsize,clip,angle=-90,bb=70 88 504 672]{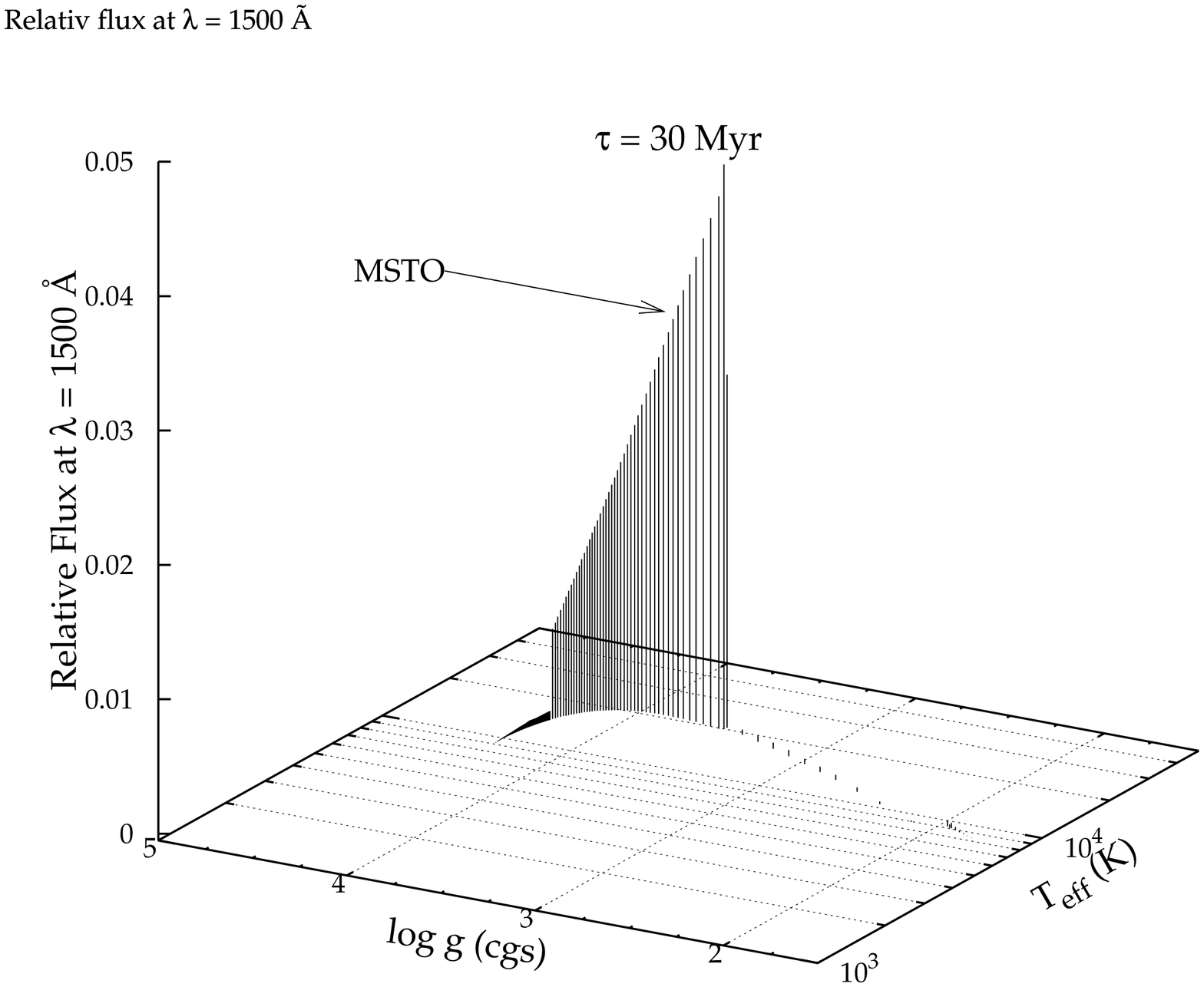}
    \includegraphics[width=.36\hsize,clip,angle=-90,bb=70 88 504 672]{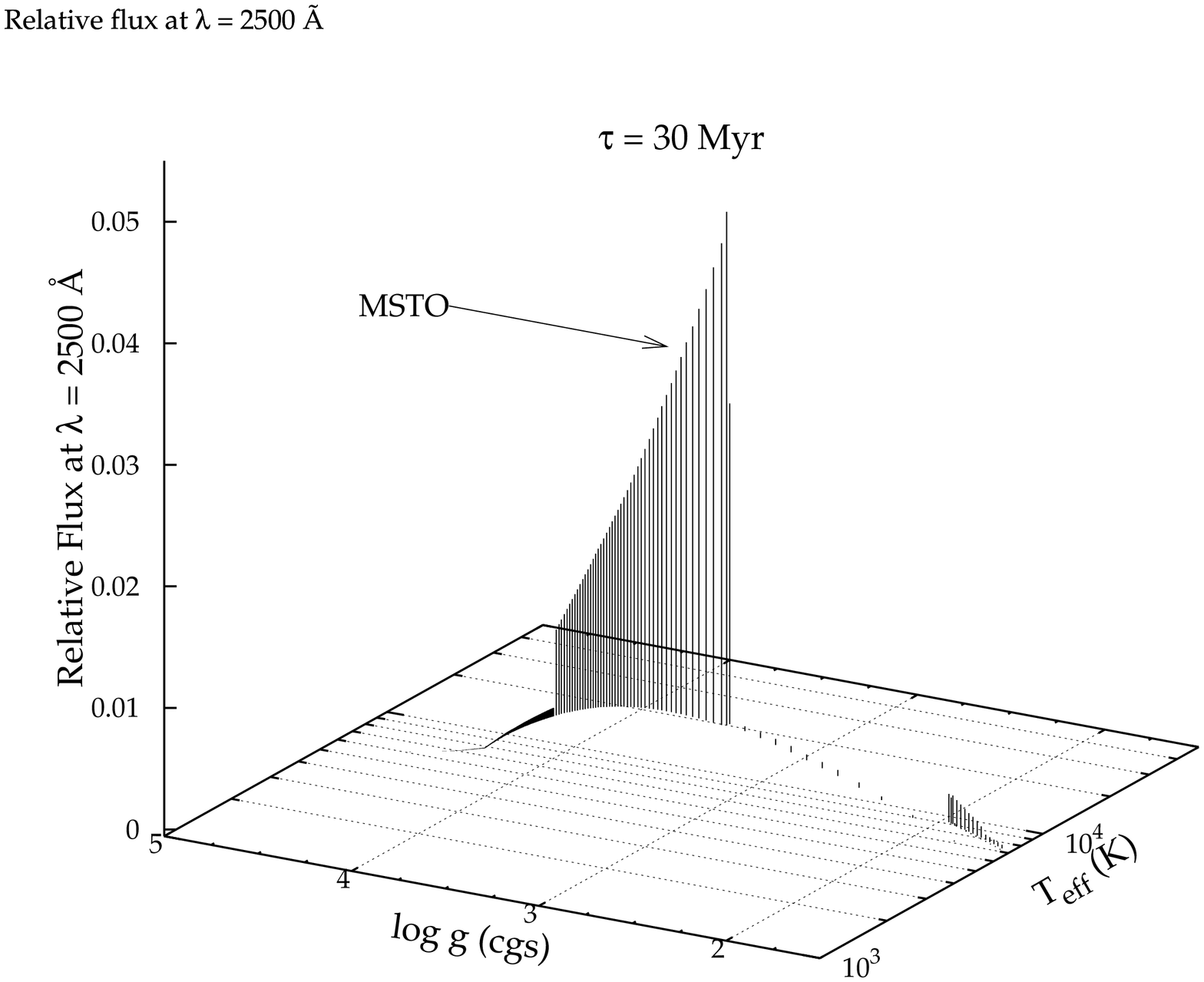}
    \caption{Contributions to the luminosity at $1500~\AA$ (left
    panel) and $2500~\AA$ (right panel), from stars of a 30
    Myr old simple stellar population with \ensuremath{Z_{\odot}}{}
    metallicity \citep{claudia_mod}, as functions of effective
    temperature $T_{\mathrm{eff}}$ and surface gravity \ensuremath{\log
    g}. Stars around the Main Sequence turnoff contribute the bulk of the
    luminosity at these wavelengths.}
    \label{fig:contr}
  \end{figure*}

  In this work we use the EPS models and code by \citet[hereafter
  \citetalias{claudia2005}]{claudia_mod, claudia2005}, in which the
  stellar tracks and isochrones are taken from the Geneva database
  \citep{gmodelsun, schaerer93} for young ages ($t\lesssim 30~ \rm Myr$),
  and from the Frascati database for older ages (see Table~1 in M05).
  The evolution of massive stars is affected by (mostly unknown)
  mass-loss, besides rotation and convective overshooting. Differences
  exist between the Geneva evolutionary tracks, and, for example, those
  from the Padova database (e.g. Girardi et al.~2000), especially in the
  post-Main Sequence evolution.  However, $UV$~spectral indices, as the
  $UV$ luminosity, mostly depend on the stars around the Main Sequence
  turnoff, hence the uncertainties in the post-Main Sequence evolution
  should have a marginal effect. As an example,
  Fig.~\ref{fig:contr} shows for a 30 Myr old, $\ensuremath{Z_{\odot}}$
  simple stellar population SSP \citep{claudia_mod} the relative
  contributions from the individual stars to the total luminosity at
  \ensuremath{\lambda\lambda 1500}~$\AA$ (left panel) and \ensuremath{\lambda\lambda 2500}~$\AA$ (right panel),
  as a function of the effective temperature ($T_{\mathrm{eff}}$) and
  the surface gravity ($\log g$). Stars around the Main Sequence turnoff
  dominate the emission, contributing more than 90\% to the total
  luminosity. To gain more insight into the effects
  of stellar tracks, we also evaluate spectral indices using stellar
  population models based on the Padova tracks (see
  Sect.~6.5).

  \section{Spectral library and line index system}
  \label{sec:lib}

  We use the library of stellar spectra assembled by
  \citetalias{fan_iv}, from \emph{IUE} low-resolution (6~\AA)
  observations of 218 stars in the solar neighbourhood. Spectra cover
  the range \ensuremath{\lambda\lambda1150\!-\!3200}~\AA\ in
  wavelength and are compiled in the ``IUE Ultraviolet spectral
  atlas'' \citep{wu, wu_i}. In its final form, the library consists of
  56 mean stellar groups, classified by spectral type (from O3 to M4)
  and luminosity class (I, III, IV and V). Out of the 56
  stellar groups, the 47 groups classified by \citetalias{fan_iv} as
  having solar metallicity were selected as input for the fitting
  functions. The ``metal-poor'' and ``metal-rich'' groups were
  excluded, due to their very low number. Each stellar group is
  constructed by averaging the spectra of a variable number of
  individual stars (between 2 and 12) with similar spectral types, colours, 
  luminosity classes and when possible direct metallicity determination. 
  When the metallicity was not known, the grouping was made by matching the spectrum. 
 The weight of each spectrum is inversely proportional to the mean signal-to-noise ratio (S/N),
  measured within a 100 \AA{} line-free window, centred at 2450, 2550
  and 2700 \AA{} for spectral types O-G4; G5-K3 and K4-M, respectively
  \citepalias[see][for more details]{fan_iv}.

\begin{figure*}[!th]
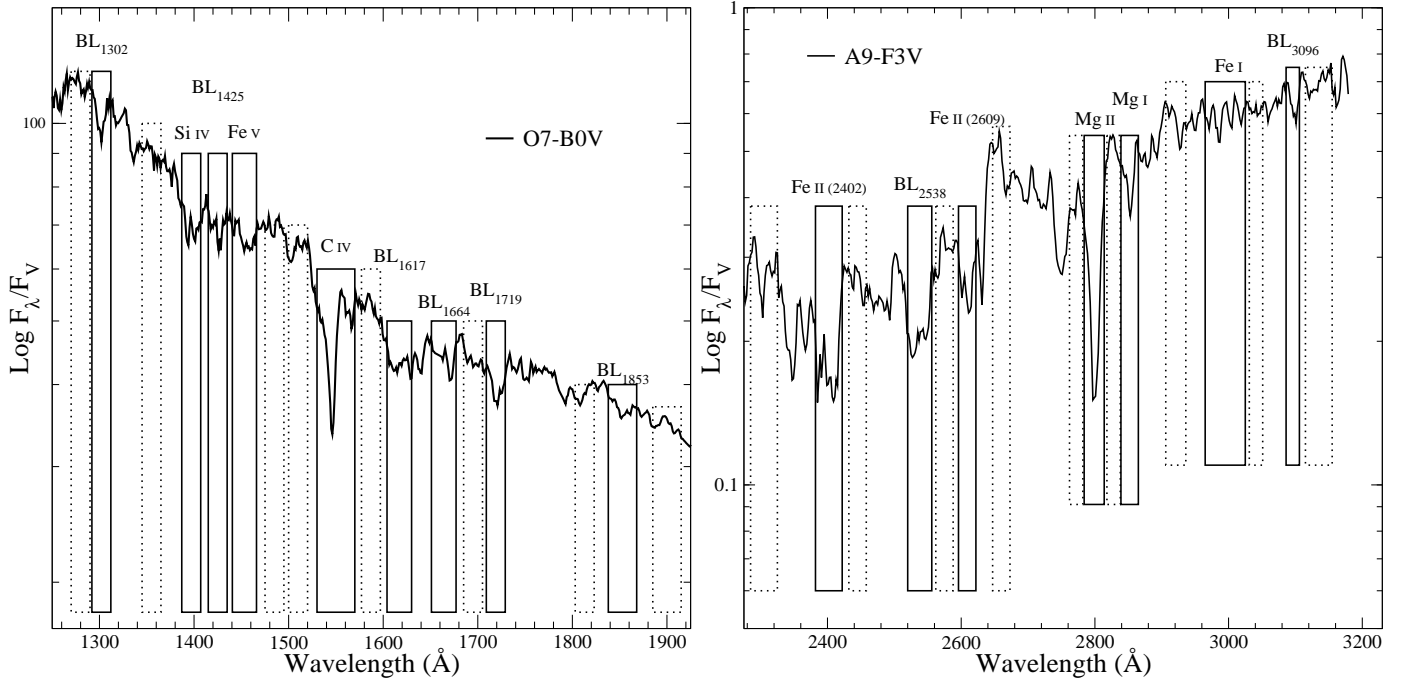

    \begin{center}
      \includegraphics[width=.49\textwidth]{6907f2a}
      \includegraphics[width=.49\textwidth]{6907f2b}
      \caption{Visualisation of index bandpasses on representative
      stellar spectra. The left-hand panel shows the far-$UV$ indices
      with the central bandpasses highlighted by solid boxes and the
      blue and red continuum bandpasses marked by dotted boxes. The
      right panel shows, in the same way, the mid-$UV$ indices. The
      bandpasses for the $\rm Mg_{\rm wide}$~index are not depicted in
      order to avoid crowding.}
      \label{fig:bandpassess}
    \end{center}
  \end{figure*}

  We adopt the line index system defined by \cite{fan_i, fan_iii,
  fan_iv}, which comprises 11 indices in the far-$UV$
  (\ensuremath{\lambda\lambda1270\!-\!1915}~\AA) and 8 indices in the
  mid-$UV$ (\ensuremath{\lambda\lambda2285\!-\!3130}~\AA). In
  Table~\ref{tab:ind} we recall the index definitions, with their
  central and bracketing sidebands, and collect relevant comments about
  the atomic species contributing to the absorption, according to
  literature identifications \citep{fan_iii, bonatto, kinney, coluzzi}.
  In Fig.~\ref{fig:bandpassess} we illustrate graphically the index
  bandpasses on stellar spectra.

  \begin{table*}[!ht]
    \centerline{%
    \begin{tabular}{rlrrrrrrl}
        \hline
        N&Name	&	\multicolumn{2}{c}{Blue
        Bandpass}&\multicolumn{2}{c}{Central
        Bandpass}&\multicolumn{2}{c}{Red Bandpass} & Comments\\
        (1) & (2) & (3) & (4) & (5) & (6) & (7) & (8) & (9)\\\hline\hline
        1 &\ensuremath{\mathrm{BL_{\scriptstyle{1302}}}} & 1270.0 &  1290.0 &  1292.0 &  1312.0 &
        1345.0 & 1365.0& \ion{Si}{iii}, \ion{Si}{ii}, \ion{O}{i}\\
        2 &\ion{Si}{iv} & 1345.0 & 1365.0 & 1387.0 & 1407.0 & 1475.0 &
        1495.0 & \ion{Si}{iv}~1393.8; 1402.8\\
        3 &\ensuremath{\mathrm{BL_{\scriptstyle{1425}}}} & 1345.0 &  1365.0 &  1415.0 &  1435.0 &
        1475.0 & 1495.0&\ion{C}{ii}~1429, \ion{Si}{iii}~1417,
        \ion{Fe}{iv}, \ion{Fe}{v}\\
        4 &\ion{Fe}{1453} & 1345.0 &  1365.0 &  1440.0 &  1466.0 &  1475.0 &
        1495.0& \ion{Fe}{v} +20 additional Fe lines\\
        5 & \ensuremath{\mathrm{C}^{\scriptscriptstyle{A}}_{\mathrm{IV}}} & 1500.0 &  1520.0 &  1530.0 &  1550.0 &  1577.0 &
        1597.0&\ion{C}{iv}~1548, in absorption\\
        6 & \ion{C}{iv} & 1500.0 & 1520.0 & 1540.0 & 1560.0 & 1577.0 & 1597.0 &
        \ion{C}{iv}~1548, central band \\
        7 & \ensuremath{\mathrm{C}^{\scriptscriptstyle{E}}_{\mathrm{IV}}} & 1500.0 &  1520.0 &  1550.0 &  1570.0 & 1577.0 &
        1597.0&\ion{C}{iv}~1548, in emission. \\
        8 & \ensuremath{\mathrm{BL_{\scriptstyle{1617}}}} & 1577.0 &  1597.0 &  1604.0 &  1630.0 &
        1685.0 & 1705.0&\ion{Fe}{iv}\\
        9 & \ensuremath{\mathrm{BL_{\scriptstyle{1664}}}} & 1577.0 &  1597.0 &  1651.0 &  1677.0 &
        1685.0 & 1705.0&\ion{C}{i}~1656.9, \ion{Al}{ii}~1670.8\\
        10 & \ensuremath{\mathrm{BL_{\scriptstyle{1719}}}} & 1685.0 &  1705.0 &  1709.0 &  1729.0 &
        1803.0 & 1823.0&\ion{N}{iv}~1718.6, \ion{Si}{iv}~1722.5; 1727.4,
        \ion{Al}{ii}\\
        11 & \ensuremath{\mathrm{BL_{\scriptstyle{1853}}}} & 1803.0 &  1823.0 &  1838.0 &  1868.0 &
        1885.0 & 1915.0&\ion{Al}{ii}, \ion{Al}{iii}, \ion{Fe}{ii},
        \ion{Fe}{iii}\\
        12 & \ion{Fe}{ii} (2402~\AA) & 2285.0 &  2325.0 & 2382.0
        & 2422.0 &  2432.0 &  2458.0\\
        13 & \ensuremath{\mathrm{BL_{\scriptstyle{2538}}}} & 2432.0 &  2458.0 &  2520.0 &  2556.0 &
        2562.0 & 2588.0& Uncertain, \ion{Fe}{i}?\\
        14 &\ion{Fe}{ii} (2609~\AA) &    2562.0 &  2588.0 &  2596.0 &  
        2622.0 &  2647.0 &  2673.0\\
        15 & \ion{Mg}{ii} &    2762.0 &  2782.0 &  2784.0 &  2814.0 &  2818.0 &  
        2838.0\\
        16 & \ion{Mg}{i} &     2818.0 &  2838.0 &  2839.0 &  2865.0 &  2906.0 &  
        2936.0\\
        17 & \ensuremath{\mathrm{Mg_{\scriptstyle{wide}}}} &      2470.0 &  2670.0 &  2670.0 &  2870.0 &  2930.0 &
        3130.0\\
        18 & \ion{Fe}{i} &     2906.0 &  2936.0 &  2965.0 &  3025.0 &  3031.0 &  
        3051.0\\
        19 & \ensuremath{\mathrm{BL_{\scriptstyle{3096}}}} &      3031.0 &  3051.0 &  3086.0 &  3106.0 &  3115.0 &
        3155.0&\ion{Al}{i}~3092, \ion{Fe}{i}~3091.6\\\hline
    \end{tabular}}
    \caption[Definition of the absorption index system]{Definition of
    the absorption index system \citepalias{fan_iv}. Columns correspond to: 
    (1) index number, (2) name, (3 and 4) blue
    passband definition, (5 and 6) central passband definition, (7 and
    8) red passband definition and (9) comments regarding the relevant
    chemical elements. Indices termed as ``BL'' are blends of several
    elements. Adaptation from \citetalias{fan_iv} (see their Tables 3.A and
    3.B)}

    \label{tab:ind}
  \end{table*}

  \subsection{Derivation of stellar parameters $T_{\mathrm{eff}}$ and \ensuremath{\log g}}
  \label{sec:trans}

  \begin{figure}[!ht]
    \centering
    \includegraphics[width=\hsize,clip]{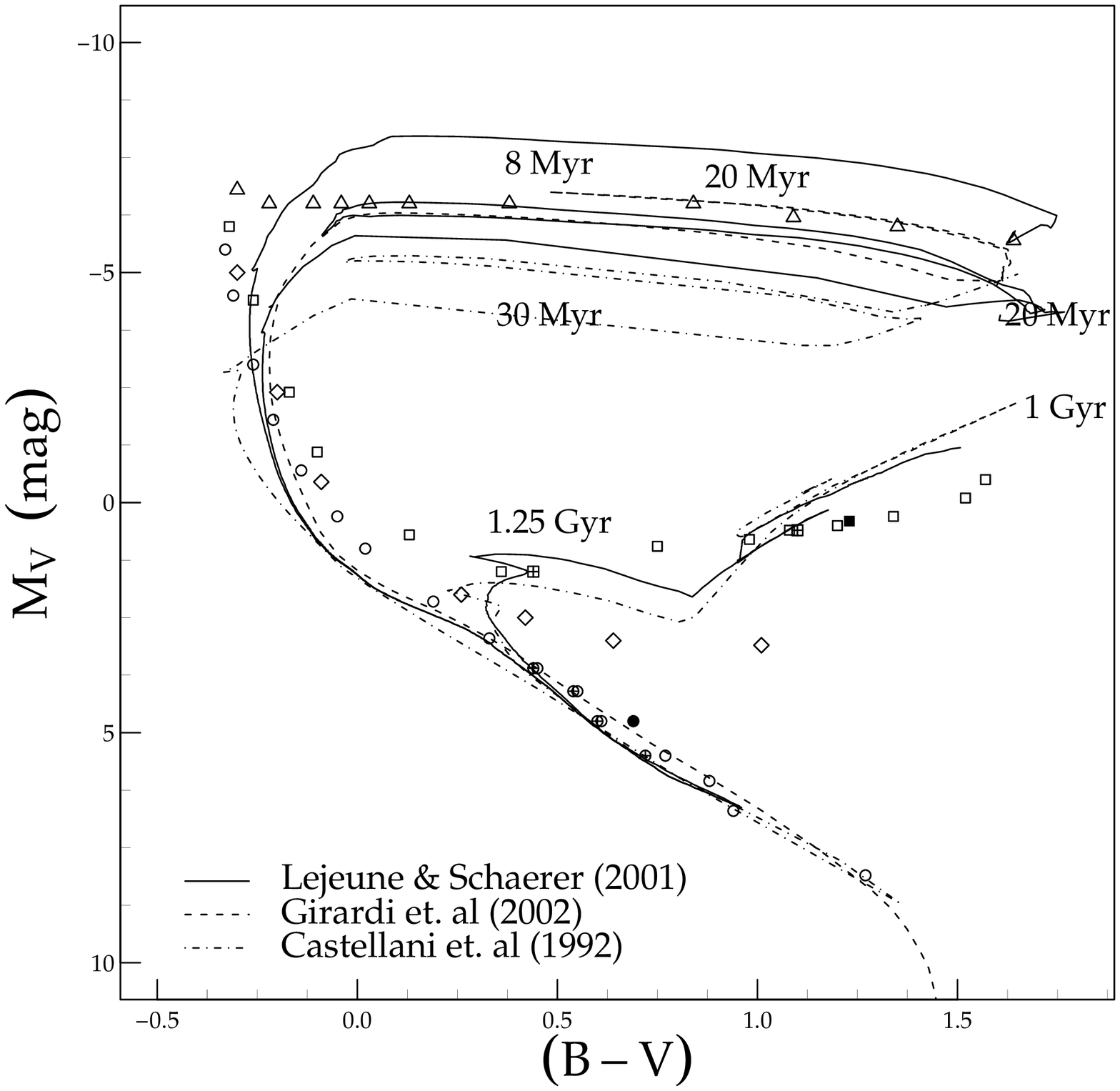}
    \caption{CMD of the stellar groups in the \citetalias{fan_iv}
    library.  Different symbols are used to denote the different
    luminosity classes, dwarfs (V, circles), giants (III, squares),
    sub-giants (IV, diamonds) and super-giants (I \& II,
    triangles). Different fill styles correspond to different
    metallicities, solar (empty),
    sub-solar (crossed) and super-solar (black-filled). Isochrones of
    8, 20\,Myr and 1.25\,Gyr (thick curve, from the compilation of
    \cite{lej01} ), with 20\,Myr (dashed curve, from \cite{girardi02}
    and 30\,Myr and 1\,Gyr (dot-dashed curve, from \cite{castellani}
    are superimposed to highlight the coverage of the library in terms
    of stellar population parameters. All isochrones have solar
    metallicity.}
  \label{fig:cmd}
  \end{figure}

  In order to construct analytical approximations which will
  describe stellar absorption indices as functions of
  $T_{\mathrm{eff}}$ and $\log g$, we need to transform the
  observed $(B-V,M_V)$ of the groups \cite[from][]{oconnell, sk, hump}
  into the theoretical $(T_{\mathrm{eff}},\ensuremath{\log g})$. This
  is done by superimposing isochrones on the colour magnitude diagram
  of stellar groups (Figure~\ref{fig:cmd}), and selecting the
  atmospheric parameters along the isochrones with the closest
  absolute magnitude and colour. Solar metallicity isochrones from
  \cite{gmodelsun,meynet94} were used, for consistency with the
  ingredients of the SSP models. The adopted calibration is given in
  Table~\ref{tab:calib}.

  \begin{table}[!tb]
    \caption[]{Adopted transformation from the observational spectral
    type (through $B-V,\ensuremath{M_V}$) to the theoretical plane
    $T_{\mathrm{eff}},\ensuremath{\log g}$. Based on isochrones from \cite{gmodelsun,meynet94}}
    \scriptsize
    \begin{tabular}{lrrrrr}\hline
      Group &	$T_{\mathrm{eff}}$~(K)	& \ensuremath{\log g}~(cgs) & Group & $T_{\mathrm{eff}}$~(K) & \ensuremath{\log g}~(cgs)\\ \hline
      O3-6V &	44157 &  4.045   &B0-2III & 25293 &  3.517 \\ 
      O7-B0V &	39811 &  4.106 &B3-6III & 18239 &  3.719 \\ 
      B1-1.5V &	25704 &  3.961 &B7-9III & 11246 &  3.617 \\
      B2-4V &	19231 &  3.972   &A3-6III &  8570 &  3.855 \\
      B5-8V &	13772 &  3.959   &A9-F6III & 7362 &  3.856\\
      B9-9.5V &	10069 &  3.939 &G0-5III &	 5058 &  2.942 \\
      A0-2V &	 9638 &  4.113   &G5-K0III & 4753 &  2.706 \\
      A5-8V &	 7907 &  4.196   &K0-2III &	 4498 &  2.376\\
      A9-F3V &	 7161 &  4.278 &K2III &	 4335 &  2.163 \\
      F5-7V &	 6501 &  4.279   &K3III &	 4102 &  1.787 \\  
      F8-9V &	 6152 &  4.320   &K4-5III &	 3899 &  1.403 \\  
      G0-5V &	 5848 &  4.419   &K7-M3III & 3681 &  0.977   \\
      G6-9V &	 5483 &  4.541   &O4-9I &	47643 &  3.913 \\  
      K0-1V &	 5152 &  4.597   &B0-2I &	21038 &  2.733 \\  
      K2-3V &	 4864 &  4.656   &B3-5I &	13122 &  2.196 \\  
      K5-M0V &	 4864 &  4.656 &B6-9I &	 9204 &  1.742 \\
      O9-B0IV &	38371 &  3.921 &A0-2I &	 7980 &  1.576 \\
      B2-5IV &	18239 &  3.719 &A5-F0I &	 7328 &  1.426 \\
      B8-9IV &	11830 &  3.877 &F2-8I &	 6486 &  1.217 \\
      A7-F0IV &	 7464 &  4.044 &G0-3I &	 5346 &  0.820 \\
      F2-7IV &	 6668 &  3.982 &G5-8I &	 4775 &  0.666 \\
      G0-2IV &	 6026 &  3.926 &K2-3I &	 4406 &  0.515 \\
      G8-K1IV &	 4898 &  3.437 &K5-M4I &	 3639 & -0.164 \\
      O5-6III & 46026 &  4.004 &  &  & \\
      \hline
    \end{tabular}
    \label{tab:calib}
  \end{table}

  \begin{figure*}[!th]
    \centering
    \includegraphics[width=.49\textwidth, clip]{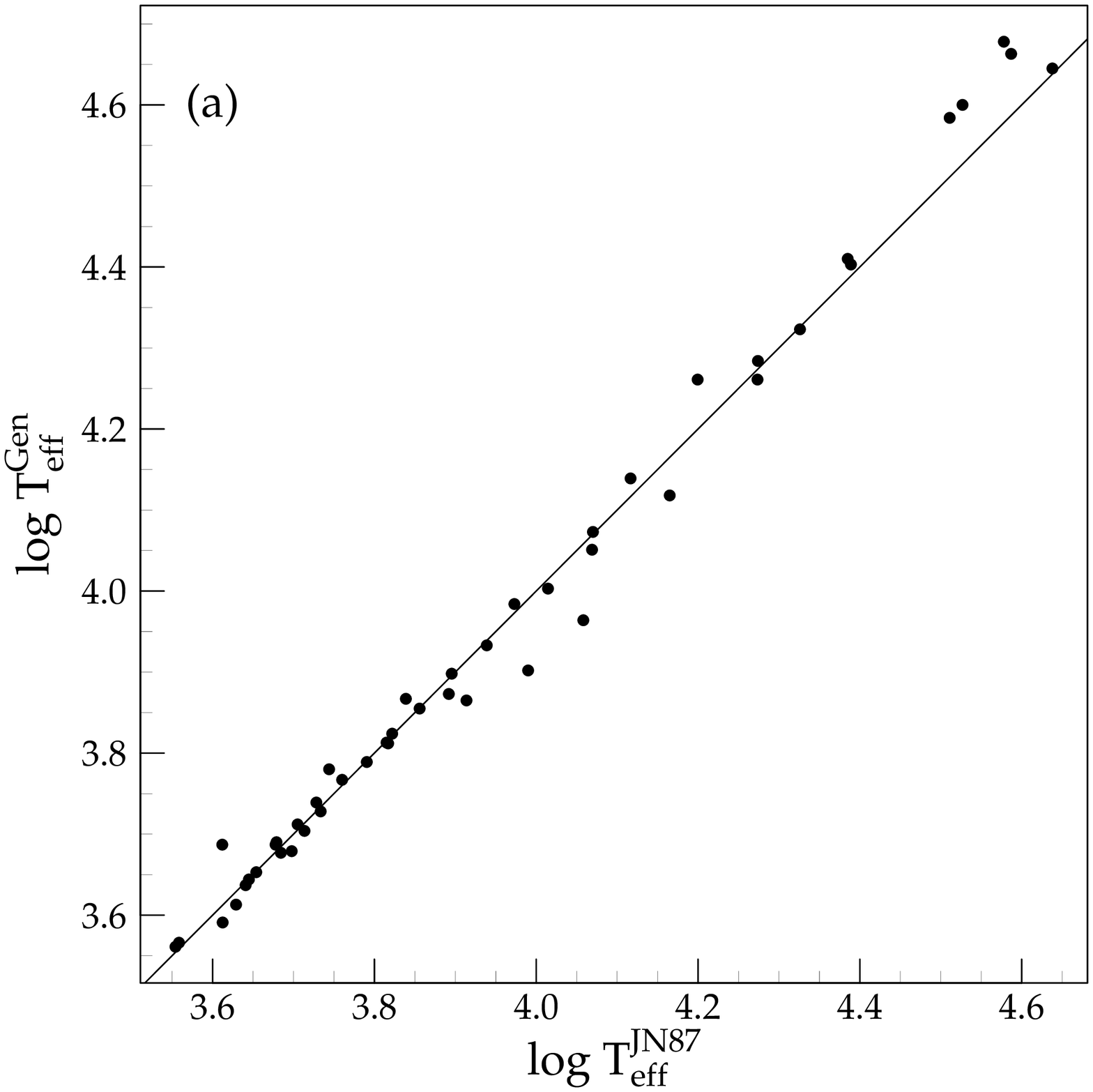}
    \includegraphics[width=.49\textwidth, clip]{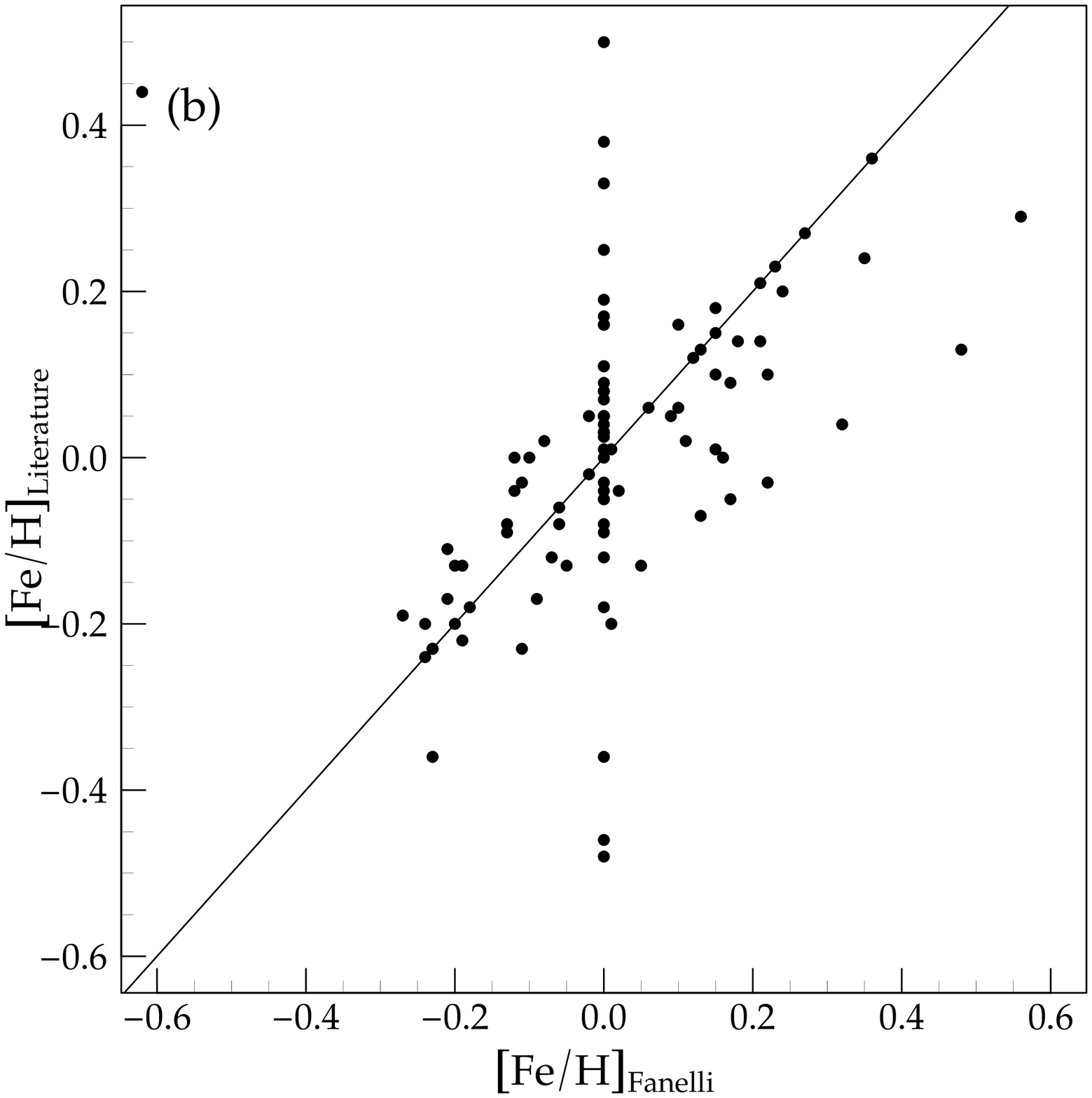}
    \caption{(a) Comparison of $T_{\mathrm{eff}}$ calibrations derived
    with two different methods. The y-axis represents effective
    temperatures derived from superposition of Geneva Group
    isochrones, while the x-axis represents effective temperatures
    derived by interpolation of the \citet{dejager} tables. The solid
    line represents the one-to-one correlation. (b) Metallicity of
    individual stars used to build the ``solar metallicity'' stellar
    groups in the \citetalias{fan_iv} library, compared to metallicity
    determinations found in the literature.}
    \label{fig:teffcomp}
  \end{figure*}

  In order to check that our adopted calibration, that is
  based on a specific set of isochrones, can be generalised to any
  arbitrary computation, we have calculated the effective temperatures
  using a completely independent calibration \citep[by][]{dejager}. The
  result is reported in Fig.~\ref{fig:teffcomp}(a). As can be seen, the
  two calibrations are consistent, the average absolute difference
  in effective temperature being small, $\langle|\log
  T_{\mathrm{eff}}^{\mathrm{Gen}} - \log
  T_{\mathrm{eff}}^{\mathrm{JN87}}|\rangle = 0.02$ dex. We have also
  checked that the corresponding effect on the FFs leaves the stellar
  population models basically unchanged. 
   
  What \citetalias{fan_iv} call the ``solar'' groups actually include stars
  with a spread in metallicity, and even stars for which there were no
  spectroscopic metallicity determinations, and the metallicity was
  assigned by visual inspection of the spectra and spectrum matching.
  We have checked whether the metallicity so determined agrees with other, more recent,
  estimates. We did the exercise for roughly half the sample (96 out of
  189 stars). Fig.~\ref{fig:teffcomp}(b) shows the result, in which the
  \citetalias{fan_iv} metallicity is plotted versus the values from the
  \citet{cayrel97} and other catalogues \citep{wallerstein59, helfer60,
  parker61, wallerstein62, bell65, conti65, alexander67, cayrel70,
  chaffee71, hearnshaw74, tomkin78, luck79, luck81, cayrel85,
  boesgaard89, boesgaard90, cayrel97}. There is no clear bias in the metallicity determination
  of \citetalias{fan_iv}. This is the case even for the stars for which
  \citetalias{fan_iv} did not have a metallicity determination at the time of
  assembly of the library (the vertical group of stars at [Fe/H] $=0$), as
  outliers are found to be evenly divided between sub and super-solar
  metallicity. Finally, and reassuringly, the tail of stars that were
  included by \citetalias{fan_iv} in the solar group though they were assigned super-solar
  metallicity at the time are found through the new determinations to indeed have
  metallicities around solar.

  \subsection{Line indices and errors for stellar groups}
  \label{sec:ew}

  The $UV$ line indices defined by \citetalias{fan_iv} are expressed as
  equivalent widths and given in \AA. We measure them on the empirical
  spectra of each group, following standard definitions 
  \citep{buretal84,fabetal85}:

	\begin{equation}
		\label{eq:ewdef}
		EW_{\lambda_{i,f}} = \int_{\lambda_i}^{\lambda_f}
		\left(1-\frac{S(\lambda)}{C(\lambda)}\right)d\lambda
	\end{equation}
  where $S(\lambda)$ and $C(\lambda)$ are the fluxes in the line and in the
  continuum, respectively. The statistical errors in the equivalent width are
  computed following \cite{cardiel}:

  \begin{equation}
    \begin{array}{ccl}
      \label{eq:cardiel}
      \sigma^2_{EW} &= & \biggl(\sum_{i=1}^N \biggl[\frac{C^2(\lambda_i)\sigma^2_i
        +
        S^2(\lambda_i)\sigma^2_{C(\lambda_i)}}{C^4(\lambda_i)}\biggr] \\
         & + &  \sum_{i=1}^N\sum_{j=1,j\ne
        i}^N\biggl[\frac{S(\lambda_i)S(\lambda_j)}{C^2(\lambda_i)C^2(\lambda_j)}(
        \Lambda_1\sigma^2_B + \Lambda_4\sigma^2_R)\biggr]\biggr)\Delta .
    \end{array}
  \end{equation}

  In equation~(\ref{eq:cardiel}) $S(\lambda_i)$, $\sigma_i$ and
  $C(\lambda_i)$, $\sigma_{C(\lambda_i)}$ are the fluxes and dispersions
  in the line and in the pseudo-continua respectively, $\Delta$ is the
  pixel size, $\sigma_B$ and $\sigma_R$ are the dispersions in the mean
  fluxes within the blue and red band-passes respectively. The
  parameters $\Lambda_{1,4}$ are defined as:

  \begin{equation}
    \label{eq:l1}
    \Lambda_1 = \frac{(\lambda_R - \lambda_i)(\lambda_R - \lambda_j)}
    {(\lambda_R - \lambda_B)^2}
  \end{equation}

  \begin{equation}
    \label{eq:l4}
    \Lambda_4 = \frac{(\lambda_i - \lambda_B)(\lambda_j - \lambda_B)}{%
    (\lambda_R - \lambda_B)^2}
  \end{equation}
  where $\lambda_B$ and $\lambda_R$ are the central wavelengths of the
  blue and red band-pasess, respectively.

  The dispersion $\sigma_i$ is assumed to be proportional, at each
  wavelength, to the mean fractional dispersion $\langle Q \rangle$
  defined by \citetalias{fan_iv} for each stellar group. The
  dispersions of the pseudo-continuum flux $\sigma_{C(\lambda_i)}$ and
  of the mean fluxes in the blue and red passbands $\sigma_B$ and
  $\sigma_R$ are calculated by error propagation. We refer to
  \cite{cardiel} for more details.

  \section{Fitting functions}
  \label{sec:ffunc}

  Our main objective is to study the behaviour of integrated $UV$
  absorption line indices of stellar population models, as functions of
  the stellar population parameters age ($t$) and metallicity
  (\ensuremath{[Z/\mathrm{H}]}). In order to include an empirical
  calibration in an evolutionary population synthesis code, one
  convenient approach followed by many authors is the construction of analytical approximations
  that describe the empirical line indices as functions of stellar
  parameters $(T_{\mathrm{eff}}, \ensuremath{\log g},
  \ensuremath{[Z/\mathrm{H}]})$ \citep[fitting functions, FF, e.g.][]{buzetal92, goretal93, worthey_lick, cenetal02, sch07}.
  
  We adopt the FF approach because FF can be easily incorporated into
  an evolutionary synthesis code in order to predict the integrated
  indices of stellar populations. Another advantage is that fluctuations in the spectra of stars with similar atmospheric parameters become averaged out.
   
  Since the observed stars have mostly solar chemical composition
  as a first step we build FFs that depend only on $T_{\mathrm{eff}}$ and \ensuremath{\log g}{}.
  We then incorporate metallicity effects by means of synthetic spectra
  (Sect.~\ref{sec:metcorr}).

It is important to point out that the specific ratios of various elements in this sample of stars may dominate the index values similarly to what has been found for optical lines \citep[e.g., ][]{buretal84, woretal92, tmb03} and this is a fact that one must bear in mind when the models are compared to extragalactic stellar populations. In this work we do not perform an element-ratio-sensitive modelling, which will be the subject of a future study.

  For the analytical form of the FFs we chose polynomials because
  these are the simplest functions, including, in principle all 
  possible terms (20 in total):

\begin{equation}
\label{eq:ff}
EW = \phi(\log T_{\mathrm{eff}},\ensuremath{\log g}) = \sum_{i=0}^3\sum_{j=0}^i a_{ij}\log T_{\mathrm{eff}}^i\ensuremath{\log g}^j.
\end{equation}

  In practice, in several cases it was not possible to accurately
  reproduce the index behaviour using only one function of the form
  given by equation~(\ref{eq:ff}). In these cases we split the
  temperature range into two regions (the ``cool'' region between
  effective temperatures $T_{a_1}$ and $T_{a_2}$, and the ``hot''
  region between effective temperatures $T_{b_1}$ and $T_{b_2}$)
  inside which the fitting procedure was carried out independently.
  Both regions are chosen to overlap and share several stellar groups
  in common. If $\phi_a(\log T_{\mathrm{eff}},\ensuremath{\log g})$
  and $\phi_b(\log T_{\mathrm{eff}},\ensuremath{\log g})$ are the
  local FFs corresponding to the first and the second region, defined
  respectively in the intervals $\log T_{\mathrm{eff}} \in
  [T_{a_1},T_{a_2}]$ and $\log T_{\mathrm{eff}} \in [T_{b_1},T_{b_2}]$
  (with $T_{b_1} < T_{a_2}$), the interpolated FF $\phi_{ab}(\log
  T_{\mathrm{eff}},\ensuremath{\log g})$ is defined as
  \citep{cenetal02}:

  \begin{equation}
  \label{eq:interp}
   \begin{array}{ccl}
      \phi_{ab}(\log T_{\mathrm{eff}},\ensuremath{\log g})& = & \omega \phi_a(\log T_{\mathrm{eff}},\ensuremath{\log g}) \\
       & + & (1 - \omega)\phi_b(\log T_{\mathrm{eff}},\ensuremath{\log g})
   \end{array}
  \end{equation}
  where $\omega$ is the weighting factor and is defined as:
 
	\begin{equation}
		\label{eq:omega}
		\omega = \cos^2\left(\frac{\pi}{2}\frac{\log T_{\mathrm{eff}} -
		\log T_{b_1}}{\log T_{a_2}-\log T_{b_1}}\right)
	\end{equation}

  The continuity of the FF is ensured by imposing that $\phi_{ab}(T_{a_2},
  \ensuremath{\log g}) = \phi_{b}(T_{a_2},\ensuremath{\log g})$ and that\hspace{-.1em}
  $\phi_{ab}(T_{b_1},\ensuremath{\log g})=\phi_{a}(T_{b_1},\ensuremath{\log g})$.
  
  The polynomial coefficients $a_{ij}$ are computed through iterative
  linear fitting, discarding in each step the statistically least
  significant coefficients. These are determined by their $p$-value,
  which is the probability that the true coefficient has a value
  greater than or equal to the computed one strictly by chance. We
  consider as significant coefficients those with a $p$-value less
  than or equal to 5\%. The fitting is recomputed after dropping the
  non-significant coefficients and the statistical significance of the
  remaining coefficients is re-evaluated. The procedure stops once all
  coefficients are statistically significant.

  The choice of the limits defining the two regions where local FFs
  are computed is arbitrary, therefore the best combination is found
  by trial and error, evaluating the effectiveness of the FFs in
  describing the global index behaviour. An important parameter we take
  into account is the lack of evident trends in the distribution of
  the residuals. In the case of existence of such trends, the local limits
  were re-adjusted or we inspected the individual group's spectra in
  order to understand if the deviations are due to particular features
  in them. 

  For each index the best fit parameters are provided together with
  the validity ranges ($3.6\lesssim
  {T_{\mathrm{eff}}/\mathrm{kK}}$$\lesssim47.5$; and
  $-0.1\lesssim\ensuremath{\log g}\lesssim4.66$), which are
  fundamental to avoid the raising of spurious results as a product of
  extrapolating the FFs outside the limits in which they where derived
  and inside which the statistical significance is guaranteed
  \citep[see][for a discussion on this issue and
  examples]{claudia_cl}.

  \subsection{Summary of index behaviour with stellar parameters}
  \label{sec:trends}

  The trends of the various line indices in the IUE sample stars as
  functions of the stellar parameters $T_{\mathrm{eff}}$,
  \ensuremath{\log g}{} and metallicity is discussed in detail by
  \cite{massa,walborn_siiv,fan_i}; \citetalias{fan_iv}.  Some of the mid-$UV$ indices 
 by Fanelli et al. have also been studied in theoretical Kurucz stellar spectra by 
 \citet{chaetal07}.
   
  Below we summarise the relevant conclusions from these works.

  In the far $UV$ the most prominent absorption features are
  \ensuremath{\mathrm{BL_{\scriptstyle{1302}}}},
  \ensuremath{\mathrm{BL_{\scriptstyle{1664}}}}, \ion{Si}{iv}
  (1400\,\AA) and \ion{C}{iv} (1550\,\AA). 

  \ensuremath{\mathrm{BL_{\scriptstyle{1302}}}} is a blend of several
  \ion{O}{i} (1302.704, 1304.858, 1306.023\,\AA), \ion{Si}{iii}
  (1298.90, 1303.30\,\AA) and \ion{Si}{ii} (1304.41\,\AA)
  \citetext{\citealt{coluzzi}; \citetalias{fan_iv}; \citealt{moore}} lines. It reaches a maximum for late B
  stars, decreasing with increasing temperature, with very little
  dependence on gravity. Similarly, \ion{Si}{iv} and \ion{C}{iv} reach
  maxima around early B - late O spectral types, but show a stronger
  gravity dependence. \ensuremath{\mathrm{BL_{\scriptstyle{1425}}}} is a
  blend of several \ion{Fe}{v}, \ion{C}{iii} and \ion{Si}{iii} lines; it
  reaches a maximum for early B spectral types ($\sim30\,$kK), but its
  temperature dependence is very weak. The blend of \ion{Fe}{v} lines at
  1453 \AA{} shows instead a stronger temperature dependence, with early
  O stars displaying the highest equivalent widths and practically
  vanishing for mid to late B stars.
  \ensuremath{\mathrm{BL_{\scriptstyle{1617}}}} is a blend of
  \ion{Fe}{iv} and \ion{Fe}{v} lines, which is strongest for mid O
  stars. \ensuremath{\mathrm{BL_{\scriptstyle{1664}}}} already shows the
  behaviour typical of mid-$UV$ absorption indices. It measures the
  combined absorption by the resonant lines of \ion{Al}{ii}
  (1670.8~\AA{}) and \ion{C}{i} (1656.9~\AA). It increases monotonously
  from  spectral types around B5 up to early A starts
  ($T_{\mathrm{eff}}\sim10$ kK).
  \ensuremath{\mathrm{BL_{\scriptstyle{1719}}}} is a blend of
  \ion{N}{iv} (1718.6~\AA), \ion{Si}{iv} (1722.5~\AA) and several
  \ion{Fe}{iv} multiplets. It shows a remarkable lack of temperature
  sensitivity, which was observed first by \citet{underhill}, and a mild
  surface gravity dependence. The last among the far-$UV$ indices is
  \ensuremath{\mathrm{BL_{\scriptstyle{1853}}}}, a blend of several
  \ion{Fe}{ii}, \ion{Al}{ii} and \ion{Al}{iii} transitions. Its
  behaviour is similar to that of
  \ensuremath{\mathrm{BL_{\scriptstyle{1664}}}}, with a strong increase
  for spectral types cooler than A0 ($T_{\mathrm{eff}}\sim10$ kK), but
  with better surface gravity separation.

  In the mid-$UV$, the most prominent absorptions features are the
  \ion{Mg}{ii}~2800\,\AA{}, the \ion{Fe}{i}~3000\,\AA{} which trace four
  neutral magnesium lines (2966.901; 2984.7302; 2999.8092 and 3021.37049
  \AA{}), the two iron \ion{Fe}{ii} features at 2402 and 2609\,\AA{} and
  \ion{Mg}{i} at 2852~\AA{}, respectively. These indices show the same
  qualitative behaviour, being close to zero until late B spectral
  types, then increasing towards earlier types reaching maxima around
  late F and early G spectral type stars. The magnesium \ion{Mg}{ii}~2800\,\AA{} shows some peculiar behaviour as a function of metallicity (Fanelli et al. 1990), and we postpone its discussion to Section 6.3.
  \ensuremath{\mathrm{BL_{\scriptstyle{2538}}}} is a blend of
  \ion{Fe}{i} and \ion{Mg}{i} which starts to rise strongly for stars
  cooler than early A-type and reaches its maximum at late F stars
  ($T_{\mathrm{eff}}\sim6$~kK), with a very clear gravity segregation
  (dwarfs have equivalent widths nearly twice as high as supergiants at
  the peak). A similar luminosity-class separation is displayed by
  \ion{Fe}{i}i{} 2402 and \ion{Fe}{i}{} 3000.
  \ensuremath{\mathrm{BL_{\scriptstyle{3096}}}}, a blend of \ion{Fe}{i}
  and \ion{Al}{i} lines, does not reach a maximum in our explored
  parameter space, showing a monotonic increase with decreasing
  temperature.

  \begin{figure*}[!htp]
     \centering%
      \includegraphics[width=0.25\hsize,clip]{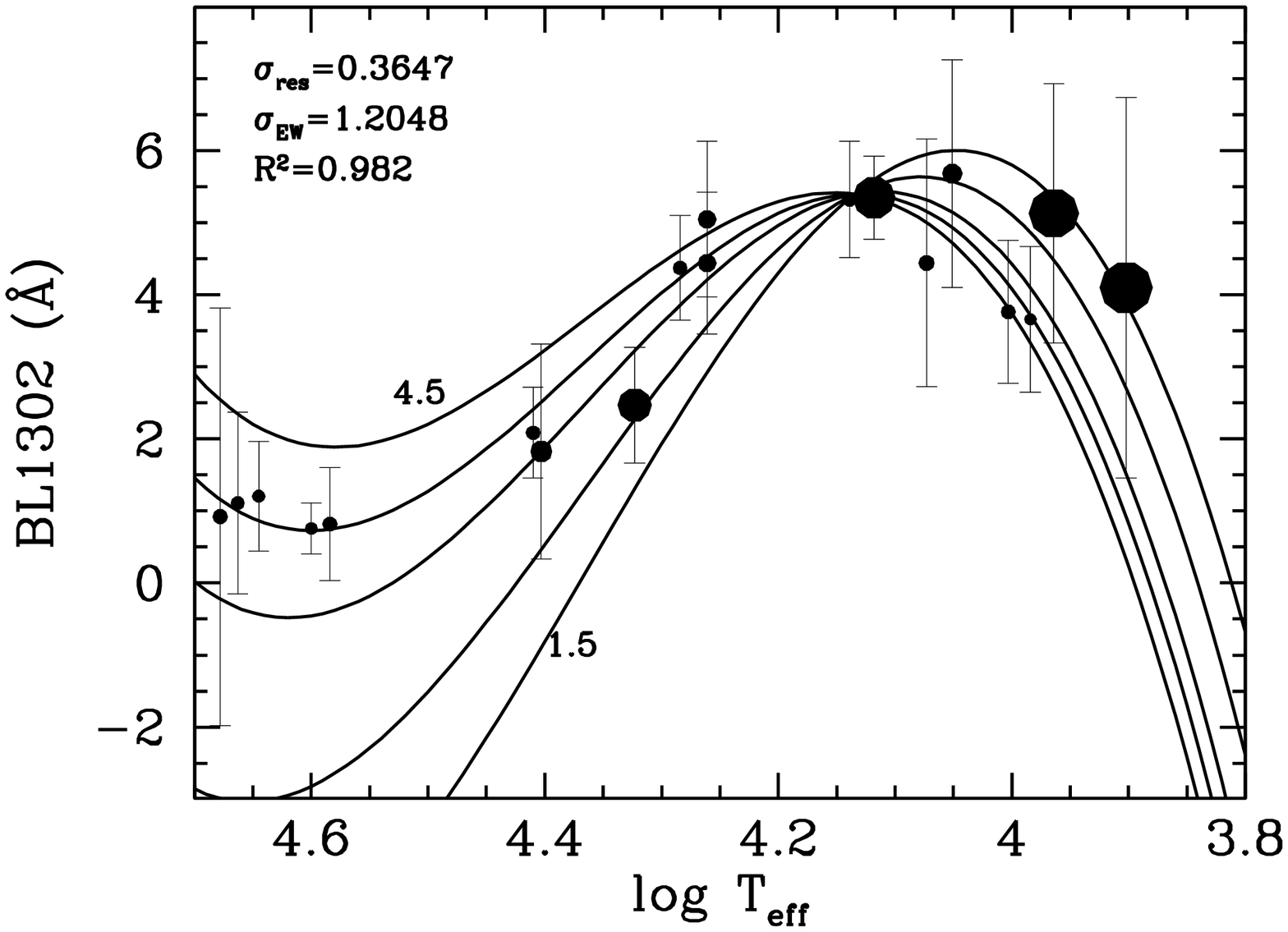}
      \includegraphics[width=0.25\hsize,clip]{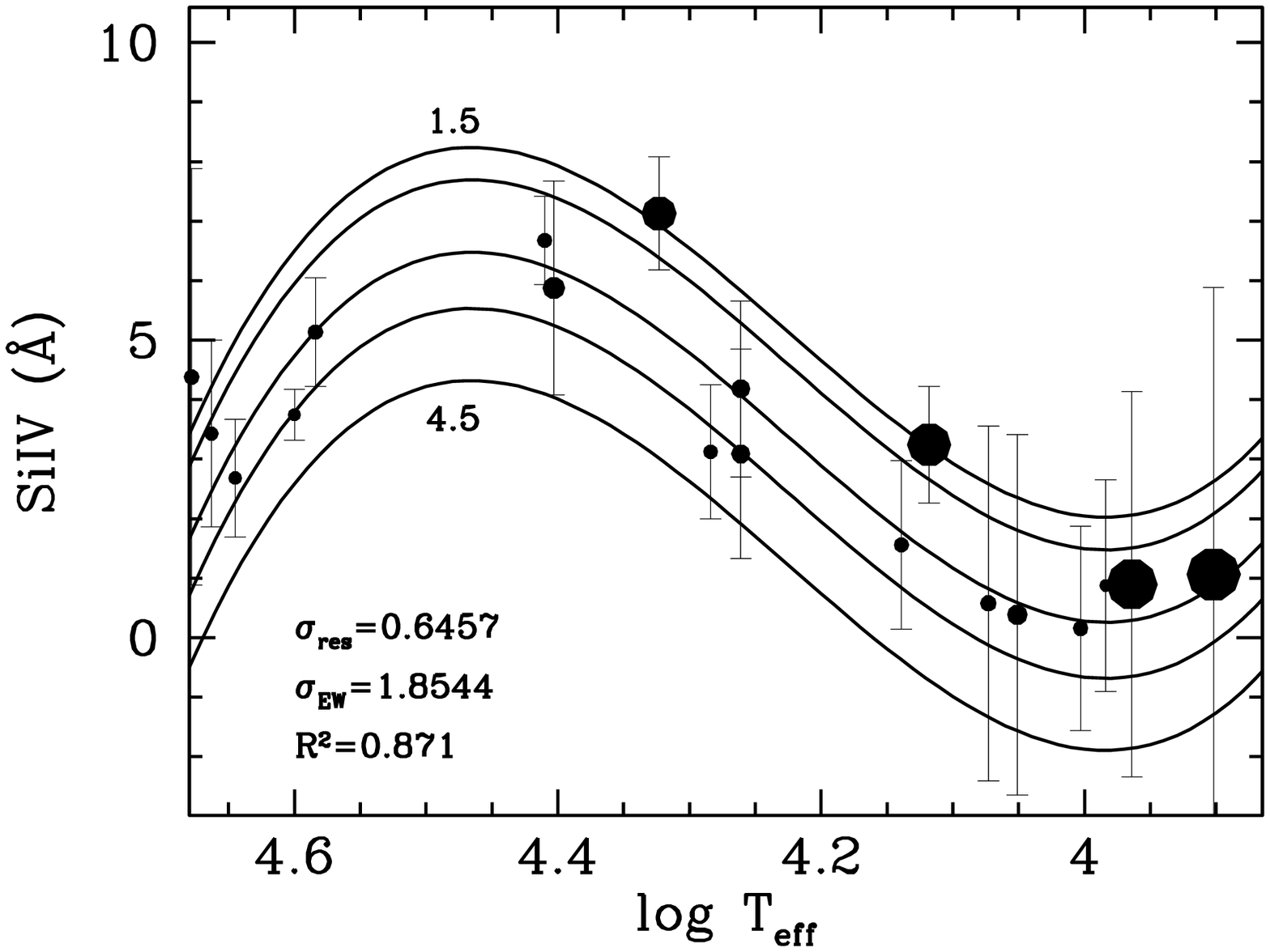}
      \includegraphics[width=0.25\hsize,clip]{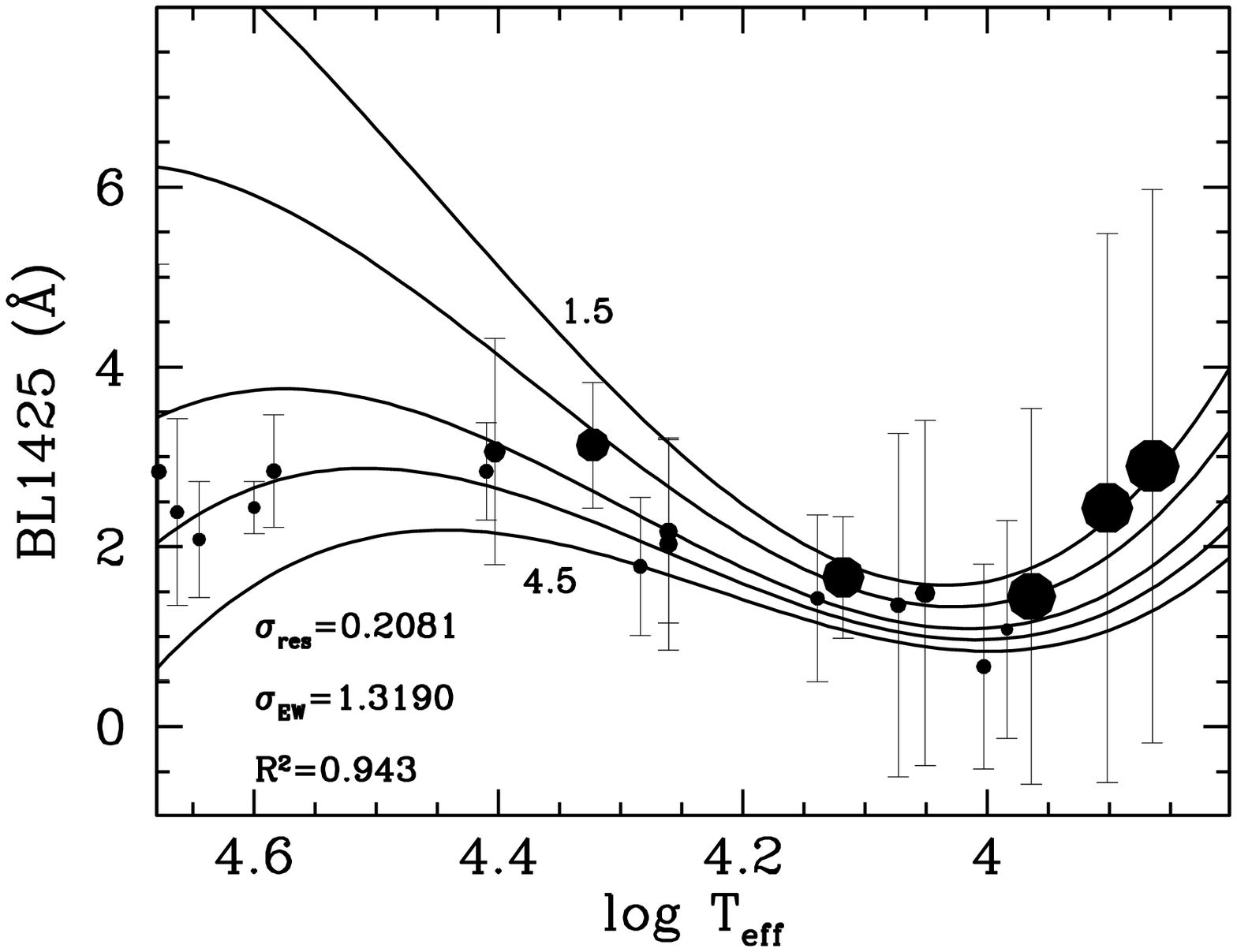}
      \includegraphics[width=0.25\hsize,clip]{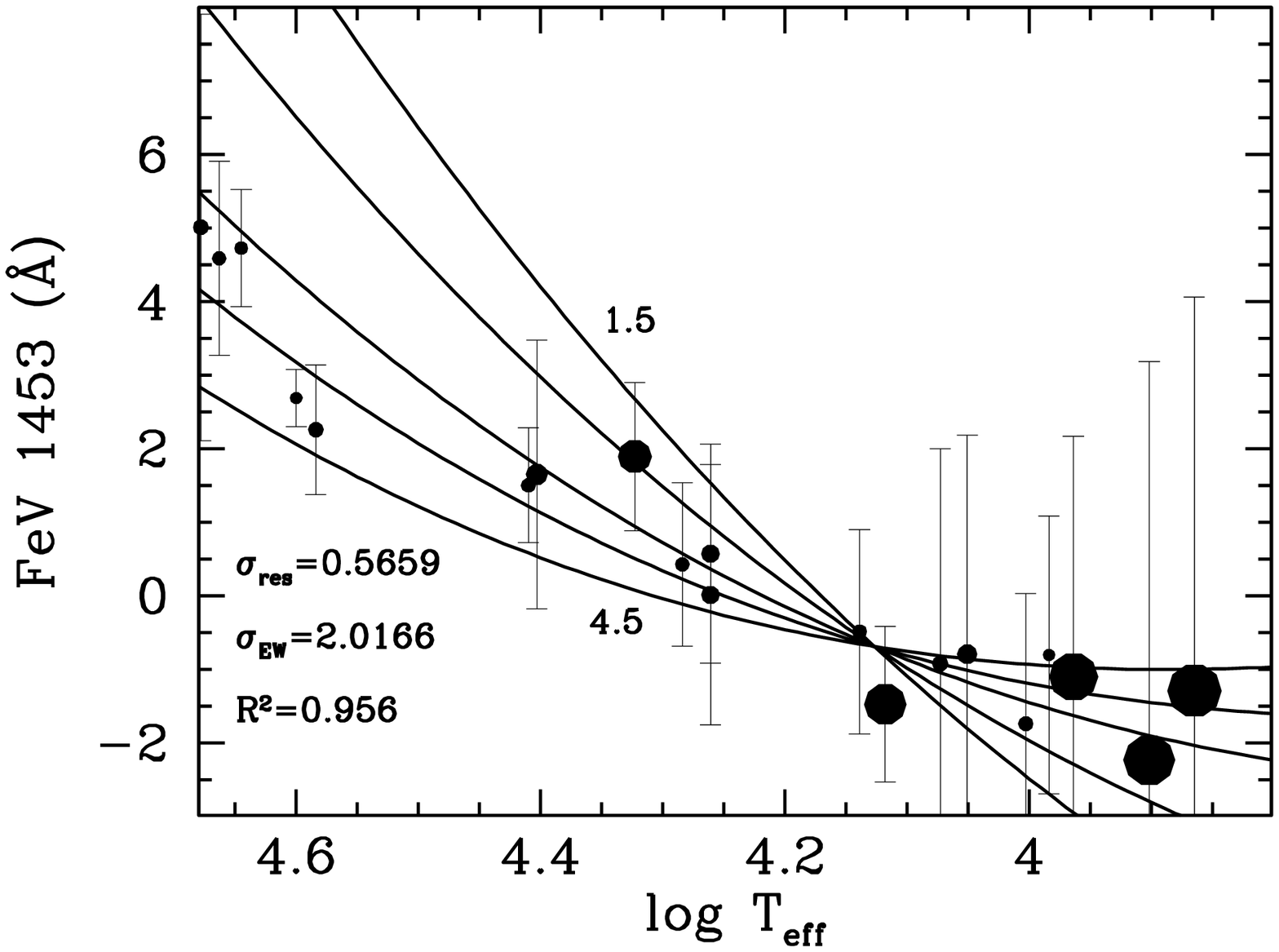}
      \includegraphics[width=0.25\hsize,clip]{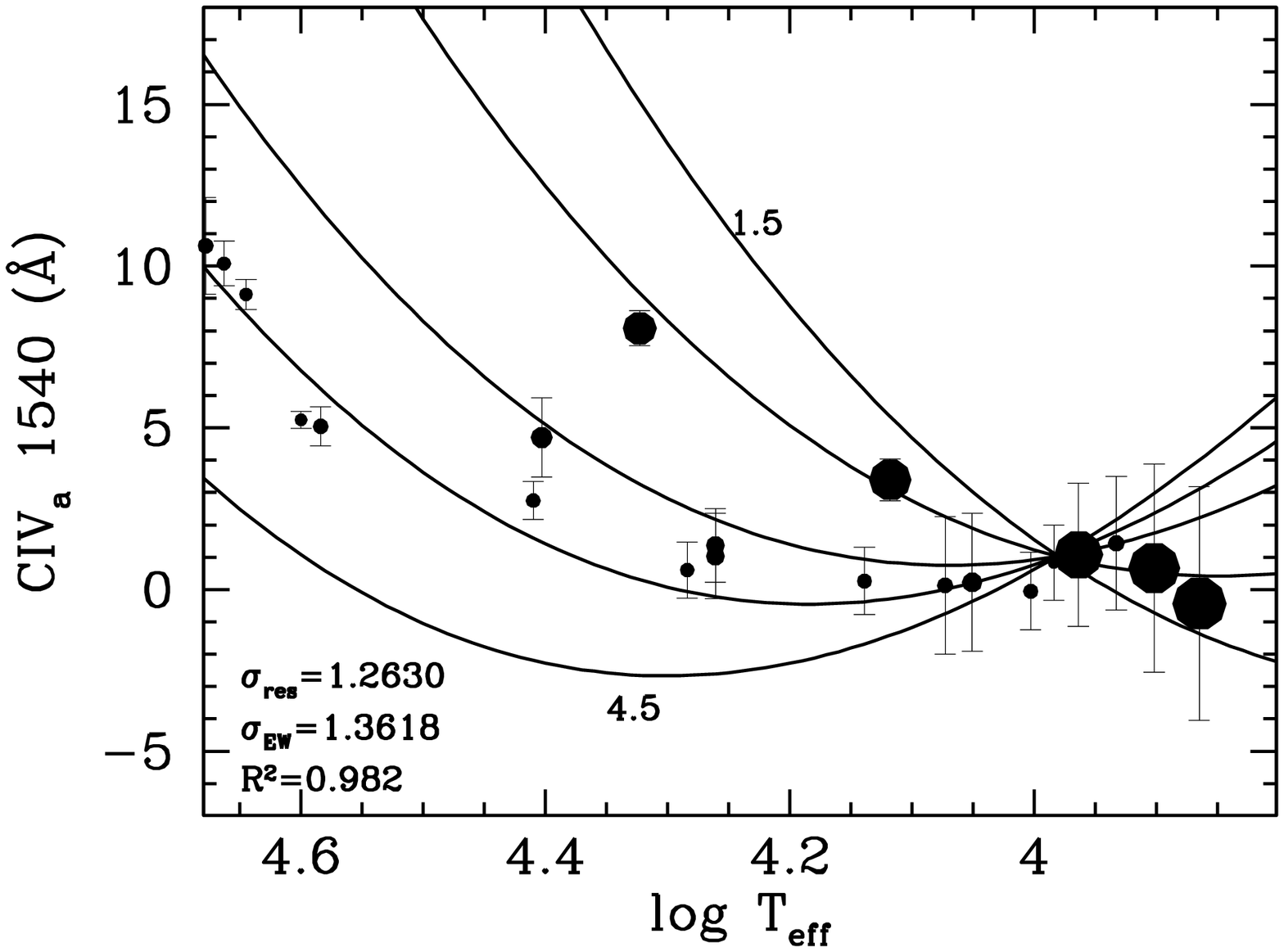}
      \includegraphics[width=0.25\hsize,clip]{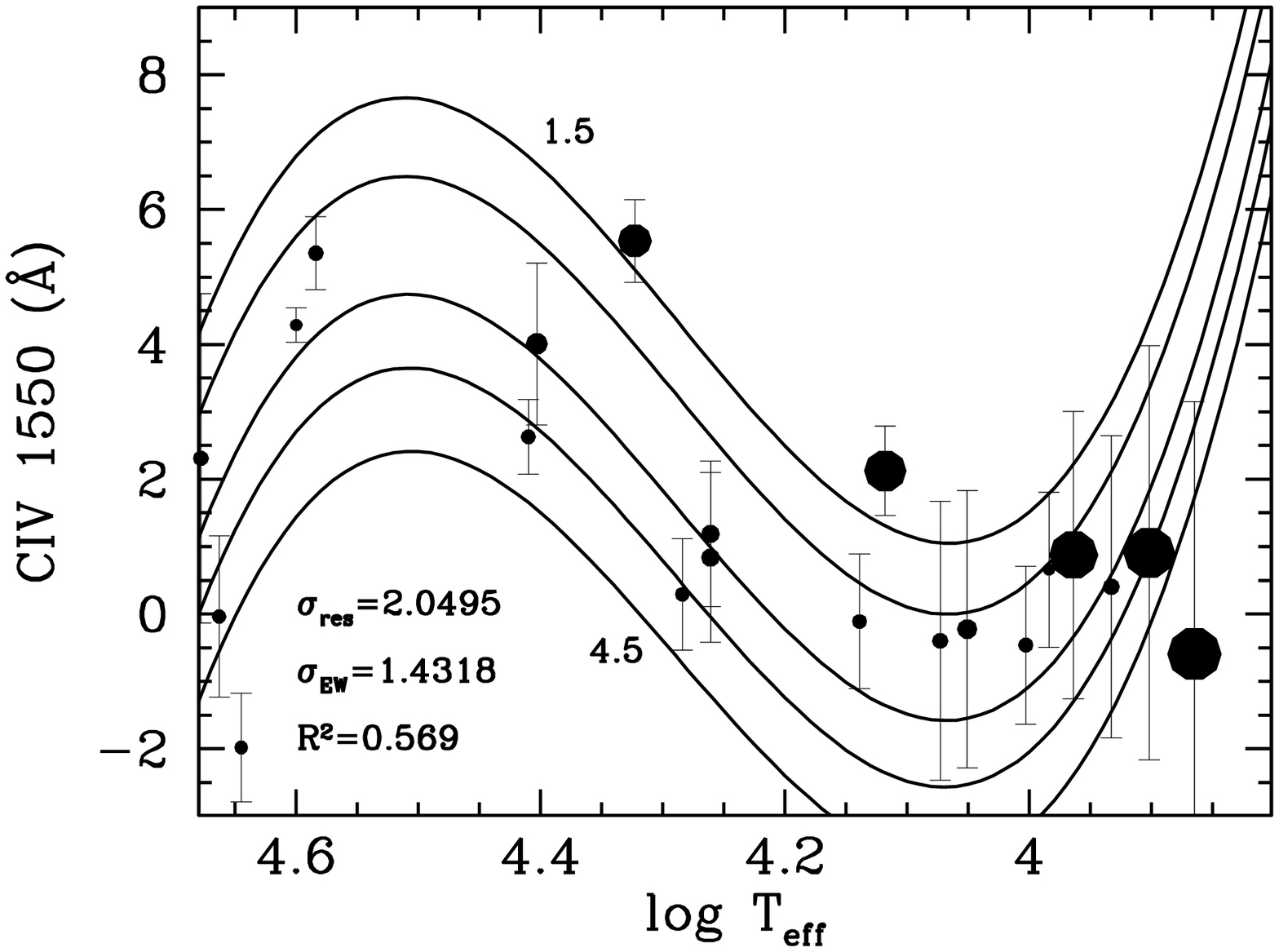}
      \includegraphics[width=0.25\hsize,clip]{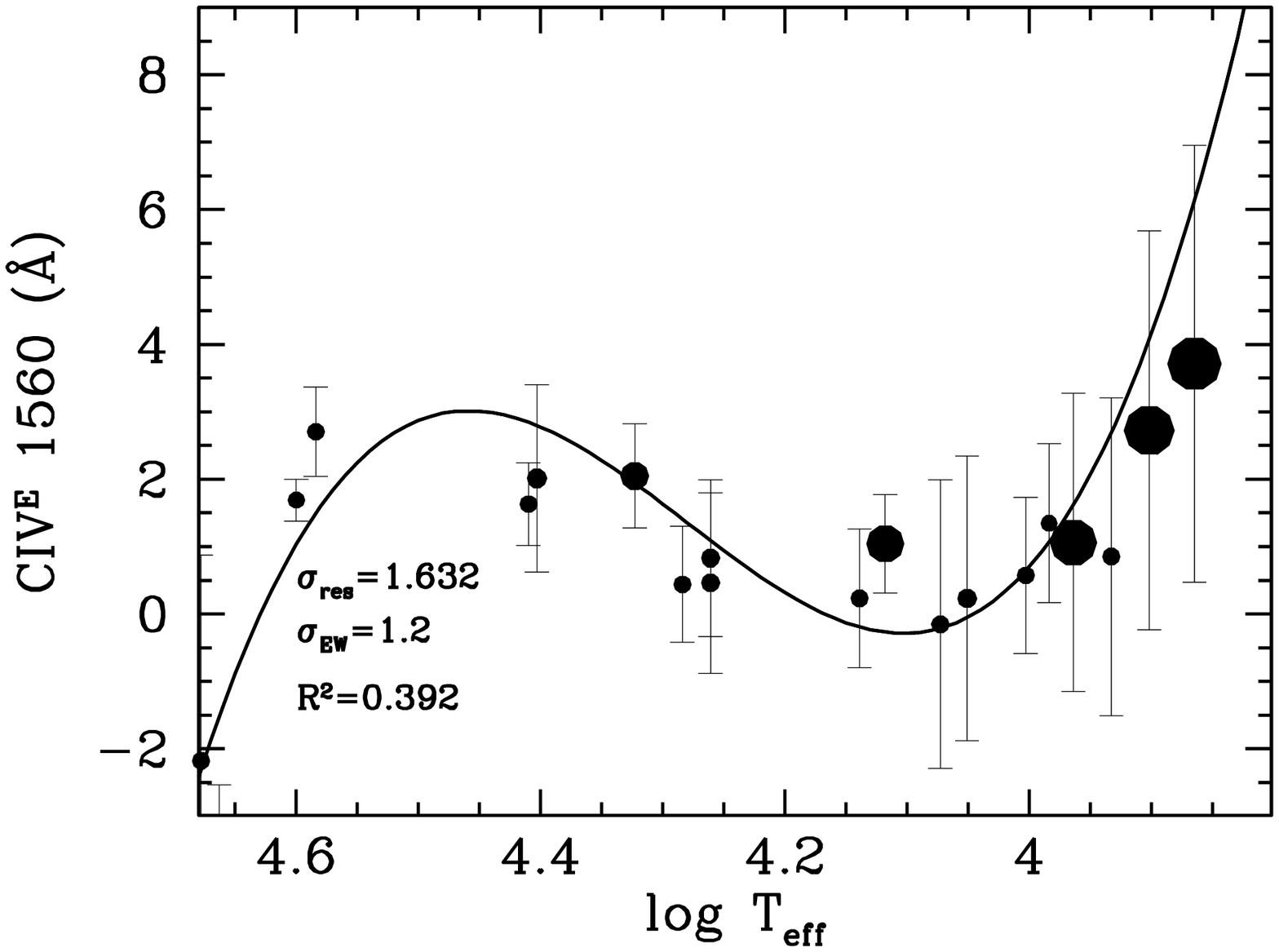}
      \includegraphics[width=0.25\hsize,clip]{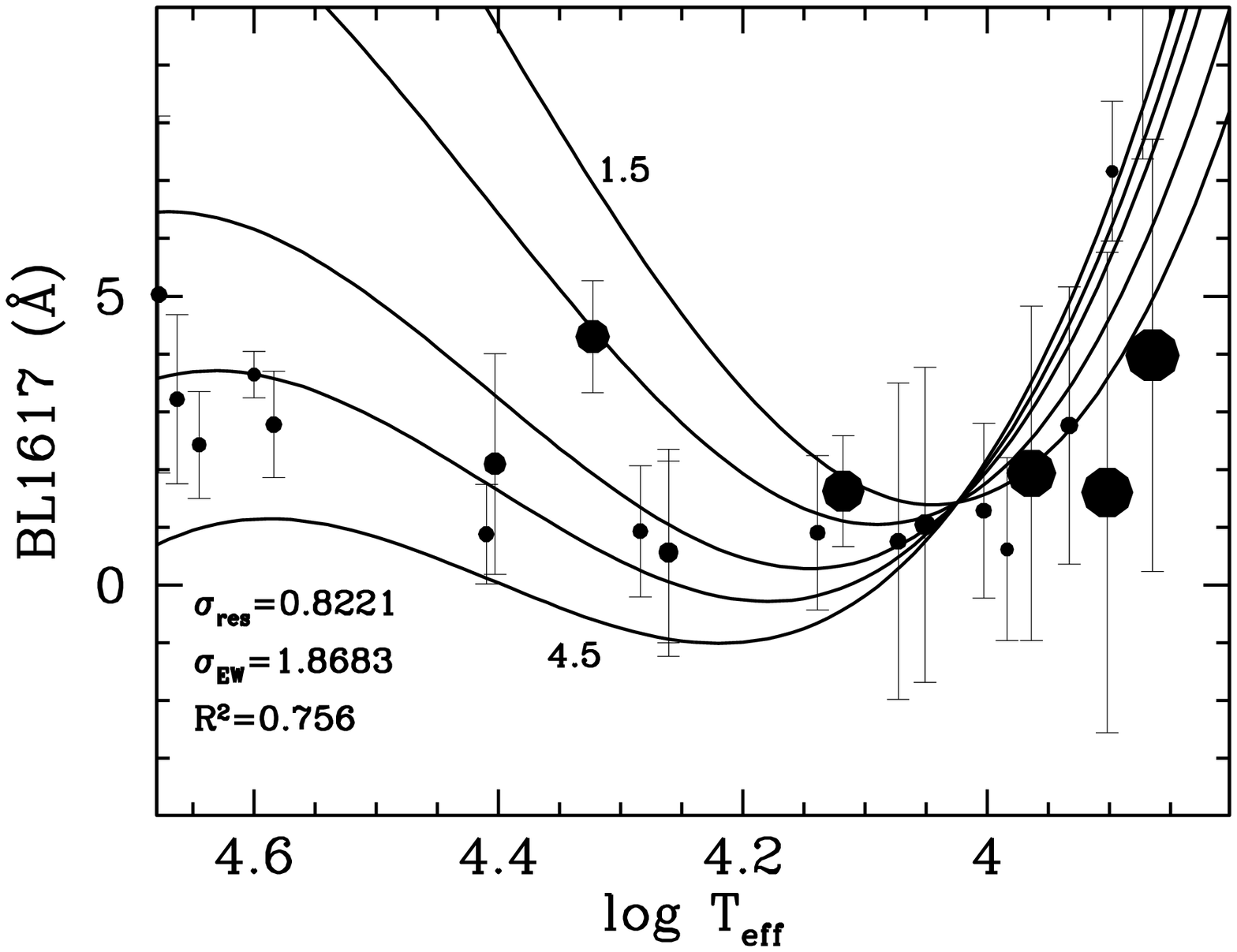}
      \includegraphics[width=0.25\hsize,clip]{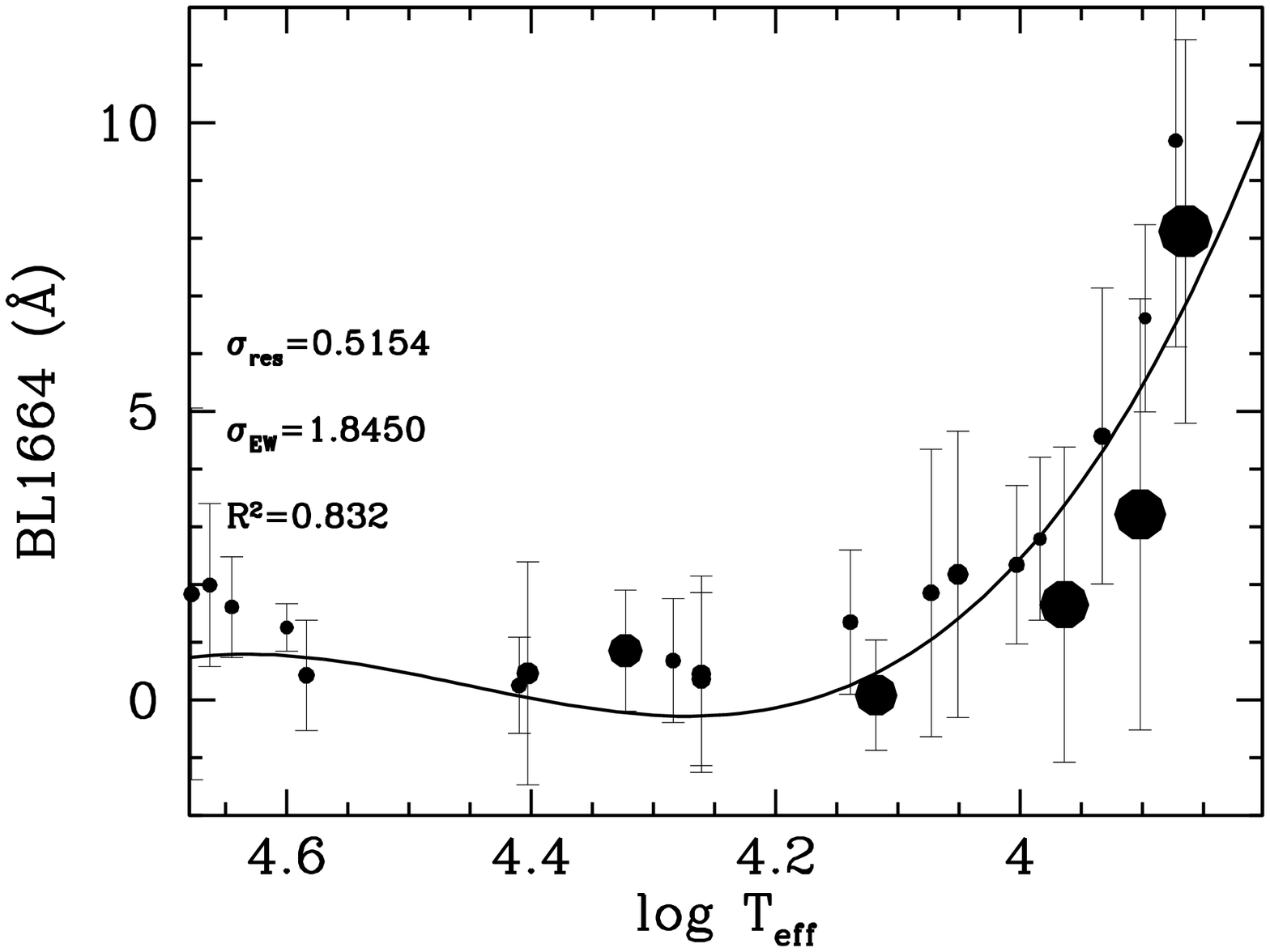}
      \includegraphics[width=0.25\hsize,clip]{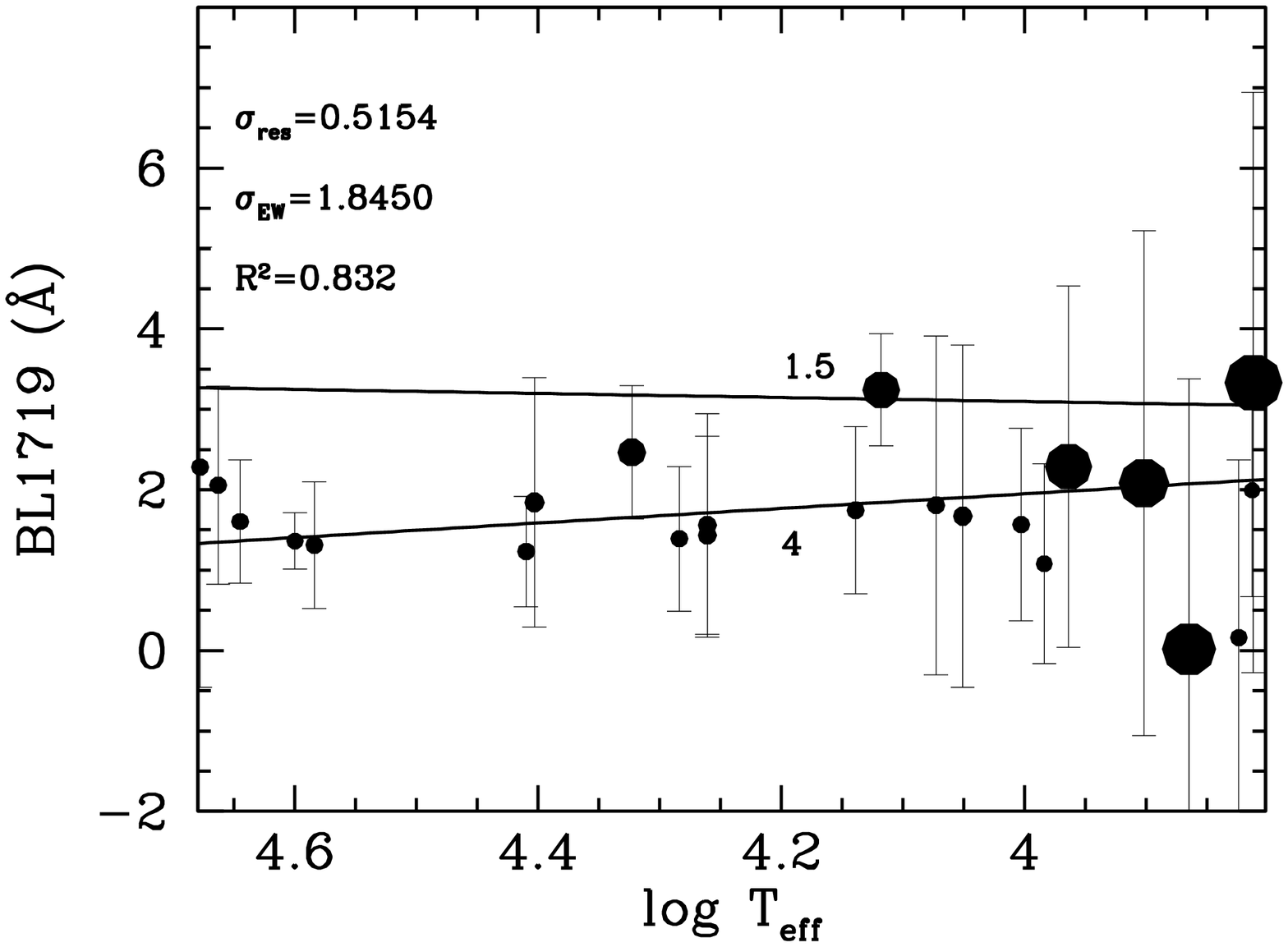}
       \includegraphics[width=0.25\hsize,clip]{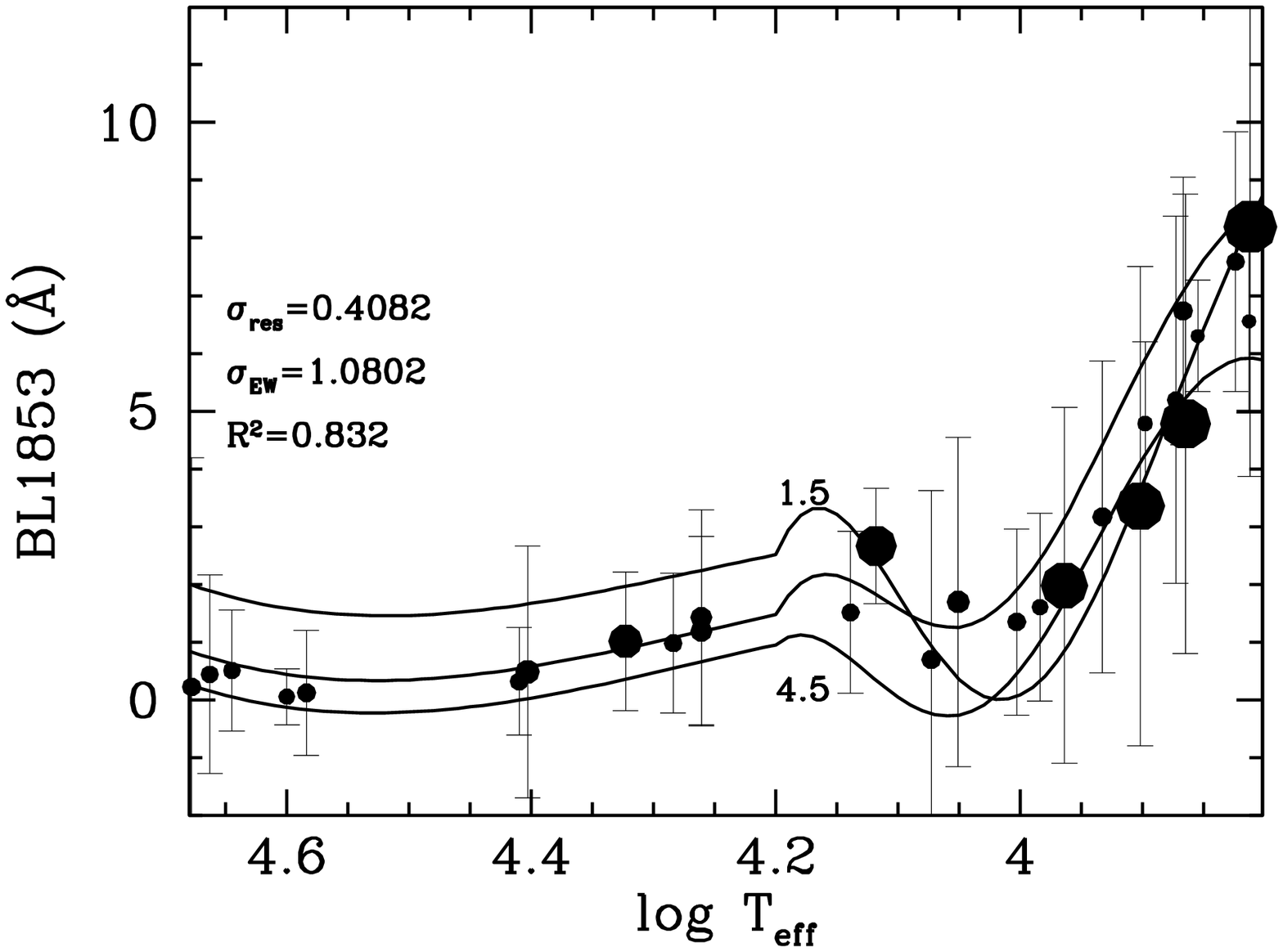}
      \includegraphics[width=0.25\hsize,clip]{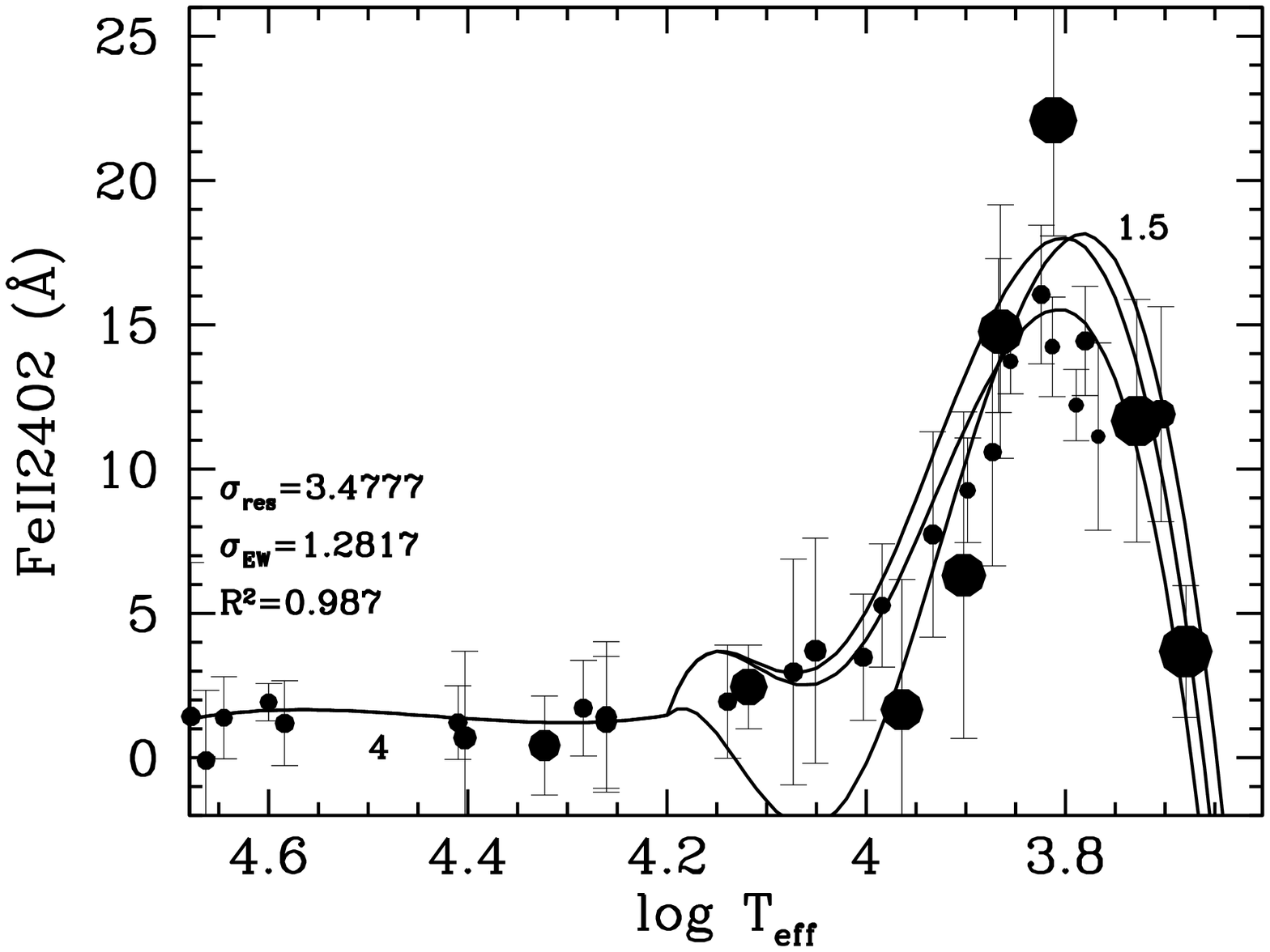}
       \includegraphics[width=0.25\hsize,clip]{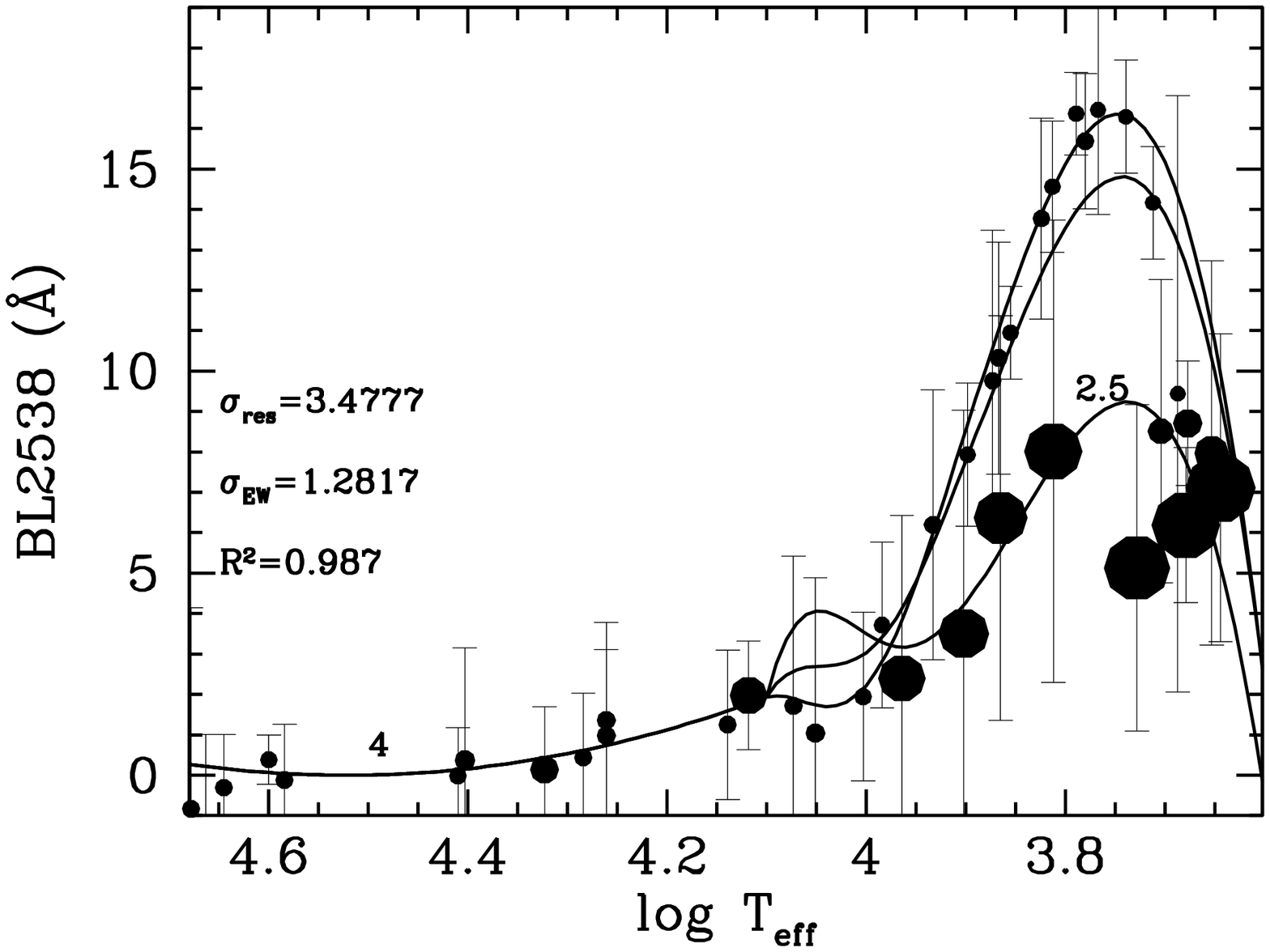}
      \includegraphics[width=0.25\hsize,clip]{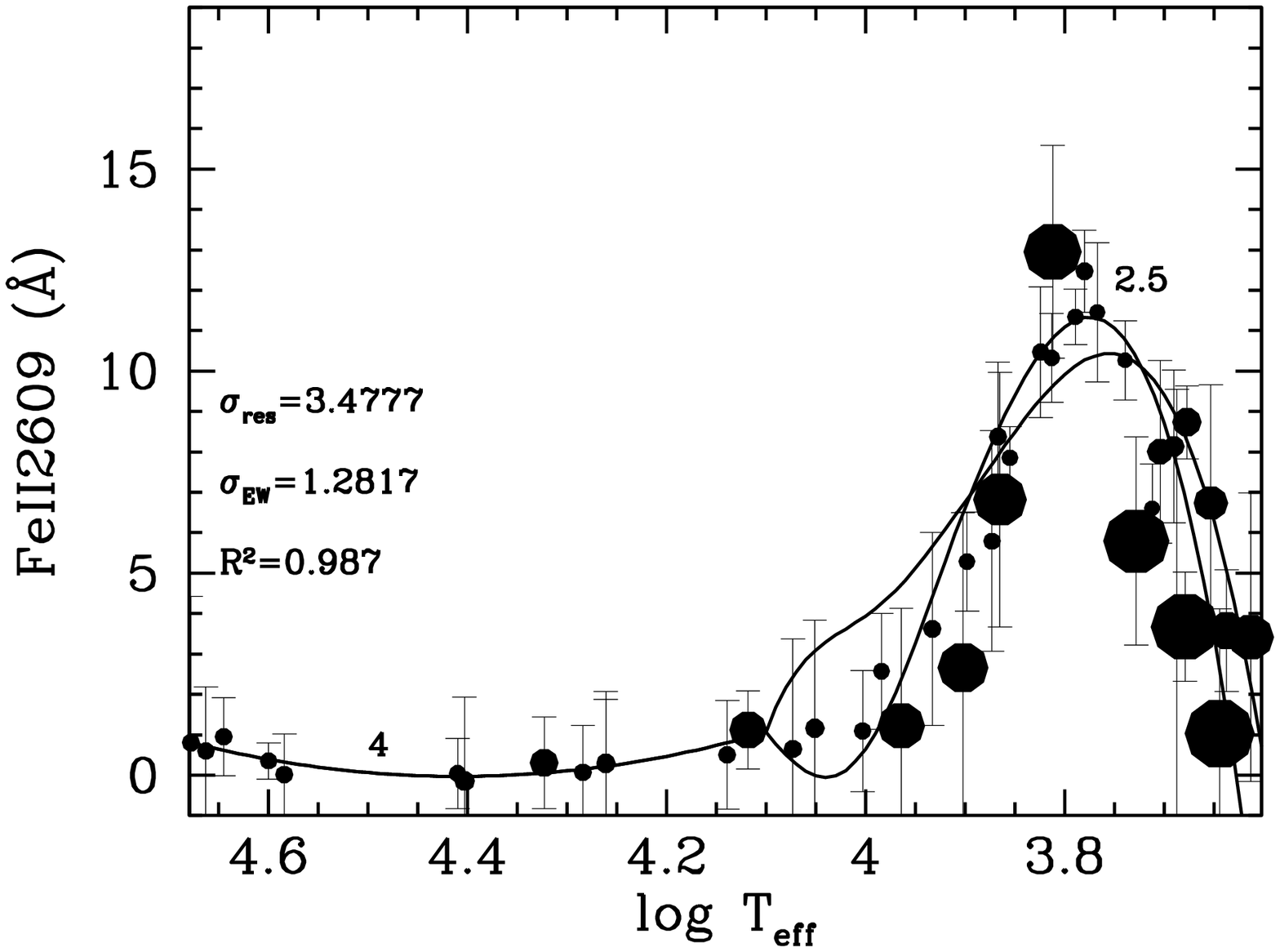}
      \includegraphics[width=0.25\hsize,clip]{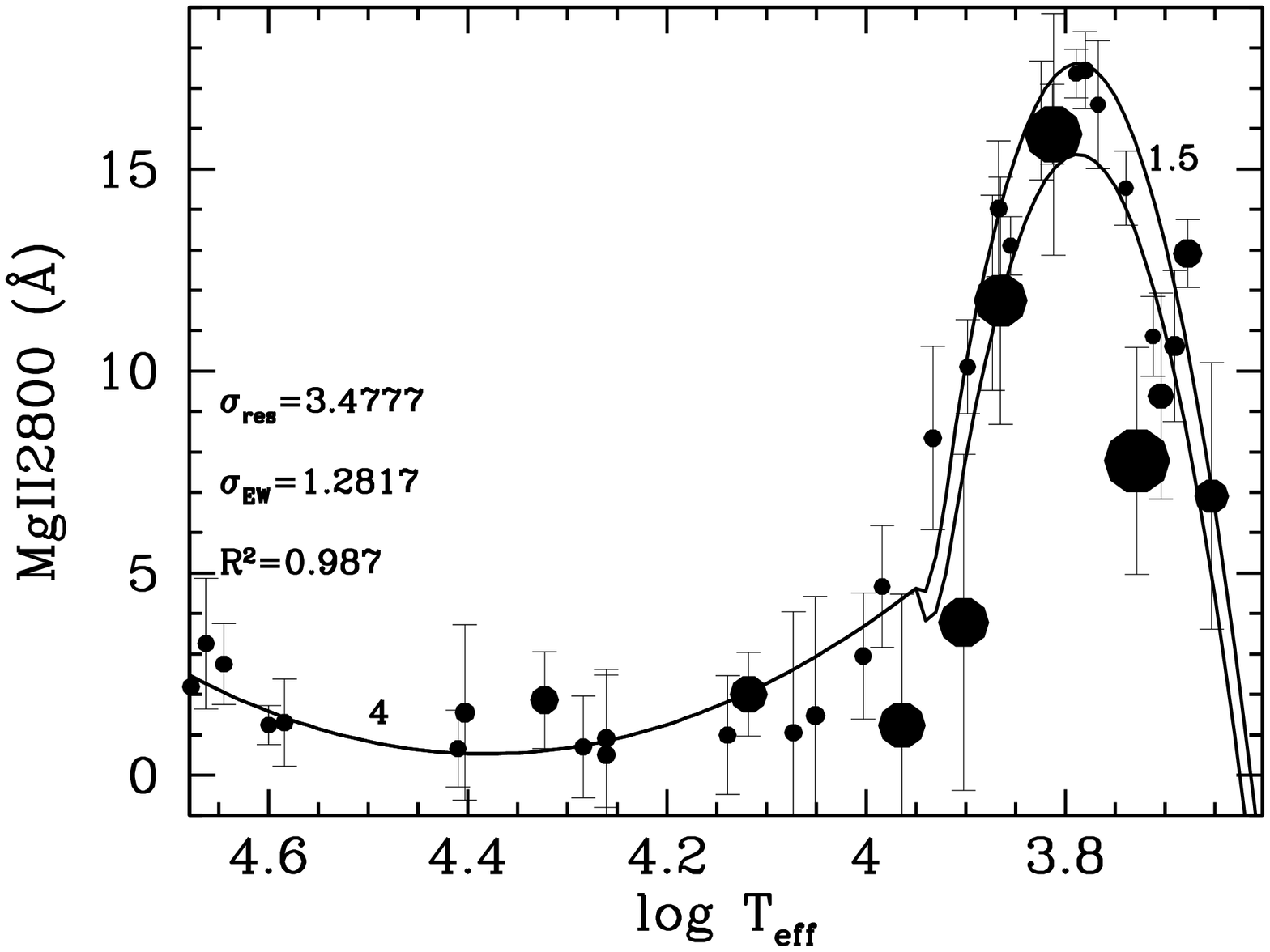}
      \includegraphics[width=0.25\hsize,clip]{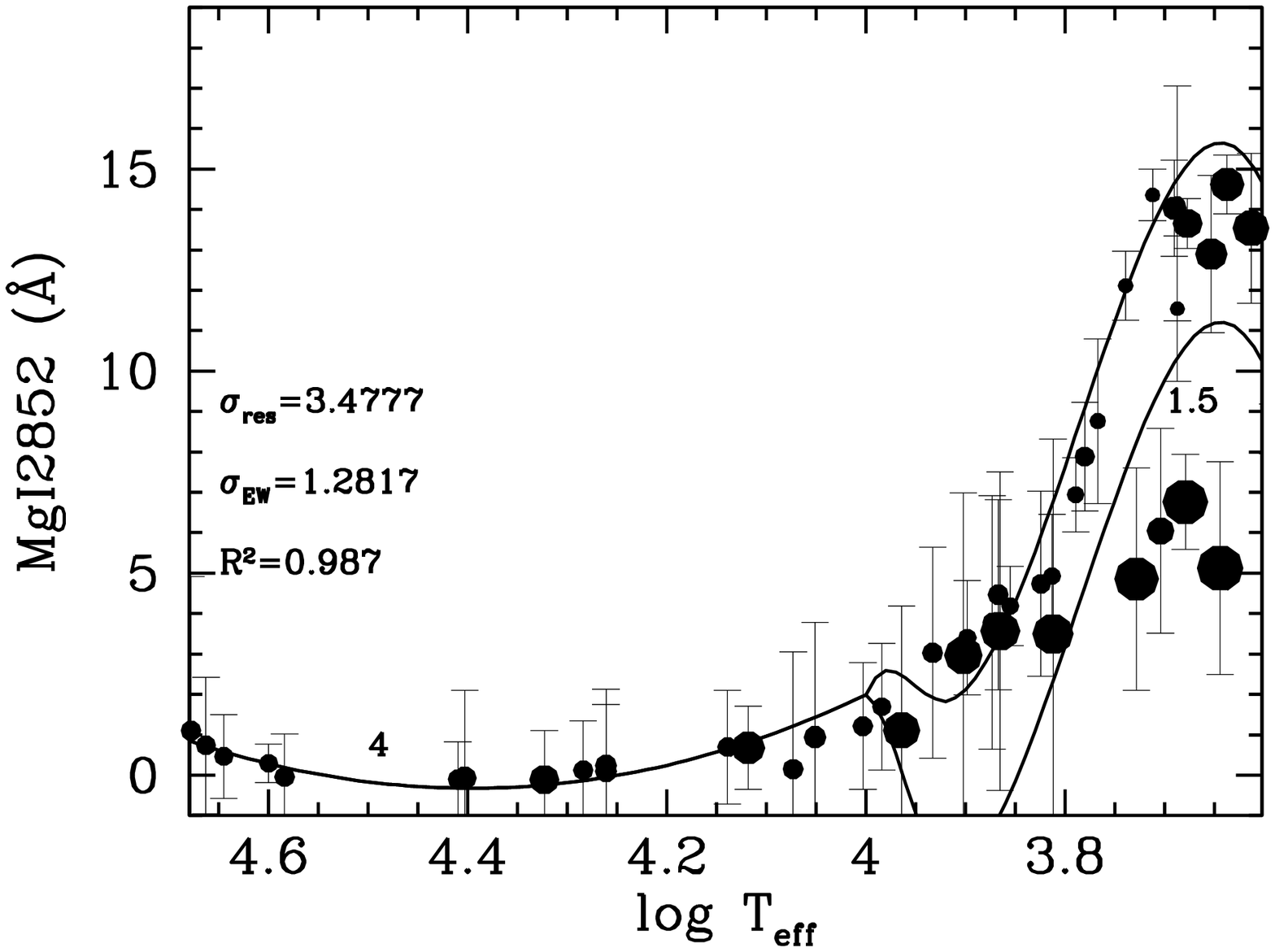}
      \includegraphics[width=0.25\hsize,clip]{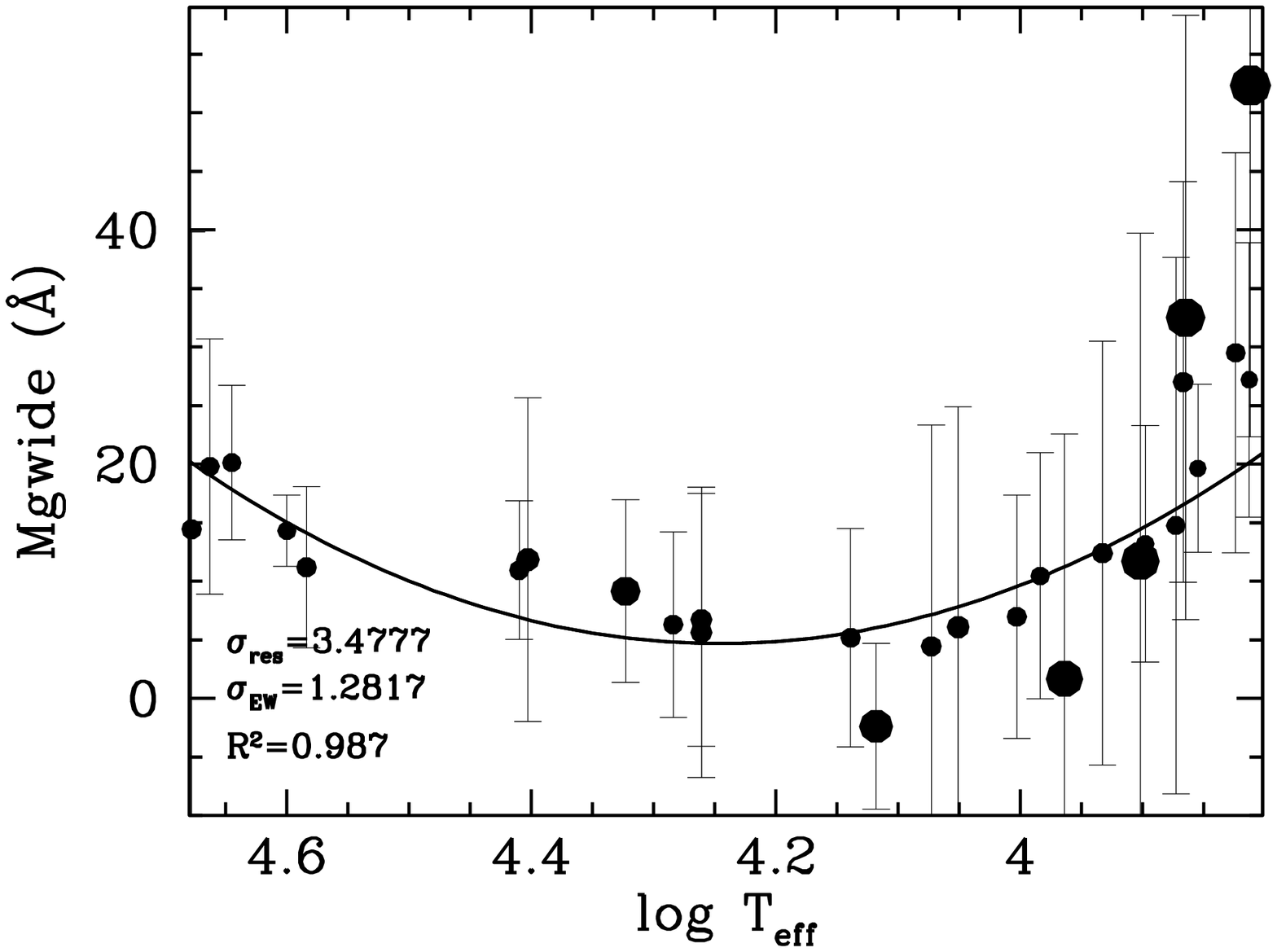}
      \includegraphics[width=0.25\hsize,clip]{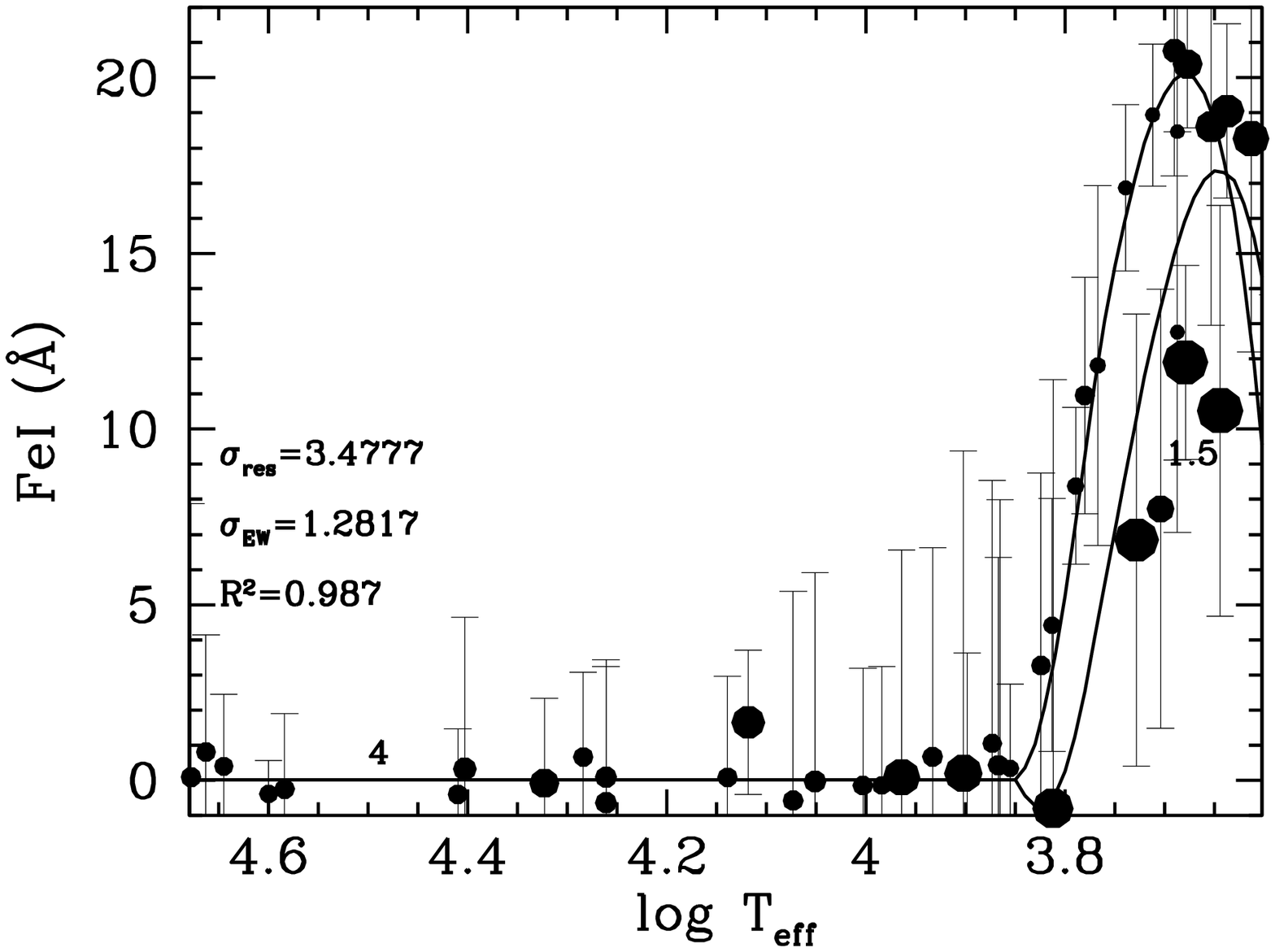}
      \includegraphics[width=0.25\hsize,clip]{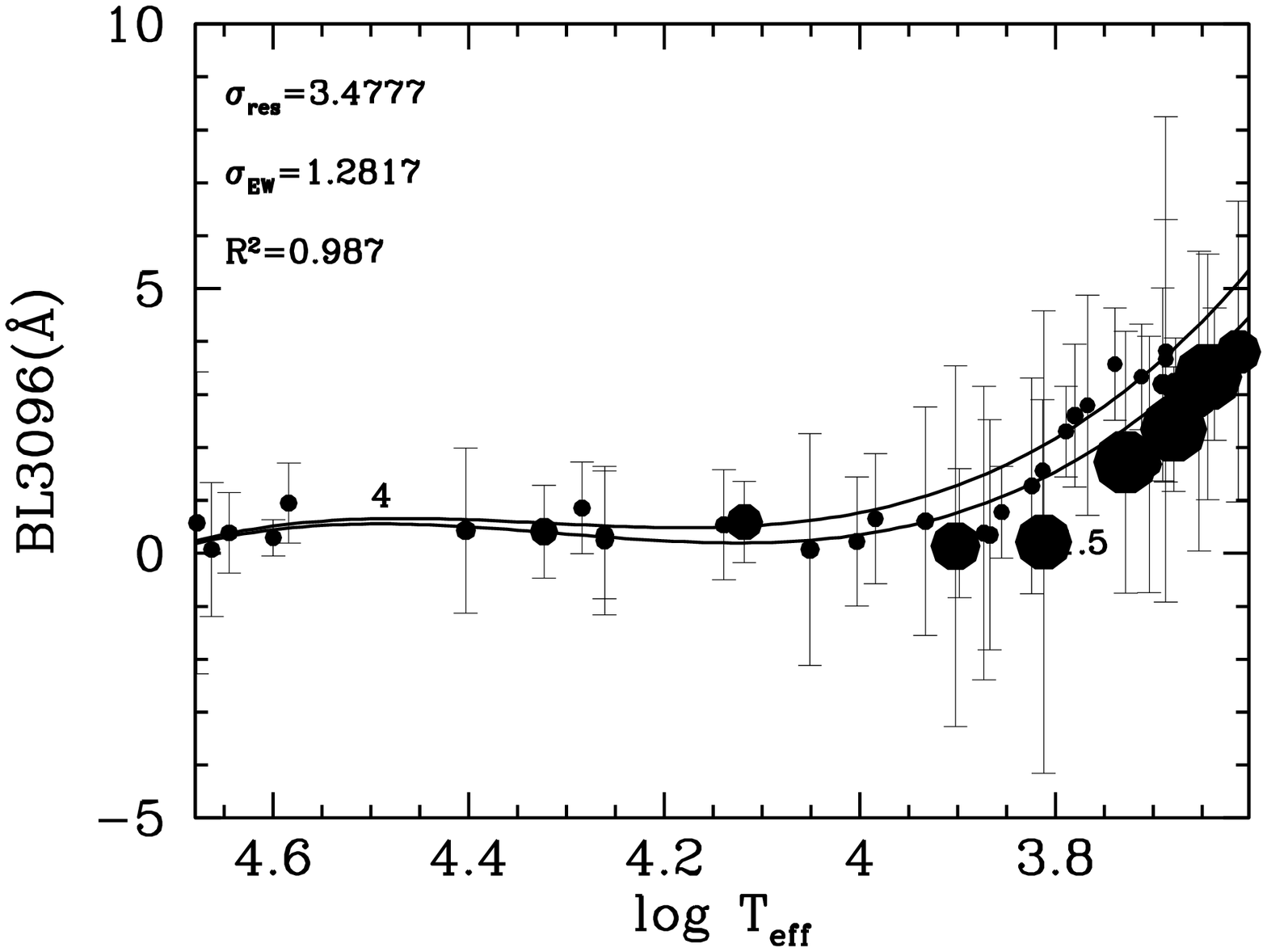}
            \caption{Fitting functions. Stellar EWs are shown as functions of the
      effective temperature.  The surface gravity is encoded in the
      symbol size (large circles for giants, small circles
      for dwarfs), using a linear function with a slope of
      2.6 and a zero-point of 0.8. FFs for various gravities (see labels) are
      plotted.  The standard deviation of the residuals
      ($\sigma_{\mathrm{res}}$) and the mean equivalent width error
      $\overline{\sigma_W}$ of all the groups included the fit are also shown.}
      \label{fig:ff}
  \end{figure*}

 In Fig.~\ref{fig:ff} we show the fitting functions as a function of temperature for different gravities, overlaid on the data of the individual stars used in the fits. Coefficients of the polynomials are provided in the Appendix. 
  
In Section~6.5, we compare our FF-based models with other models using the Fanelli et al. library directly and find that the results are 
 consistent. This indicates that the fitting function procedure has not introduced spurious effects.

  \subsection{Inclusion of metallicity effects}
  \label{sec:metcorr}

  As shown in Fig.~\ref{fig:teffcomp}(b), the spectral library of
  \citetalias{fan_iv} is mostly composed of stars of solar
  metallicity. On the other hand, the consideration of metallicity
  effects on the integrated spectrophotometric properties of stellar
  populations is very important, as it allows us to gain insights into
  the galaxy formation process and the global star formation history
  of the universe.

 In order to include the dependence on metallicity in the FFs we use the library of high-resolution synthetic Kurucz spectra calculated by \cite{rodetal05} (see Section 5).

 We consider model spectra with $\ensuremath{\log g}=4$, since the most
  important contributors to the $UV$ light are turnoff stars and
  gravity does not vary appreciably around the Main Sequence
  turnoff.
 
  \begin{figure*}[!htp]
     \centering
       \includegraphics[width=\hsize,clip]{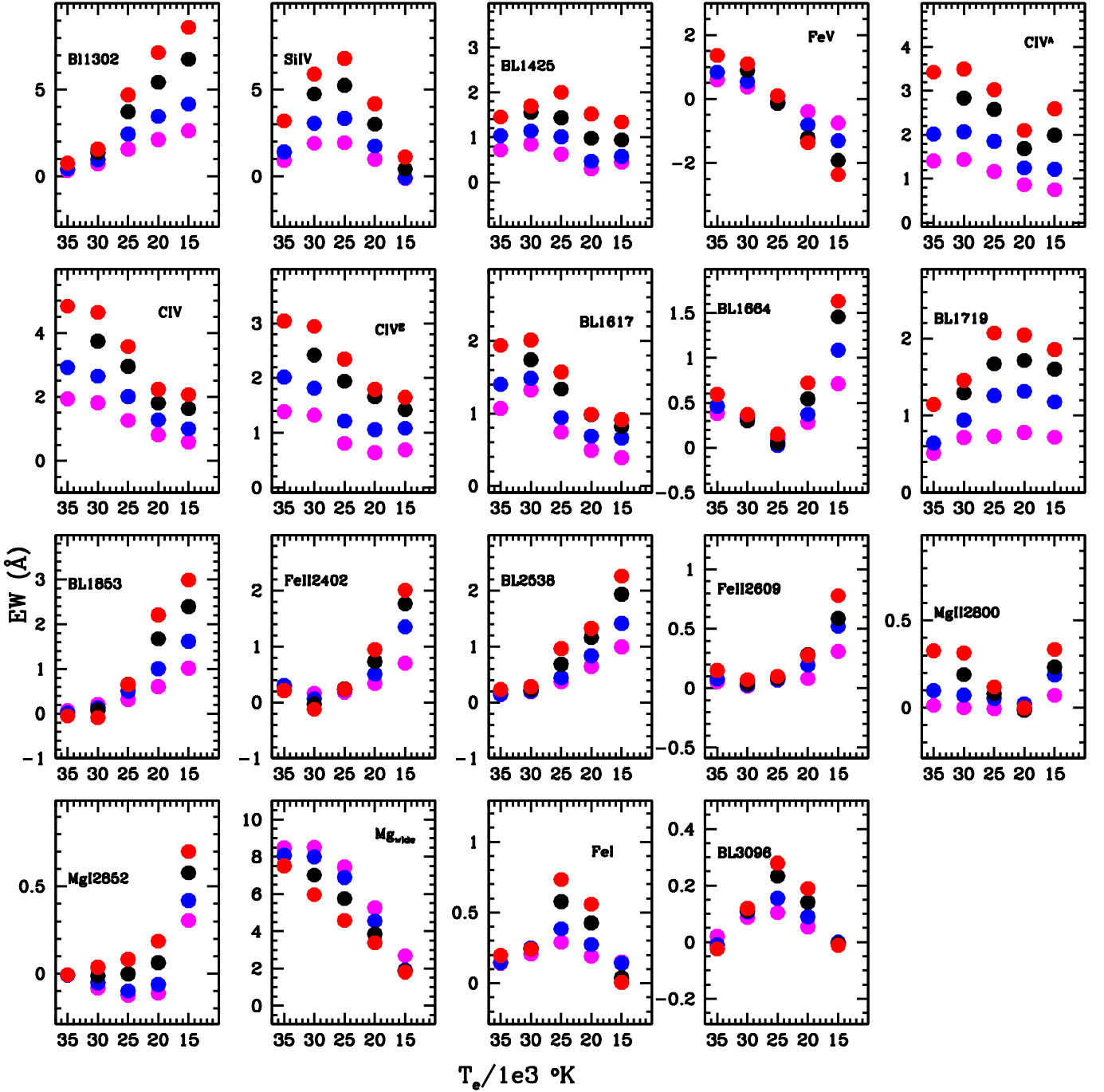}
     \caption{Metallicity dependence of individual indices calculated on the Kurucz high-resolution synthetic spectra (smoothed to the IUE resolution of $6~\AA$) with gravity $\ensuremath{\log g}=4$ and various temperatures. Colours code the metallicity, with red, black, blue and magenta displaying [Fe/H]=+0.3,  0, -0.5, -1, respectively.}
      \label{fig:kurucz_stars_metals}
  \end{figure*}
  
Figure~\ref{fig:kurucz_stars_metals} shows the effect of metallicity on the EWs of individual indices, as a function of temperature, for a constant gravity of $logg=4$. Indices have been calculated on the high-resolution Kurucz spectra of \cite{rodetal05}, after smoothing the spectra to the IUE resolution.
We briefly comment on the most relevant trends as a wider discussion can be found in \citet{chaetal07}, where the focus is on stellar indices, while the focus of this paper is on stellar population models.

Indices generally react to metallicity with higher metallicites (red) displaying stronger EWs. There are however some interesting exceptions. The mid-$UV$ index BL3096 is almost insensitive to metallicity, as already pointed out by \citet{chaetal07}, which makes it in principle a potential age-indicator. The indices FeV and $\rm Mg_{wide}$ display stronger EWs at the lowest metallicities, the index FeV showing such inversion around a temperature of 25,000 K. The effect on the $\rm Mg_{wide}$ index is very small and confined to the very high temperatures.
 
Metallicity effects will be further commented on using the computed SSP models in the next Section.
  
 In order to estimate fractional metallicity corrections we consider
  the separate effects of abundance changes on the absorption feature
  and on the pseudo-continuum fluxes, which are expressed by the
  multiplicative factors $\alpha(T_{\mathrm{eff}})$ and
  $\beta(T_{\mathrm{eff}})$:
  
  \begin{equation}
    \label{eq:alpha_beta}
    \begin{array}{lll}
      S_{NS} & = & \alpha(T_{\mathrm{eff}}) S_{\odot} \\
      C_{NS} & = & \beta(T_{\mathrm{eff}}) C_{\odot}
    \end{array}
  \end{equation}
  $\alpha(T_{\mathrm{eff}})$ and $\beta(T_{\mathrm{eff}})$ are the
  ratios of the non-solar and solar fluxes in the line and the
  continuum, respectively, and are calculated from theoretical spectra. 

    Taking equation~\ref{eq:alpha_beta} into consideration, the non-solar
  FF ($\phi_{\mathrm{NS}}$) is written as:
  
  \begin{equation}
    \label{eq:altfluxes}
    \phi_{NS} = \left(1 - \frac{ \alpha(T_{\mathrm{eff}}) S_{\odot} }{
    \beta(T_{\mathrm{eff}}) C_{\odot}}\right) \Delta \lambda.
  \end{equation}
  which we can express  in terms of the solar FF ($\phi_{\odot}$) as
  follows:

  \begin{equation}
    \label{eq:altff}
    \phi_{NS} = \left[ 1 - \frac{\alpha}{\beta}\left( 1 -
    \frac{\phi_{\odot}}{\Delta\lambda} \right)\right]\Delta\lambda
  \end{equation}

  The FF calculated this way are plugged into the
  \citetalias{claudia2005} EPS code to produce integrated indices for
  metallicities 2, $1/2$ and $1/20Z_{\odot}$.

  An alternative way to estimate the metallicity corrections consists of
  calculating the differential effect of abundance changes on the
  equivalent widths, computed again from the Kurucz theoretical
  spectra, and applying this differential at each temperature to the
  solar FF:

  \begin{equation}
    \label{eq:corr}
    \phi_{NS} = \phi_{\odot} + \Delta EW_K
  \end{equation}

  In this case we would be assuming that second order metallicity effects
  $O(\Delta[\mathrm{Z}/\mathrm{H}]^2)$ are negligible. The comparisons
  we ran (not shown here for space considerations) showed that both
  approaches yield virtually the same results. 
 
  \section{Integrated line indices of stellar populations models}
  \label{sec:ssp}

  We calculate the integrated line indices of stellar population models by
  following two approaches as described in the next two
  subsections.
  Synthetic line indices are obtained for SSP, i.e. instantaneous and chemically
  homogeneous bursts, with ages $t \gtrsim 1$ Myr and various
  metallicities (2\ensuremath{Z_{\odot}}, \ensuremath{Z_{\odot}},
  1/2\ensuremath{Z_{\odot}}{} and $1/20\ensuremath{Z_{\odot}}$), and for
  composite stellar populations of solar metallicity. The IMF of the
  models is the \cite{salp} one. We have checked that the values of the
  line indices remain almost unchanged whether we use the
  \citet{Kroupa2001} or a top-heavy IMF, with exponent 1 in the notation
  in which the Salpeter one is 2.35.

 In both cases, the model line indices will be tested
  with globular cluster data (Section~6).

  \subsection{Empirical indices of stellar population models}
  \label{sec:ffbased}

  We use the FFs described in Section 4 to assign empirically-based
  line indices to each star of the synthetic population. These are
  then added in order to obtain the integrated line index of
  the whole population. The integrated line index $EW_{P}$ can be
  expressed for an SSP as the sum of the continuum-flux weighted
  line indices of the stars \citep{claudia_cl}:

  \begin{equation}
    \label{eq:fint}
    EW_P = \sum_i\beta_i {EW}_i .
  \end{equation}
  
  $\beta_i$ is the weighting factor, which takes into account the relative
  contribution of the $i$-th star of the population to the continuum flux.
  The uncertainty on the integrated equivalent width follows as:

  \begin{equation}
    \label{eq:errint}
    \sigma^2_{\mathrm{\scriptscriptstyle{P}}} = \sum_i\beta_i\sigma^2_{EW_i}
  \end{equation}
  where $\sigma^2_{EW_i}$ is the uncertainty on the equivalent width of the
  $i-th$ star.

  Since by construction the residuals of the fitting procedure are randomly
  distributed in the $T_{\mathrm{eff}}$ and \ensuremath{\log g}{} space, we can assume that the
  uncertainty is equal to the standard error of the residuals \ensuremath{\sigma_{\mathrm{\scriptstyle{res}}}}{}  
  defined as:

  \begin{equation}
    \label{eq:sres}
    \ensuremath{\sigma_{\mathrm{\scriptstyle{res}}}}=\sqrt{\frac{1}{N-1}\sum_j^N(R_j -\bar{R})^2}
  \end{equation}
  where $R_j = EW^*_j - EW_j$ is the difference between the equivalent width
  predicted by the FF ($EW^*_j$) and the measured one ($EW_j$), and $N$ is the
  number of stellar groups contributing to the fit. Therefore
  $\sigma^2_{EW_i}=\ensuremath{\sigma_{\mathrm{\scriptstyle{res}}}}^2$ can be taken out from the sum in
  equation~(\ref{eq:errint}) and, since by definition $\sum_i\beta_i = 1$, we
  get $\sigma_{\mathrm{\scriptscriptstyle{P}}} \equiv\ensuremath{\sigma_{\mathrm{\scriptstyle{res}}}}$.

  \subsection{Theoretical indices of stellar population models}
  \label{sec:directbased}

  In parallel to the FF-based approach, we calculate the integrated
  line indices via direct integration on the synthetic spectral energy
  distribution (SED) of the M05 stellar population models. We computed a version of the M05 stellar population models using as input the $UV$BLUE library by \citet{rodetal05}. The $UV$BLUE is based on LTE calculations
carried out with the ATLAS9 and the SYNTHE codes developed by Kurucz (1979), and spectra are
provided for a wide range of chemical compositions, gravities and temperatures. The wavelength
range is 850-4700 $\AA$.  The spectral resolution is very high $\lambda / \Delta\lambda \sim 50,000$.
The authors compare the library with IUE spectra of stars with known 
atmospheric parameters and find an overall good agreement for stars of types B to G5 and for most
spectral features. Exceptions are the CIV line for O, B and to a lesser extent A-type stars and the spectral region 2400 to 2700 $\AA$ in the mid-$UV$ of F to G stars. In general the authors note that
the discrepancy is worse in giants than in dwarfs. 
  
\begin{figure*}[!htp]
     \centering
       \includegraphics[width=\hsize,clip]{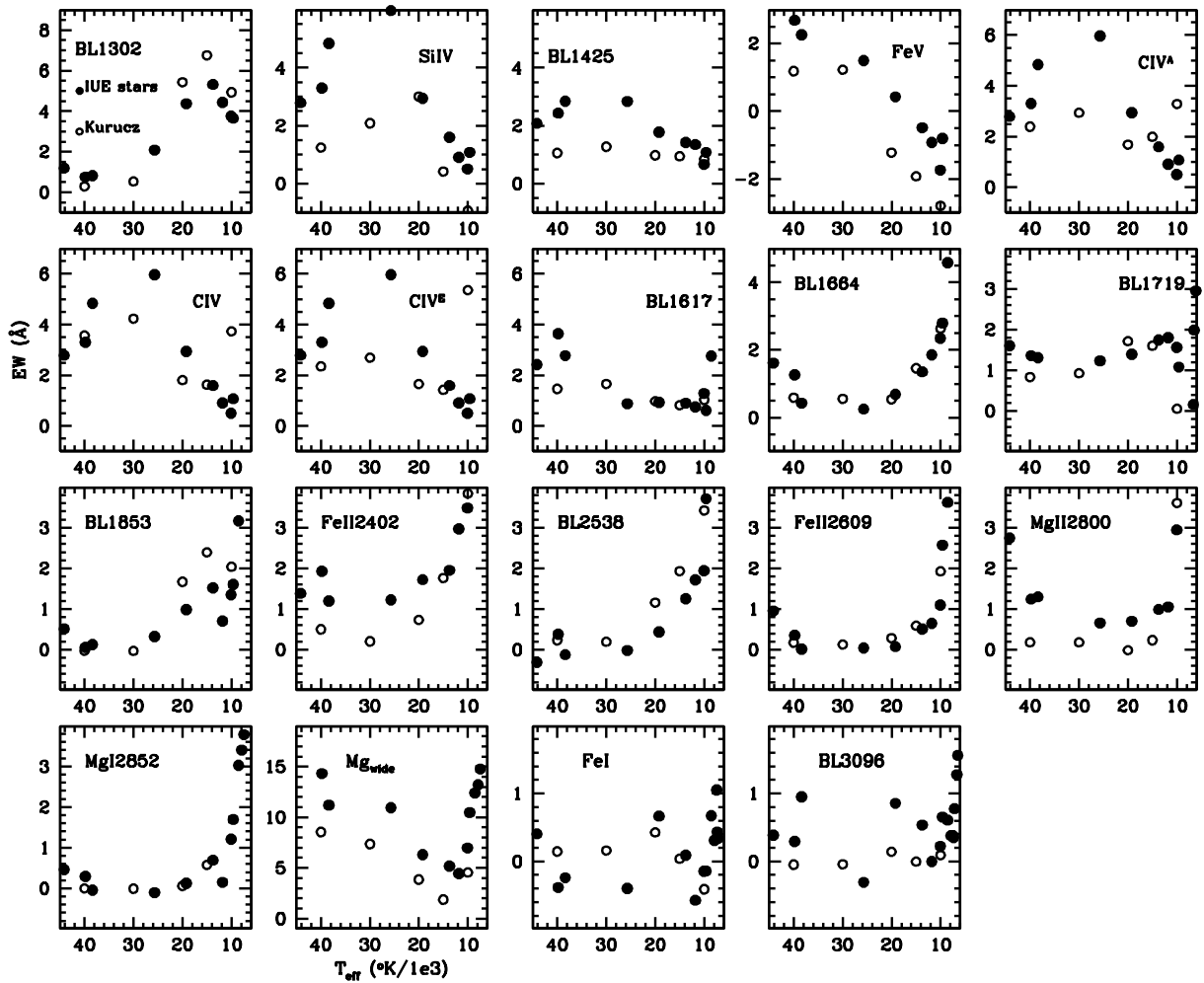}
     \caption{Comparison between the indices of individual stars from the F92 IUE library (filled symbols) and those computed on solar metallicity Kurucz-$UV$BLUE synthetic spectra (smoothed to the IUE resolution of $6~\AA$) with gravity $logg=4$ and various temperatures (open symbols).}
      \label{fig:comp_stars}
  \end{figure*}

In Figure \ref{fig:comp_stars} we compare the $UV$ indices as measured on the individual IUE observed stars and Kurucz stellar spectra, as a function of temperature for a fixed gravity ($logg=4$ in the Kurucz spectra and $\geq 3.8$ for the IUE stars). The figure shows that several indices are consistent between the real and the synthetic stars, while others such as BL1302, SiIV (for some temperatures) and BL1425 are discrepant, a fact that will propagate into the stellar population models. When the indices are discrepant, the real stars often, but not always, have stronger indices. Some of these effects may originate from abundance ratio effects, others from a wrong stellar parameter assignment \citep[as discussed by][]{rodetal05} or by incomplete line-lists in the theoretical models.
For the far-$UV$ indices the discrepancies may arise from non-LTE effects, stellar winds, etc. \citep{adi,doretal93,chaetal07}. 

  We have computed the stellar population models using the original high-resolution $UV$BLUE library and then smoothed them using a Gaussian filter convolution in order 
  to match the 6 $\AA$ resolution of the IUE spectra. We refer to these as Kurucz-high-resolution based 
  M05 models.
  
  The SED-based line indices will give us insights into how well the Kurucz spectra reproduce stellar
  absorptions in the $UV$.  
  It is interesting to note that we resorted to this approach after having found for some
  indices severe discrepancies between the LMC GCs data and the 
  Milky Way FF-based models.

  \subsection{Age and metallicity sensitivity of individual indices}
  \label{sec:metaleffect}

 We use the models computed using the high-resolution Kurucz library to determine the age and metallicity sensitivity of the individual line indices.
 
  \begin{figure*}[!ht]
    \begin{center}
      \includegraphics[clip,width=\textwidth]{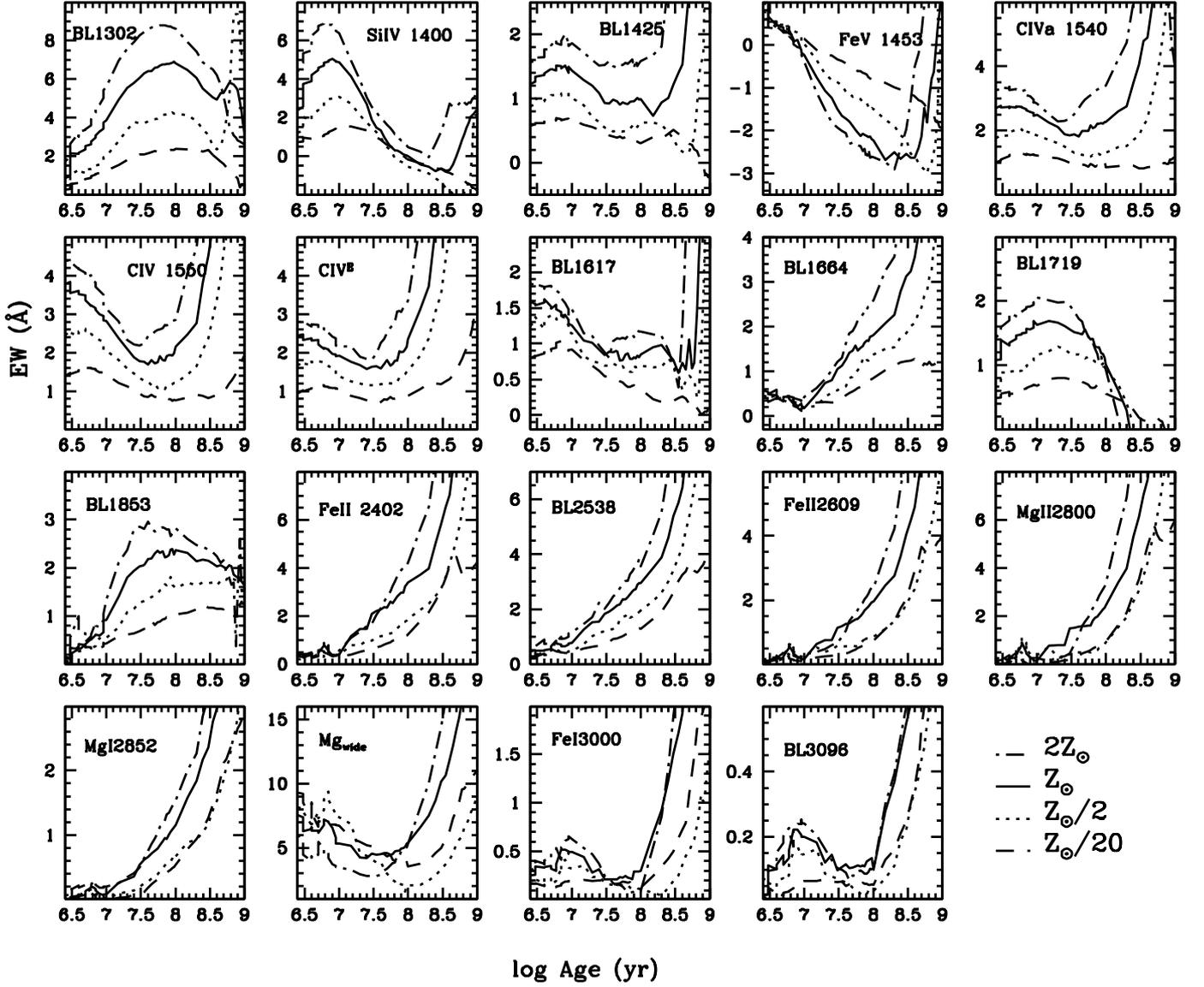}
       \caption[Effect of age and metallicity on the line strength
      evolution]{Effect of age and metallicity on line indices of SSP models. }
      \label{fig:met1} 
    \end{center}
  \end{figure*}
 
 Fig.~\ref{fig:met1} shows the time evolution
  of synthetic line indices of SSP for different total metallicities
  (2, 1, 1/2 and $1/20\ensuremath{Z_{\odot}}$).
  
  The dependence on age of indices is straightforward: all indices evolve strongly with age, which is due to the fast evolution
  of the temperature of the turnoff mass at these low ages.
 
  Metallicity effects are sometimes more complex, as in some cases
  temperature effects on the continuum and the effects of the actual
  absorbers in the lines cancel each other and produce an
  insensitivity of the line indices to metallicity.  
  In the optical region,
  metal line indices decrease in strength with decreasing metallicity, and
  vice versa. In the $UV$, the behaviour of the integrated indices
  depends more strongly on the turnoff temperature, whose effect can
  dominate over the abundance effect on the line itself (see also discussion in Chavez et al. 2007). At
  a given temperature, lower metallicity stars tend to display weaker
  absorption lines due to lower opacity. At a given age, on the other hand, stars
  around the MSTO become hotter as metallicity decreases, which may
  strengthen the lines in some indices and temperature ranges (remember that MSTO stars are the main
  contributors to the integrated indices), thereby balancing the pure
  abundance effect. Which effect dominates depends on the index and
  on the temperature range (and therefore on the age of the SSP). In
  general far-$UV$ indices show higher equivalent widths for higher
  temperatures, while mid-$UV$ indices display the opposite behaviour.
     
  In general, lower EWs correspond to sub-solar chemical compositions,
  but the temperature dependence is complex. The line indices
  \ion{C}{iv}{}, $\mathrm{C}^{\scriptscriptstyle{A}}_{\mathrm{IV}}$ and
  BL$_{2538}$ behave regularly along the whole range of temperatures
  calculated, while the \ion{Si}{iv}{} and BL$_{1425}$ lines do so until
  $T_{\mathrm{eff}}\sim15$~kK (mid B spectral type), while at lower
  temperatures metallicity effects vanish. The line
  \ensuremath{\mathrm{BL_{\scriptstyle{1617}}}} is attributed to
  transitions of \ion{Fe}{iv} and \ion{Fe}{v} \citep{dean} however it
  shows little variation with the total chemical composition. We have
  confirmed this trend (for $T_{\mathrm{eff}}>20$~kK), by computing
  synthetic spectra models with the WM-basic software \citep{adi} for
  solar and half-solar metallicity. The difference $\Delta EW$ was found
  to be less than 0.2~\AA\ and comparable with the results obtained with
  the Kurucz spectra. The line
  \ensuremath{\mathrm{BL_{\scriptstyle{1664}}}} behaves similarly to
  BL$_{1617}$, but is very strong at high metallicity. Most likely for these two lines the effect of increasing
  metallicity mainly works in depressing the continuum, which has the
  effect of lowering the line-strength.

  \ion{Si}{iv}{} and \ion{C}{iv}{} indices display the strongest
  sensitivity to metallicity at ages below 50 Myr, while
  \ensuremath{\mathrm{BL_{\scriptstyle{1853}}}} is a good metallicity
  indicator at older
  ages. \ensuremath{\mathrm{BL_{\scriptstyle{1617}}}} and FeV show a
  remarkable insensitivity to metallicity at young ages. Though only
  up to ages around 60 Myr and 20 Myr respectively, these lines are
  potentially powerful age indicators. 
  FeV is the only index that becomes stronger in lower metallicity populations at ages larger then 20 Myr.

  \section{Testing the indices with real data}
  \label{sec:cal}

  \subsection{Globular cluster templates}
  \label{sec:cal}

  Before using the models to study complicated stellar populations
  like galaxies, it is necessary to test them on
  globular clusters (GCs) with independently known ages,
  metallicities and abundance ratios \citep[e.g.,][]{renfus88, wor94, claudia_cl, leoros03, beaetal02, proetal04, leewor05, sch07, koletal08, grasch08}. GCs are the ideal templates since they are
  prototypical simple stellar populations. 

  We use \emph{IUE} observations \citep{cassatella} of 10 GCs in the
  Large Magellanic Cloud (LMC), which have independent age and
  metallicity determinations (Table~\ref{tab:clusters2}). Ages and
  metallicities were obtained from different sources. Both metallicity
  and the age of NGC~1711 were taken from \citet{dirsch}, the former was
  determined by using metallicity to Str\"omgrem colour empirical
  calibrations, the latter by fitting CMD to Geneva \citep{schaerer93}
  and Padova \citep{bertelli94} isochrones. Age estimates for NGC~1805
  and NGC~1818 were taken from the review by \cite{degrijs} and
  metallicities from \citet{johnson} (who used fits to near-infrared HST
  CMDs). Ages for all the other clusters come from \cite{elson} and
  \cite{elson_fall} and were estimated using optical CMDs, while
  metallicities were estimated by matching stellar spectral models to
  medium resolution optical spectroscopy of individual stars for
  NGC~1850, NGC~1866, NGC~2004 and NGC~2100 \citep{jasn}; infrared
  spectroscopy through equivalent width measurements of the 1.62 $\mu$m
  near-IR feature for NGC~1984 and NGC~2011 \citep{oliva}; and from
  optical HST CMDs for NGC~2164 \citep{mackey}.

\begin{table}[!ht]
  \begin{center}
    \begin{tabular}{lrrr}
      \hline
      Name  & $\log(\mathrm{age})$ (yr) &
      $[\mathrm{Z}/\mathrm{H}]$ &
      $E(B-V)\footnotemark[10]$ \\ \hline\hline
NGC 1711 &  $7.70\pm0.05$ \footnotemark[1] &                                   
$-0.57\pm0.17$\footnotemark[1] & 0.14 \\         
                                                                                               
NGC 1805 & 	$7.00\pm0.05$\footnotemark[2] &                                 
$-0.40\lesssim[Z/\mathrm{H}]\lesssim0.0^{2,5}$ & 0.10\\
                                                                                                
NGC 1818 &\mbox{\ensuremath{7.40\pm0.30}}\footnotemark[2] &                                    
$-0.40\lesssim\ensuremath{[Z/\mathrm{H}]}\lesssim0.0^{2,5}$ & 0.07\\

NGC 1847 & $ 7.42\pm0.30$\footnotemark[4] & $-0.37$\footnotemark[9] & 0.09 \\
NGC 1850 &\mbox{\ensuremath{7.40\pm0.20}}\footnotemark[3] & \mbox{\ensuremath{-0.12\pm0.20}}$^6$
& 0.09\\

NGC 1866 &\mbox{\ensuremath{8.12\pm0.30}}\footnotemark[4] &
\mbox{\ensuremath{-0.50\pm0.10}}$^6$ & 0.07 \\

NGC 1984 & \mbox{\ensuremath{7.06\pm0.30}}\footnotemark[4] &
\mbox{\ensuremath{-0.90\pm0.40}}$^8$ & 0.14 \\ 

NGC 2004 & 	\mbox{\ensuremath{7.30\pm0.20}}\footnotemark[3] &
\mbox{\ensuremath{-0.56\pm0.20}}$^6$ & 0.09\\

NGC 2011 &	\mbox{\ensuremath{6.99\pm0.30}}\footnotemark[4] &
\mbox{\ensuremath{-0.47\pm0.40}}$^8$ & 0.08\footnotemark[11]\\ 

NGC 2100 &	\mbox{\ensuremath{7.20\pm0.20}}\footnotemark[3] &
\mbox{\ensuremath{-0.32\pm0.20}}$^6$ & 0.19\\ 
\hline 
\end{tabular}
\end{center}
\caption[Relevant parameters of
  the LMC clusters used in the comparison]{Ages, metallicities and
  colour excess $E(B-V)$ of the sample clusters. References:
  \footnotemark[1]\cite{dirsch}; \footnotemark[2]\cite{degrijs};
  \footnotemark[3]\cite{elson}; \footnotemark[4]\cite{elson_fall};
  \footnotemark[5]\cite{johnson}; $^6$\cite{jasn}; $^7$\cite{hill};
  $^8$\cite{oliva}; $^9$\cite{mackey}; \footnotemark[10]Colour excess
  from: \citet{cassatella}, with the exception of NGC 2011;
  \footnotemark[11]\cite{persson}. For NGC~1847 no error on metallicity is given and we have conservatively assumed 0.2, a value common to most globular clusters.} 
  \label{tab:clusters2}
\end{table}

  On the GC spectra, we measure the line indices listed in
  Table~\ref{tab:ind}. These values are given in Tables~\ref{tab:ews} and \ref{tab:ews2}
  in Appendix~\ref{sec:ffcoeff} together with estimates of the
  associated errors. The uncertainties of the measured equivalent
  widths are calculated in the same way as we estimated the errors for
  the measurements on the stellar group spectra (see
  Sect.~\ref{sec:ew}), but using the error spectra provided directly
  by the IUE observations. Since the
  maximum age of the template GCs is 130 Myr and mid-$UV$ indices are
  significantly strong only for populations older than this limit, we are
  not able to check most of the synthetic mid-$UV$ line indices. 
  This would be possible in the future, provided that good quality
  observations of older clusters are carried out in the $UV$.

   \subsection{Far-$UV$ indices}
  \label{sec:selection}

  \begin{figure*}[!ht]
    \begin{center}
      \includegraphics[clip,width=\textwidth]{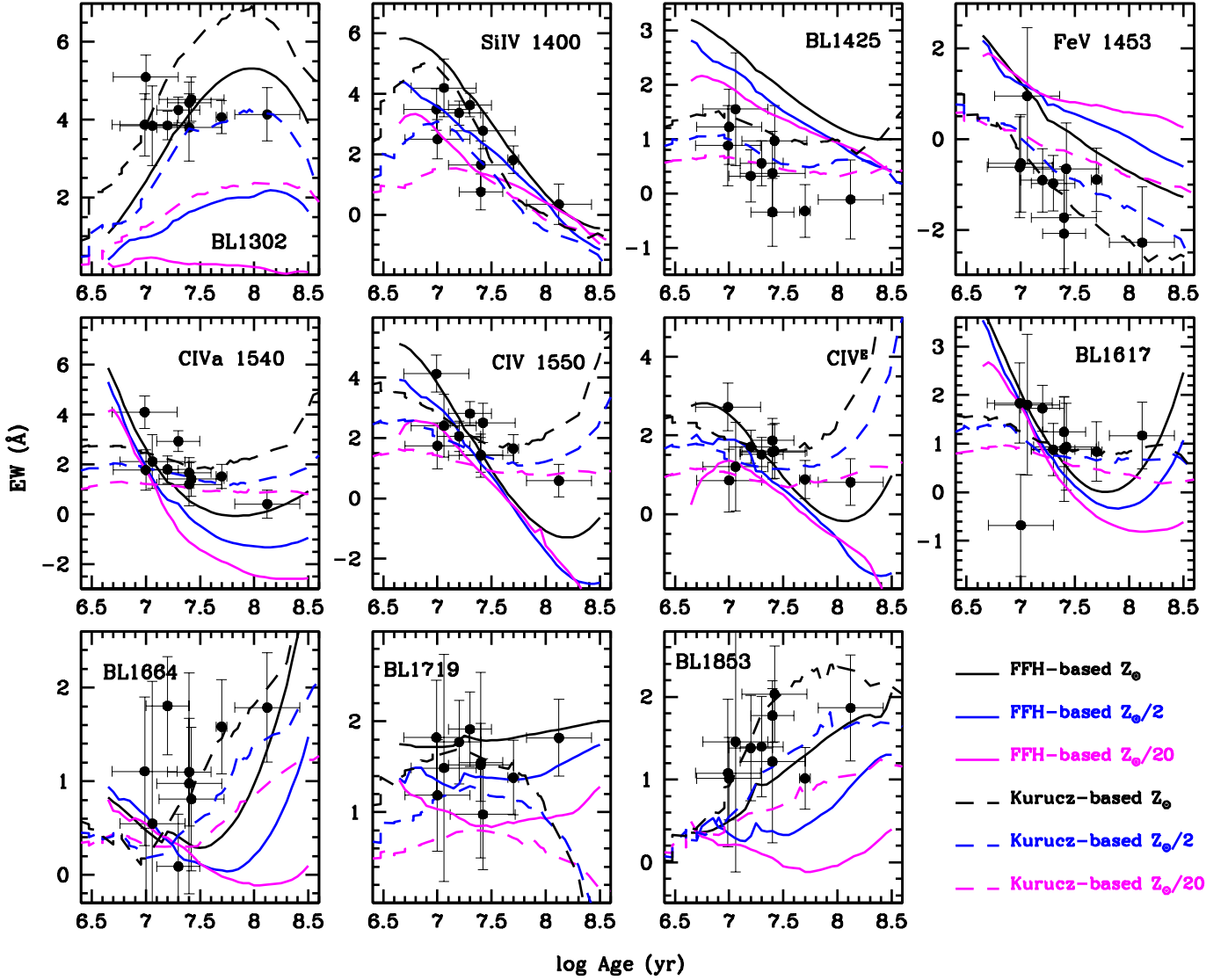}
         \caption[Time evolution of synthetic line indices]{Time evolution
      of synthetic far-$UV$ line indices of simple stellar population (SSP) models, overlaid on
      data of LMC GCs with ages taken from the literature (symbols with error bars).
      Solid lines are models based on the IUE-FF while dashed lines are models based on 
      high-resolution Kurucz model atmospheres. Black and blue lines refer to solar and half-solar metallicity, 
      the magenta lines to $Z~1/20~Z_{\odot}$. }
      \label{far$UV$}
    \end{center}
  \end{figure*}

  Figure~\ref{far$UV$} shows the comparison of SSP synthetic line
  indices in the far-$UV$ with GCs data. For three metallicities, solar, half-solar and $Z~1/20~Z_{\odot}$, two sets of models are shown,
  those based on the empirical-IUE FFs (solid lines) and those on the M05 SEDs using
  as input the high-resolution Kurucz-based synthetic stellar library.  
   
 Strictly speaking, due to the sub-solar metallicity of the LMC star clusters, only sub-solar models can be meaningfully checked with these data. 
  However, since the solar-metallicity models are those uniquely locked to the IUE-based
  FFs and are not dependent on our recipe for inserting metallicity effects, we plot them as well. When we calculate the metallicity and the age of the LMC clusters explicitly from the indices we use the full grid of models (see Section~6.4).
  
Looking at Figure~\ref{far$UV$} we see that theoretical and empirical models agree reasonably well and reproduce GCs data for most indices. These results, especially for \ion{C}{iv}, are remarkable, since these line indices have been discussed as being poorly determined due to blending with interstellar lines, dust effects, and IMF effects \citep[see discussion in][]{rix2004}. Also 
 \cite{rodetal05} quote the \ion{C}{iv} line as one of the most poorly described by the $UV$BLUE library.

Exceptions are BL1425 and FeV 1453 for which the empirical models lie above the data. 

Errors in the fitting functions can be
excluded, since the same discrepancy is found in the stellar
population models of \cite{bc03} that use as input spectral
library the F92 without passing through the fitting function procedure (see Sect.~\ref{sec:comp}, Figure~16). 

Abundance effects, for example an under abundance of iron in the LMC GCs as compared to the IUE MW stars, also seem unlikely as neither the solar-scaled Kurucz-based models nor the Starburst99 models
based on a higher-resolution version of the IUE library show the same discrepancy (see
Section~6.5). In addition, other iron-sensitive line-indices like Bl1617 do not behave the same way. For the same reasons an iron enhancement in the IUE stars can also be ruled out.
  
An effect from interstellar lines appears unlikely for several reasons. Interstellar lines exist at these wavelengths \citep[see e.g.][]{casetal84} but they are known also to affect the SiIV and CIV indices \citep{casetal84, robert1993} whereas such a sizable discrepancy is not evident for the other indices in Figure~\ref{far$UV$}. In addition, the Starburst99 models do not show the same discrepancy and they have also not been corrected for interstellar lines \citep{robert1993}. Also it could be argued that interstellar lines affect the GCs in a similar way as the empirical models that are constructed using real stars, and if at all it should be Kurucz-based models that are discrepant, as theoretical spectra do not incorporate the effect of interstellar lines.

 Finally, note that in case of BL1302 and BL1853 the empirical models would fit the data for metallicities above solar, which is too high. We come back to this point later when we use the indices to derive quantitatively metallicities and ages.
 
   \subsection{mid-$UV$ indices}
  \label{sec:selection}

  \begin{figure*}[!ht]
    \begin{center}
      \includegraphics[clip,width=\textwidth]{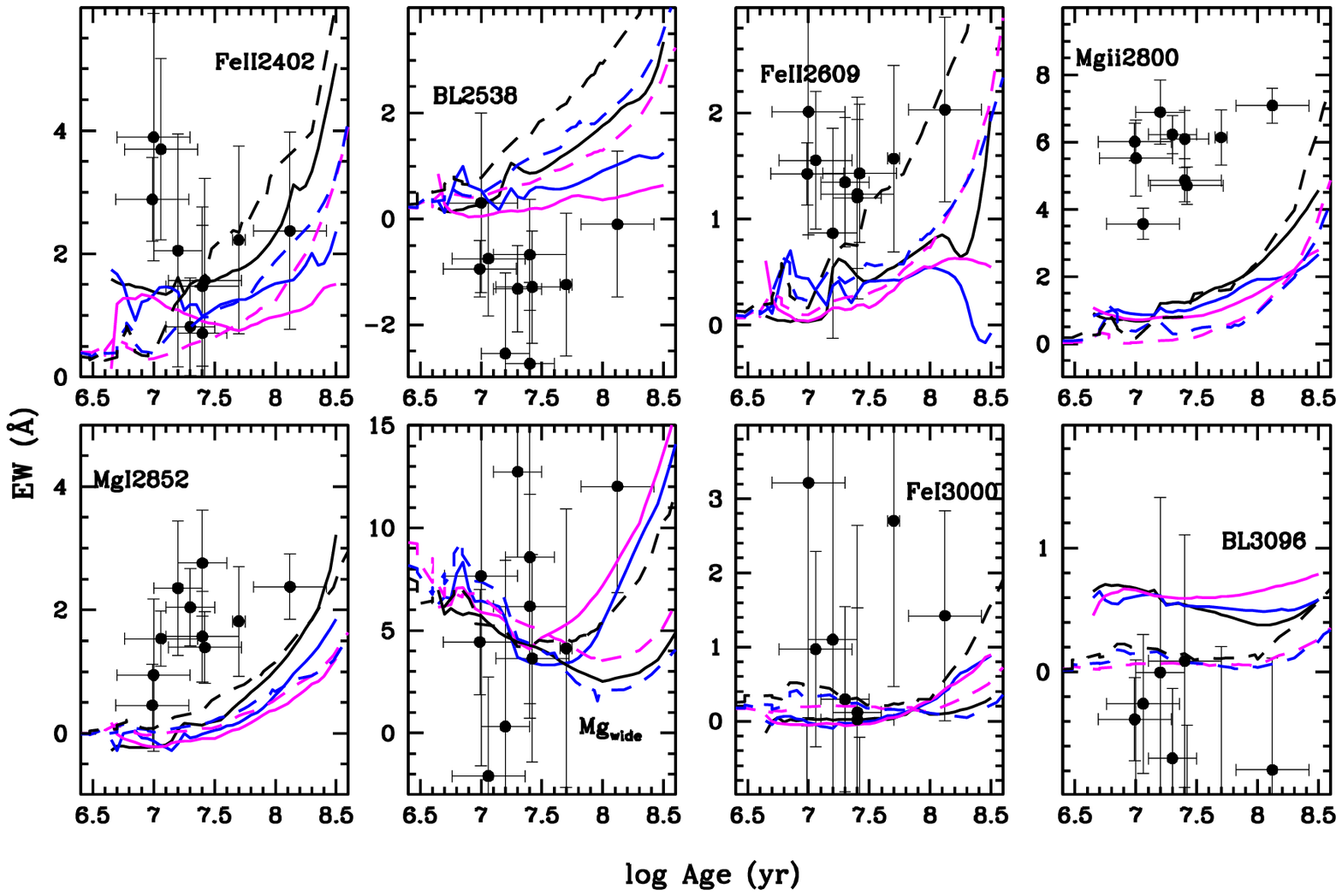}
         \caption[Time evolution of synthetic line indices]{Same as Figure \ref{far$UV$} for mid-$UV$ indices.}
      \label{mid$UV$}
    \end{center}
  \end{figure*}

Figure~\ref{mid$UV$} is the analogue of Figure~\ref{far$UV$} for indices in the mid-$UV$ ($\lambda > 2000~\AA$).

 The validation of the mid-$UV$ indices is less conclusive since the
  ages of the \citeauthor{cassatella} GCs are limited to 130 Myr, while
  older ages are required to set constraints on
  this spectral region.
  
In spite of the age limitation, glancing at Figure~\ref{mid$UV$} one sees that most synthetic indices are discrepant to those measured in the GCs, the FeII2402 and the $Mg_{\rm wide}$ being perhaps the only indices for which, given the large error bars and scatter, one could argue that the models are not clearly offset.
  
The reasons for these discrepancies could be either on the model side or on the GC data side.
 From the GC data side one cannot exclude problems in the observed spectra in the mid-$UV$.
 
From the model point of view, problems with the FF approach can be excluded as the comparison with the BC03 models confirms the discrepancies (Figure 16).

One could argue that element abundance ratio effects start affecting the indices in the mid-$UV$ as we know they affect most optical indices. For example an overabundance of Magnesium in the LMC GCs may push the data off the models. On the other hand, the iron lines are sometimes above and sometimes below the models and no clear pattern is detectable. Clearly a modelling of abundance ratio effects in individual lines is required here, which is the scope of a future investigation.

Finally, contamination by interstellar lines that are known to affect the region of the $Mg$~absorption at 2800~$\AA$ 
\citep{moretal75,casetal84} remain the most likely possibility (see below). 

For this work we have carefully checked the case of the MgII line at 2800~$\AA$, which we discuss below as an example of a discrepant index. We choose this index as it is often used in studies of high-redshift galaxies and the adjacent spectral region is employed to measure spectroscopic redshifts for high-$z$~galaxies (Daddi et al. 2005).

   \begin{figure}[!th]
    \begin{center}
      \includegraphics[width=.70\hsize,clip]{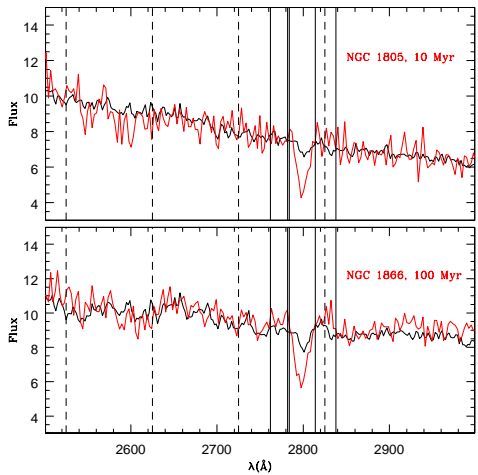}
      \caption{Zoom of the spectral region containing the
      \ion{Mg}{ii}{} 2800 index. The observed spectra of two LMC
      clusters from the Cassatella et al. sample are shown in red, with
      overlaid template spectra from BC03 based on the Fanelli et
      al. library (black) at the age of the observed clusters
      (labeled in the panels). Vertical solid lines show the line and
      continua windows defining the \ion{Mg}{ii}{} 2800 index
      (cf. Table 1); dashed lines refer to the
      \ensuremath{\mathrm{Mg}_{UV}}{} of Daddi et al. (2005).}
      \label{fig:mg_spectrum}
    \end{center}
  \end{figure}
  
  \begin{figure}[!th]
    \begin{center}
      \includegraphics[width=.70\hsize,clip]{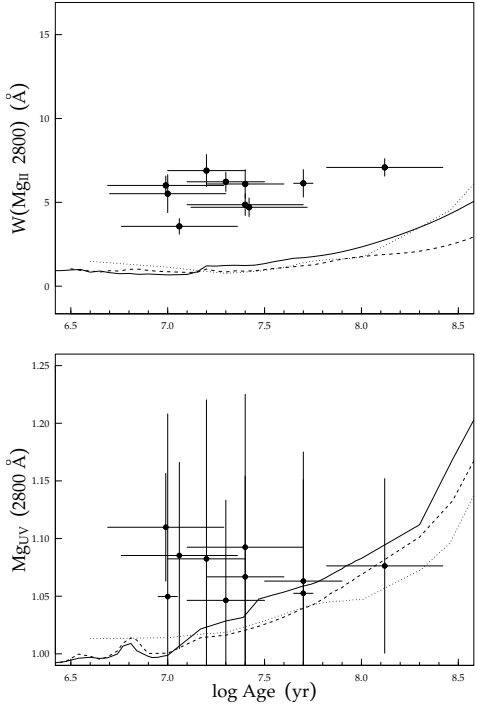}
      \caption{Comparison between the \ion{Mg}{ii}{} 2800 from Fanelli
      et al. and the \ensuremath{\mathrm{Mg}_{UV}}{} index defined in
      Daddi et al. (2005). In both panels, solar and half-solar
      metallicity models (solid and dashed lines, respectively) are
      plotted together with the GCs data. Also plotted are
      the indices as derived from \citet{bc03} solar metallicity
      models (dotted line), which use the same spectral library we use
      in the $UV$ and give very similar results to ours.}
      \label{fig:mgii_mg$UV$}
    \end{center}
  \end{figure}
  
  \begin{figure}[!th]
    \begin{center}
      \includegraphics[width=.70\hsize,clip]{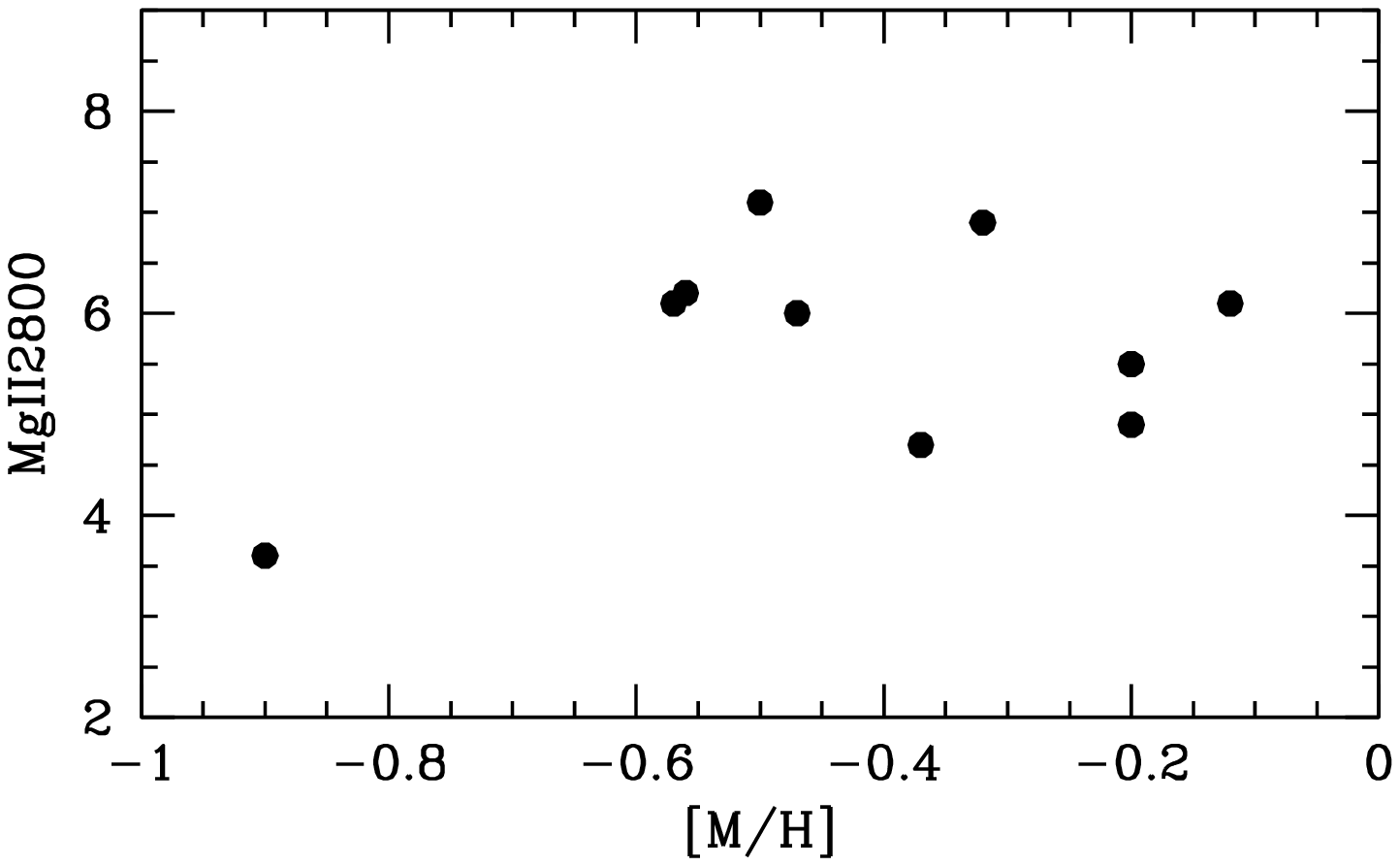}
      \caption{The \ion{Mg}{ii}{} 2800 index of the LMC GCs as function of the metallicity as determined in the literature (see Table D.1 in the Appendix)..}
      \label{fig:mg_mh}
    \end{center}
  \end{figure}
  
  As we see in Figure~\ref{mid$UV$}, both the empirical models and the Kurucz-based ones 
  display significantly lower equivalent widths than the GCs. The same is true when we calculate the index on the BC03 models incorporating the F92 library direclty (dotted line in Figure~\ref{fig:mgii_mg$UV$}, upper panel).
   The mismatch is somewhat reduced, instead, when we apply broadband indices, such as the
  \ensuremath{\mathrm{Mg}_{\rm UV}}{} defined by Daddi et al. (2005)\footnote{$ \ensuremath{\mathrm{Mg}_{UV}} = \frac{2
  \int_{2625}^{2725} f_{\lambda} d\lambda}{\int_{2525}^{2625}
  f_{\lambda} d\lambda + \int_{2725}^{2825} f_{\lambda} d\lambda}$ },
  or the \ensuremath{\mathrm{Mg}_{\mathrm{wide}}} (see
  Fig.~10).

  In order to understand the discrepancy we have compared the
  individual cluster spectra to the BC03 templates that include the
  Fanelli et al. library (see Fig.~\ref{fig:mg_spectrum}). As it can be
  seen the GC spectra display a much stronger magnesium absorption
  than the templates based on the MW stars. This explains the
  discrepancy seen in Figure~\ref{fig:mgii_mg$UV$}. The fact that the
  Mg$UV$ appears to be in better agreement with the data outlines
  that these indices do not trace the absorption feature but the
  continuum shape around those wavelength. The Mg$UV$ cannot be used
  to measure the magnesium abundance.

  Fanelli et al.~(1990; 1992) find that the \ion{Mg}{ii}{} (2800~\AA)
  is very sensitive to temperature and is stronger in metal-poor stars
  than metal-rich ones at given effective temperature (see
  e.g. their Figure 6-o), a behaviour attributed to a combination of
  effects, including chromospheric emission.

  The case here is more complicated as the net effect is due to a
  stellar population and not to a single star. The clusters of our
  sample span a range in metallicity and some have the same age, hence
  the most-metal poor ones typically should have higher
  temperatures. However if one compares the \ion{Mg}{ii}{} (2800~\AA)
  indices of these clusters (see Figure~\ref{fig:mg_mh}), there is no indication that
  the most metal-poor have stronger indices.

 A true abundance effect,
  with the LMC clusters being enriched in magnesium with respect to
  the MW stars of which the Fanelli et al. library is composed, finds support from the work of Beasley et
  al. (2002), which reports an indication of a possible magnesium overabundance
  in other LMC clusters of similar age from the analysis
  of Lick indices. The caveat we have with this interpretation is that if an abundance effect is responsible for the deep line present in the GC spectra, the overabundance of magnesium would need to be very large.
    In a future work we intend to study the abundance effect on the
  MgII2800 index as this is a potentially very powerful chemical
  abundance indicator for high redshift galaxies (Heap et al.~1998; McCarthy et al. 2004).
 
Contamination by interstellar lines originating from the warm neutral interstellar medium of the LMC \citep{weletal99} remains the most likely reason for the strong absorptions around the Mg region. It is interesting that the IUE stars do not show such contamination, which may be due to the much lower extinction of galactic sitelines.
 
  \subsection{Using the indices to derive age and metallicity}
  \label{sec:disentangle}

Here we derive quantitatively the ages and the metallicities of the 
LMC clusters using the indices and compare them to the values we compiled
from the literature.

For this exercise we use all eleven far-$UV$ indices in case of the Kurucz-based models and 
seven indices, i.e. the far-$UV$ indices minus BL1302, BL1425, FeV, BL1853 in case of the 
FF-based models. The latter choice is motivated by the clear discrepancy between the GC 
data and the empirical models for those indices, as we have discussed previously (see Figure~\ref{far$UV$}). Mid-$UV$ indices are not considered as the agreement between models and data is generally poor. Also, the template GCs are very young and the use of the far-$UV$ is more appropriate.

Ages and metallicities are derived by minimizing the quadratic distance between the measured EWs and the SSP models for the selected set of indices simultaneously.
 
  \begin{figure*}[!ht]
    \begin{center}
      \includegraphics[width=.48\textwidth,clip]{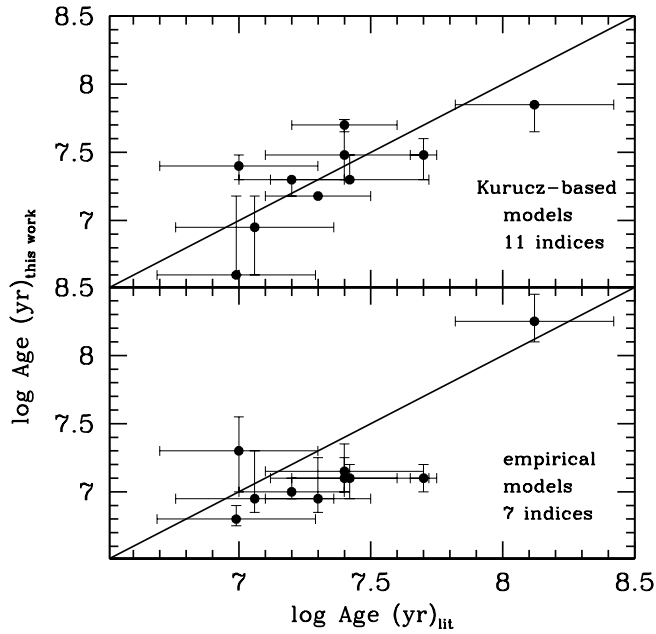}
      \caption{Age estimates derived from all 11 far-$UV$ indices for the Kurucz-based models (upper panel)
   and from seven indices for the empirical models compared to the literature values.}
      \label{fig:age_est}
    \end{center}
  \end{figure*}

 \begin{figure*}[!ht]
    \begin{center}
      \includegraphics[width=.48\textwidth,clip]{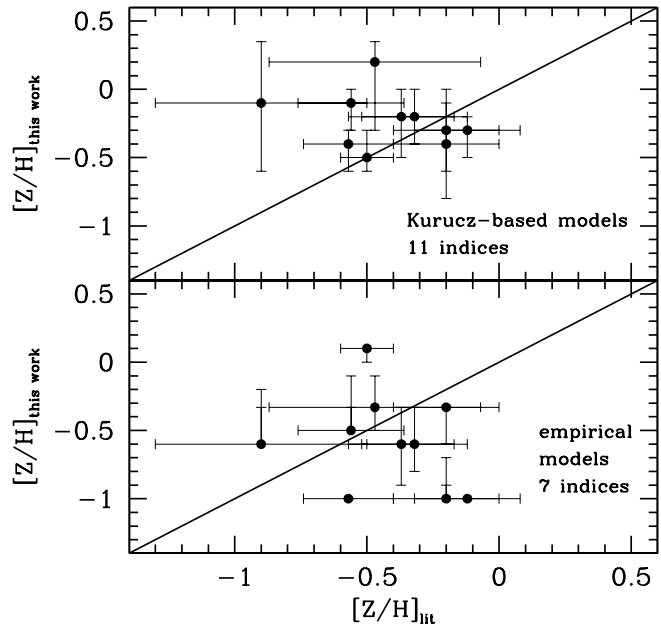}
      \caption{Same as in Figure~\ref{fig:age_est} for metallicity.}
      \label{fig:met_est}
    \end{center}
  \end{figure*}

  The ages and metallicities we derive for the GCs from the indices are compared with literature values in Fig.~\ref{fig:age_est}  and ~\ref{fig:met_est}  (values are given in Table~\ref{tab:agemetest} in the Appendix). 
  
 The ages we derive from the indices are in good agreement with those compiled from the literature. This is the case for both sets of models.

The metallicities are also in satisfactory agreement considering that the metallicity dependence of the $UV$ spectrum is not as straightforward as in the optical (see also discussion in Chavez et al. 2007) and that the metallicity determinations for the GCs are not homogeneous. Here the models behave differently. 
The Kurucz-based models give a very good determination for more than half the sample with a tendency to lie above the literature values for three objects. The empirical models display a larger scatter, but no systematic trend. For the empirical models we checked that had we also used the BL1302 and BL1853 indices we would have obtained systematically higher metallicities, for the reasons discussed in Section~6.2. When we include the two indices BL1425 and FeV for which the models are above the data, this systematic effect is rectified because the discrepancy pushes the fit to lower metallicities. 

In conclusion we recommend the use of seven indices, namely SiIV, $\rm CIV^{A}$, CIV, $\rm CIV^{E}$, BL1617, BL1664 and BL1719, in the case of empirical models and eleven indices namely BL1302, SiIV, BL1425, FeV, $\rm CIV^{A}$, CIV, $\rm CIV^{E}$, BL1617, BL1664, BL1719, BL1853 for the Kurucz-based models.

It will be interesting to apply this age and metallicity tool to distant galaxies for which many rest-frame $UV$ spectra are available in the literature.

  \subsection{Models based on other tracks or spectral libraries}
  \label{sec:comp}

  No other model in the literature provides all the line indices we
  present here. Still, we can use the integrated spectral energy
  distributions of other population synthesis models and obtain the
  line-indices by direct integration. Two models are interesting in
  this context, the \citet{bc03} and \emph{Starburst 99}
  \citep{leitherer99} models. The reason is that the former adopt the
  Padova tracks and the same spectral library we use
  \citepalias{fan_iv}. This allows us to check the joint effect of stellar tracks 
  and our fitting function procedure. Though we cannot separate the two effects, 
  we should be able to use this comparison to detect gross discrepancies due to the 
  fitting function procedure.
  On the other hand, \emph{Starburst 99} uses the same
  stellar tracks as in the M05 models, but a different spectral
  library \citep[described in][]{robert1993}, mostly based on
  high-resolution \emph{IUE} spectra from \citet{howarth89}. Hence with
  the \emph{Starburst 99} models we can check the effect of the spectral
  library.~\footnote{To allow for a meaningful comparison, we have smoothed
  the SEDs of \emph{Starburst 99} to the same resolution as the
  \citetalias{fan_iv} library ($6$~\AA).}

  \begin{figure*}[!ht]
    \begin{center}
      \includegraphics[width=\textwidth]{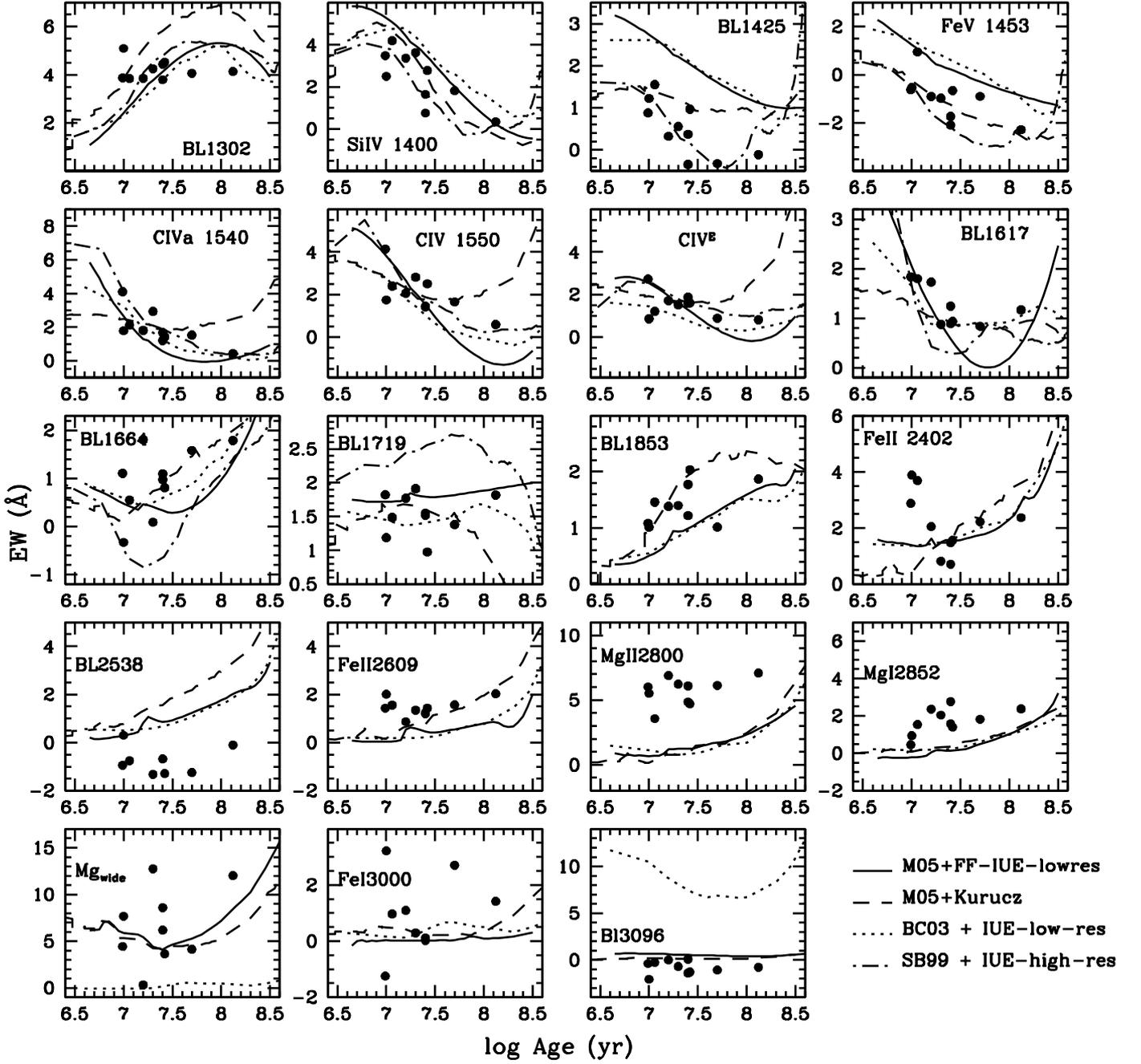}
      \caption{Comparison between several model indices at solar metallicity.
       Our FF-based (solid)
      and Kurucz-based (dashed) indices are compared to those we calculate on the \cite{bc03}
      (dotted) and \cite{leitherer99} (dashed-dotted) models.  \cite{bc03} use
      \citetalias{fan_iv} spectral library and \cite{leitherer99} use
      a higher-dispersion version of \emph{IUE} spectra in addition to the F92 library
      \citep{robert1993}.}
      \label{fig:literature}
    \end{center}
  \end{figure*}

  Figure~\ref{fig:literature} shows the results of our comparison. In general, the
  indices behave similarly in most cases, suggesting that our
  conclusions are robust against different stellar tracks and
  different empirical libraries.
  
The discrepancies between the FF-based models of the indices \ensuremath{\mathrm{BL_{\scriptstyle{1425}}}} and FeV 1453 and the template GCs is found also when the \cite{bc03} SEDs are used (dotted line). This excludes that the origin of the discrepancies lies in an error of the
  FF approach. The fact  that the \emph{Starburst 99}-based indices do not display the same discrepancy
  suggests that the origin of the mismatch lies in the O-star IUE-low-resolution library.
  
 All the discrepant indices in the mid-$UV$ remain discrepant also for the BC03 models in which the F92 library is used without passing through FF, again excluding a major problem in the FF procedure.

  \subsection{The effect of mass-loss}
  \label{sec:massloss}

  Very massive stars are known to suffer from strong mass-loss, a
  process driven essentially by radiation pressure which is very difficult
  to predict by stellar evolution models. For this reason we have
  investigated the effect, if any, of varying the recipes used in the
  stellar tracks for the mass-loss rates on the upper part of the
  IMF. We computed solar metallicity models, using as inputs
  isochrones from the Geneva group \citep{meynet94}, which implement a
  higher mass-loss rates (twice the one used in their standard tracks)
  for the very massive stars ($\gtrsim15\mathcal{M}_{\odot}$).

  \begin{figure}[!ht]
    \begin{center}
      \includegraphics[width=0.6\hsize,clip]{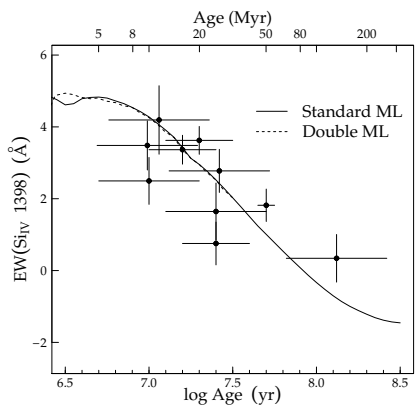}
      \caption{Comparison between models adopting Geneva tracks with a
      standard (solid line) and twice the standard (dashed line)
      mass-loss rate for massive stars ($\gtrsim15\mathcal{M}_{\odot}$). No effect
      is appreciable.}
      \label{fig:siiv_hml}
    \end{center}
  \end{figure}

  We show in Fig.~\ref{fig:siiv_hml} the case of the \ion{Si}{iv}{} 1400
  absorption index, as an illustration of the general case. The doubling
  of the mass-loss rate has a negligible effect on the
  indices. 

  \section{Summary and conclusions}
  \label{sec:concl}

  We have exploited
  the low-resolution (6 \AA) absorption line index system in the far
  and mid-$UV$ as defined by Fanelli et al. (1992) based on the IUE 
  low-resolution stellar library. Though the latter work was published sixteen years ago,
  it remains the most comprehensive study of $UV$ stellar absorption
  lines that is based on real stars. Surprisingly, it
  has not yet been used for population synthesis studies.  Our aim is to construct integrated line-indices
  of stellar population models that can be used as age and metallicity
  indicators of young stellar populations in the
  local Universe or at high redshift. In total we model 19 indices in the wavelength range $\sim 1200~\AA$ to 
  $\sim 3000~\AA$. 
  
  We follow a two-fold strategy. 
  First, we have
  constructed analytical approximations (fitting functions) to the
  empirical line indices measured in single stars, as functions of the
  stellar parameters gravity and temperature. This approach has been used as the FFs
  can be conveniently inserted in population synthesis models and because fluctuations in the spectra of stars with similar atmospheric parameters are averaged out.\footnote{The fitting functions are given in the Appendix.}
  The IUE empirical
  library contains mostly stars with solar metallicity. We
  then inserted a metallicity dependence in the fitting functions
  by means of Kurucz synthetic spectra. As a next step, the fitting
  functions were
  plugged into an evolutionary synthesis code in order to compute
  integrated line indices of stellar population models as a function of
  the age and metallicity of the stellar population. In this way we have calculated semi-empirical model indices of stellar populations.
  
 In parallel we have used a high-resolution version of the Kurucz theoretical spectral library (Rodriguez-Merinos et al. 2005) to construct high-resolution SEDs of the M05 stellar population models and computed the $UV$ indices directly on these SEDs. This gives us theoretical model indices of stellar populations. \footnote{Both types of model indices as well as the high-resolution model SEDs are available at
  http://www.maraston.eu} 
  
  We have checked both models by comparing the
  synthetic line-indices with IUE data of template globular clusters
  in the Magellanic Clouds with independently known ages - from 10 Myr to 0.13 Gyr - and
  metallicities.
  
 Most of the far-$UV$ indices are found to be in satisfactory agreement with the models.
The lines of \ion{C}{iv}{} and \ion{Si}{iv}{} already have been studied
  extensively in the literature \citep{storchi, leitherer95,
  leitherer99, heckman98, rix2004}. We add here their validation with
  GCs. The other indices instead are modelled and compared with globular clusters in this work for the first time.  

 The models of most mid-$UV$ indices, instead, fail to reproduce the GC data. Given that the GC are young ($< 0.13~\rm Gyr$) and mid-$UV$ indices are strong in older populations, one cannot rule out a problem in the mid-$UV$ part of the GC observed spectra. We shall pursue this aspect in a future paper.
  
 For a quantitative test we have calculated the ages and metallicities we derive for the GCs using both empirical and Kurucz-based models. For the empirical models we use seven indices (SiIV, $\rm CIV^{A}$, CIV, $\rm CIV^{E}$, BL1617, BL1664, BL1719) whereas for the Kurucz-based models we could use all eleven indices in the far-$UV$ (see Figure~\ref{far$UV$}).
 
 For these selected sets of indices we find good agreement between the ages and metallicities we derived based on the indices and the values given in the literature. These sets can now be applied to age and metallicity derivations for distant galaxies.

  It is important to note that the theoretical models presented
  here refer to solar-scaled abundance ratios of chemical
  elements, while the empirical models refer to the pattern of the Milky Way stars out of which the input library is constructed. On the other hand,
  some evidence has been reported in the literature of enhancement
  of $\alpha$-elements with respect to iron in Magellanic Cloud
  globular clusters \citep{oliva, beaetal02}.
  Given the satisfactory agreement we obtain in the derived ages and metallicities of the LMC GCs
  one tends to conclude that, at least when using our selected set of far-$UV$ indices, possible chemical abundance anomalies do not matter.  Nonetheless, we are following up the modelling of $UV$
  indices for various abundance ratios that will hopefully clarify this issue.
  \begin{acknowledgements}
   We are grateful to the anonymous referee for
    prompt replies and constructive comments that
    significantly improved the paper, and to Christy Tremonti and Guinevere Kauffmann for clarifying comments. CM and DT acknowledge the ospitality of the European Southern Observatory Visitor Programme during which this work was completed.
    CM is a Marie Curie Excellence Grant Team Leader and
    ackwnoledges grant MEXT-CT-2006-042754 of the European Community.
  \end{acknowledgements}

  \begin{appendix}

  \section{Coefficients of the fitting functions}
  \label{sec:ffcoeff}

  Here we provide the numerical coefficients of the FFs
 and the ranges in temperature and gravity
  inside which the fitting functions are valid (Table~\ref{tab:coeff}) . The fitting functions should not be
  extrapolated outside these regions. The equivalent width at all
  temperatures and surface gravities inside the validity range is the
  third degree polynomial given by the coefficients listed in
  Table~\ref{tab:coeff} and equation~\ref{eq:ff} in Sect.~\ref{sec:ffunc}.
  When a two-region fit is necessary to construct the FF,
  each local FF (cool and hot) has to be applied in its corresponding
  region. Equation~\ref{eq:interp} applies in the overlapping region.

\begin{landscape}
  \begin{table}
	\centering
    \caption{Coefficients of the empirical fitting functions. Some indices required fits
    performed in adjacent temperature regions (see Sect.~4), in such
    cases the first row refers to the cool region and
    the second one to the hot. The continuity between the two functions is obtained using a cosin function (see Eq.~\ref{eq:interp}).
    Validity limits in \ensuremath{\log g}{} and $\log
    T_{\mathrm{eff}}$ are
		defined by the coolest and hottest stellar groups, and
		the maximum and minimum values of surface gravity
		available in the library that was included in the
		fit.  These are: $\log
    T_{\mathrm{eff}}=4.67-3.86$ and \ensuremath{\log g}{}$=1.4-4.1$ for far-$UV$ indices, 
    and  $\log T_{\mathrm{eff}}=4.67-3.78$ and \ensuremath{\log g}{}$=1.2-4.3$ for mid-$UV$ indices.}
    \label{tab:coeff}
  \begin{tabular}{lrrrrrrrrrr}
    \hline
    Index & $a_0$ & $\log T_{\mathrm{eff}}$ & \ensuremath{\log g} & $\log T_{\mathrm{eff}}^2$ &  $\log T_{\mathrm{eff}}\ensuremath{\log g}$ &
    $\ensuremath{\log g}^2$& $\log T_{\mathrm{eff}}^3$ & $\log T_{\mathrm{eff}}^2\ensuremath{\log g}$ & $\log T_{\mathrm{eff}}\ensuremath{\log g}^2$
    &$\ensuremath{\log g}^3$\\\hline\hline
\ensuremath{\mathrm{BL_{\scriptstyle{1302}}}}      
&-7.410092e+03 & 5.161229e+03 &-2.104500e+01 &-1.190241e+03 & 5.089514e+00 & 0.000000e+00 & 9.087288e+01 & 0.000000e+00 & 0.000000e+00 & 0.000000e+00\\
\ion{Si}{iv}        & 8.329603e+03 &-5.950179e+03 & 0.000000e+00 & 1.413073e+03 & 0.000000e+00 & 0.000000e+00 & -1.114978e+02 & 0.000000e+00 & 0.000000e+00 &-4.464140e-02\\
\ensuremath{\mathrm{BL_{\scriptstyle{1425}}}}           & 2.801138e+03 &-1.887279e+03 &-1.124998e+02 & 4.198010e+02 & 5.522990e+01 & 0.000000e+00 &-3.073181e+01 &-6.792400e+00 & 0.000000e+00 & 0.000000e+00\\
\ion{Fe}{v}         & 0.000000e+00 & 0.000000e+00 & 0.000000e+00 & -5.370300e+00 & 4.218601e+00 & 0.000000e+00 & 1.291487e+00 & -1.022240e+00 & 0.000000e+00 & 0.000000e+00\\
\ensuremath{\mathrm{C}^{\scriptscriptstyle{A}}_{\mathrm{IV}}}             & 0.000000e+00 & 9.475584e+01 & 0.000000e+00 &-5.981724e+01 & 1.591892e+01 & 0.000000e+0.0 & 9.063165e+00 &-3.998758e+00 & 0.000000e+00 & 0.000000e+00\\
\ion{C}{iv}         & 1.160656e+04 &-8.160319e+03 & 0.000000e+00 & 1.908096e+03 & 0.000000e+00 & 0.000000e+00 & -1.483084e+02 & 0.000000e+00 &-6.465476e-02 & 0.000000e+00\\
\ensuremath{\mathrm{C}^{\scriptscriptstyle{E}}_{\mathrm{IV}}}  & 1.146391e+04 &-8.060925e+03 & 0.000000e+00 & 1.886312e+03 & 0.000000e+00 & 0.000000e+00 & -1.468829e+02 & 0.000000e+00 & 0.000000e+00& 0.000000e+00\\
\ensuremath{\mathrm{BL_{\scriptstyle{1617}}}}    & 7.195609e+03 &-4.969059e+03 & 1.637501e+01 & 1.134789e+03 & 0.000000e+00 &0.000000e+00  & -8.554261e+01 & -1.010919e+00 & 0.000000e+00 & 0.000000e+00\\
\ensuremath{\mathrm{BL_{\scriptstyle{1664}}}}           & 3.99173e+03 &-2.69840e+03 & 0.000000e+00 & 6.070700e+02 & 0.000000e+00 & 0.000000e+00 &-4.545000e+01 & 0.000000e+00 & 0.000000e+00 & 0.000000e+00\\
\ensuremath{\mathrm{BL_{\scriptstyle{1719}}}}    & 0.000000e+00 & 0.946425e+00 & 1.39515e+00 & 0.000000e+00 & -0.463588e+00 & 0.000000e+00 & 0.000000e+00 & 0.000000e+00 & 0.000000e+00 & 0.000000e+00\\
\ensuremath{\mathrm{BL_{\scriptstyle{1853}}}}           &-3.992238e+04 & 3.132500e+04 &-8.032732e+02 &-8.174011e+03 & 4.072430e+02 & 0.000000e+00 & 7.093685e+02 &-5.124554e+01 &-2.300320e-01 & 0.000000e+00\\
                    &-1.741241e+03 & 1.236778e+03 & 0.000000e+00 &-2.912705e+02 &-1.240263e-01 & 0.000000e+00 & 2.278486e+01 & 0.000000e+00 & 0.000000e+00 & 0.000000e+00\\
\ion{Fe}{ii}~2402   &-1.025711e+05 & 7.846243e+04 & 0.000000e+00 &-1.997183e+04 &-1.152896e+01 & 0.000000e+00 & 1.691469e+03 & 3.738713e+00 &-5.328224e-01 & 0.000000e+00\\
  & 4.187948e+03 &-2.837521e+03 & 0.000000e+00 & 6.404977e+02 & 0.000000e+00 &  0.000000e+00 & -4.814913e+01 & 0.000000e+00 & 0.000000e+00 & 0.000000e+00\\
\ensuremath{\mathrm{BL_{\scriptstyle{2538}}}}      &-5.765329e+04 & 4.594702e+04 &-1.525047e+03 &-1.220363e+04 & 8.150653e+02 & 0.000000e+00 & 1.080533e+03 &-1.101304e+02 & 2.225463e+00 &-9.983048e-01\\
 & 2.215483e+02 &-9.798301e+01 & 0.000000e+00 & 1.083353e+01 & 0.000000e+00 & 0.000000e+00 & 0.000000e+00 & 0.000000e+00 & 0.000000e+00 & 0.000000e+00\\
\ion{Fe}{ii}~2609   &-4.290130e+04 & 3.423995e+04 &-1.540460e+03 &-9.101116e+03 & 8.113314e+02 & 0.000000e+00 & 8.056931e+02 &-1.065136e+02 & 0.000000e+00 &-1.165127e-01\\
&2.225659e+02       &-1.009843e+02 & 0.000000e+00 & 1.145332e+01 & 0.000000e+00 & 0.000000e+00 & 0.000000e+00 & 0.000000e+00 & 0.000000e+00 & 0.000000e+00\\
\ion{Mg}{ii}        & -8.369101e+03 & 4.425491e+03 & 0.000000e+00 & -0.584735e+03 &  2.583200e+00 & 0.000000e+00 & 0.000000e+00 & 0.000000e+00 & -0.426400e+00 & 0.000000e+00\\
                    & 4.220988e+02 &-1.924251e+02 & 0.000000e+00 & 2.195847e+01 & 0.000000e+00 & 0.000000e+00 & 0.000000e+00 & 0.000000e+00 & 0.000000e+00 & 0.000000e+00\\
\ion{Mg}{i}         &-6.560245e+04 & 5.213871e+04 & 5.667943e+00 &-1.379378e+04 & 0.000000e+00 &-7.080052e-01 & 1.214714e+03 & 0.000000e+00 & 0.000000e+00 & 0.000000e+00\\
                    & 2.833468e+02 &-1.290402e+02 & 0.000000e+00 & 1.467496e+01 & 0.000000e+00 & 0.000000e+00 & 0.000000e+00 & 0.000000e+00 & 0.000000e+00 & 0.000000e+00\\
 Mg$_{\mathrm{wide}}$    & 1.493626e+03 &-7.013927e+02 & 0.000000e+00 & 8.260315e+01 & 0.000000e+00 & 0.000000e+00 & 0.000000e+00 & 0.000000e+00 & 0.000000e+00 & 0.000000e+00\\
\ion{Fe}{i}         &-2.324729e+05 & 1.873365e+05 &-2.620267e+03 &-5.027795e+04 & 1.396493e+03 &-1.061878e+00 & 4.494049e+03 &-1.854398e+02 & 0.000000e+00 & 0.000000e+00\\
                    &1.522971e-02 & 0.000000e+00  &  0.000000e+00 & 0.000000e+00 & 0.000000e+00 & 0.000000e+00 & 0.000000e+00 & 0.000000e+00 & 0.000000e+00 & 0.000000e+00\\
 \ensuremath{\mathrm{BL_{\scriptstyle{3096}}}}     & 1.153169e+03 &-8.051949e+02 & 0.000000e+00 & 1.870753e+02 & 0.000000e+00 & 0.000000e+00 &-1.449478e+01 & 1.972870e-01 &-3.151360e-01 & 1.554005e-01\\
\hline
\end{tabular}
  \end{table}
\end{landscape}

  \section{Absorption line-indices of LMC globular clusters}
  \label{sec:ewgc}

  In Table~\ref{tab:ews} and \ref{tab:ews2} we list our measurements of the
  $UV$ absorption line-indices in the LMC globular cluster spectra. 

\setlength{\tabcolsep}{4pt}
\begin{table*}[!ht]
  \centering
  \begin{tabular}{lrrrrrrrrrrrrrrrrrrrrrr}
    \hline
    Cluster & \multicolumn{2}{c}{\ensuremath{\mathrm{BL_{\scriptstyle{1302}}}}} & \multicolumn{2}{c}{\ion{Si}{iv}}
    & \multicolumn{2}{c}{\ensuremath{\mathrm{BL_{\scriptstyle{1425}}}}} & \multicolumn{2}{c}{\ion{Fe}{v}} &\multicolumn{2}{c}{\ensuremath{\mathrm{C}^{\scriptscriptstyle{A}}_{\mathrm{IV}}}}
   &  \multicolumn{2}{c}{\ion{C}{iv}}  & \multicolumn{2}{c}{\ensuremath{\mathrm{C}^{\scriptscriptstyle{E}}_{\mathrm{IV}}}} &
    \multicolumn{2}{c}{\ensuremath{\mathrm{BL_{\scriptstyle{1617}}}}} & \multicolumn{2}{c}{\ensuremath{\mathrm{BL_{\scriptstyle{1664}}}}} &    \multicolumn{2}{c}{\ensuremath{\mathrm{BL_{\scriptstyle{1719}}}}} &
    \multicolumn{2}{c}{\ensuremath{\mathrm{BL_{\scriptstyle{1853}}}}} \\ \hline\hline
NGC 2011 & 3.9 &   0.8 &  3.5 &  0.7 & 0.9 &  0.7 &  -0.6 &  1.1 &  4.1 &   0.6 &  4.1 &   0.6 &  2.7 & 0.6 & 1.8 &   0.8 &  1.1 &   0.8 &  1.8 & 0.6 & 1.1 &   0.9 \\ 
 NGC 1805 & 5.1 &   0.6 &  2.5 &  0.7 & 1.2 &  1.7 & -0.5 &  1.1 &  1.8 &   0.8 &  1.7 &   0.7 & 0.9 & 0.8 & -0.7 &   1.1 & -0.3 &   0.8 &  1.2 & 0.6 & 1.0 &   0.5 \\  
 NGC 1984 & 3.9 & 1.0 &  4.2 &  0.9 &  1.6 &  1.0 &  1.0 & 1.5 &  2.1 &   1.1 &  2.4 &   1.0 &  1.2 & 1.1 & 1.8 &   1.5 &  0.5 &   1.5 &  1.5 & 1.3 & 1.5 &   1.6 \\ 
  NGC 2100 & 3.9 &   0.4 &  3.4 &  0.4 & 0.3 &  0.5 &  -0.9 & -0.7 &  1.8 &   0.5 &  2.1 &   0.5 &  1.7 & 0.4 & 1.7&    0.5 &  1.8 &   0.5 &  1.8 & 0.5 & 1.4 &   0.6 \\ 
NGC 2004 & 4.2 &   0.3 &  3.7 &  0.4 & 0.6 &  0.4 & -1.0 &  0.6 &  2.9 &   0.4 &  2.8 &   1.5 & 0.4 & 0.4 &   0.9 &   0.5 &  0.1 &   0.6 & 1.9 & 0.4 & 1.4 &   0.6 \\
NGC 1818 & 3.8 &   0.9 &  1.6 &  0.8 & 0.4 & 0.8 & -1.7 &  1.1 &  1.2 &   0.9 &  1.4 &   0.7 &  1.6 & 0.7 & 0.9 &   1.1 &  0.9 &   1.2 &  1.5 & 1 & 1.2 &  1.0 \\
NGC 1847 & 4.5 &   0.6 &  2.8 &  0.6 & 0.9 & 0.7 & -0.7 & 1.0 &  1.4 &   0.7 &  2.5 &   0.6 &   1.6 & 0.7 & 0.9 &   0.9 &  0.8 &   0.8 & 1.0 & 0.6 &  2.0 &   0.6 \\
NGC 1711 & 4.1 &   0.4 &  1.8 &  0.5 & -0.3 &  0.5 & -0.9 &  0.7 &  1.5 &   0.5 &  1.6 &   0.4 &   0.9 & 0.5 & 0.8 &   0.6 &  1.6 &   0.5 & 1.4 & 0.4 &  1.0 &   0.4 \\
NGC 1866 & 4.1 &   0.7 &  0.3 &  0.7 & -0.1 & 0.7 & -2.3 & 1.2 &  0.4 &   0.6 &  0.5 &   0.5 &  0.8 & 0.6 & 1.2 &   0.7 &  1.8 &   0.6 & 1.8 & 0.4 & 1.9 &   0.6 \\ 
NGC 1850 & 4.4 &   0.5 &  0.8 &  0.6 & -0.3 & 0.6 & -2.1 &  1.0 &  1.7 &   0.6 &  1.4 &   0.6 &  1.9 & 0.6 & 1.2 &   0.7 &  1.1 &   0.6 &  1.5 & 0.4 & 1.8 &   0.3 \\ 
\hline
\end{tabular}
  \caption{Equivalent widths and estimated errors (both in \AA) of
  far-$UV$ indices as measured in the IUE spectra of LMC
  globular clusters.}
  \label{tab:ews}
\end{table*}

  \begin{table*}[!ht]
    \centering
    \begin{tabular}{lrrrrrrrrrrrrrrrr}
      \hline
      Cluster &  
  \multicolumn{2}{c}{\ion{Fe}{ii}~2402} &
  \multicolumn{2}{c}{\ensuremath{\mathrm{BL_{\scriptstyle{2538}}}}} &
      \multicolumn{2}{c}{\ion{Fe}{i}i2609} & \multicolumn{2}{c}{\ion{Mg}{ii}2800} &
      \multicolumn{2}{c}{\ion{Mg}{i}2852} &  \multicolumn{2}{c}{\ensuremath{\mathrm{Mg}_{\mathrm{wide}}}} &\multicolumn{2}{c}{\ion{Fe}{i}3000} &
      \multicolumn{2}{c}{\ensuremath{\mathrm{BL_{\scriptstyle{3096}}}}}\\ \hline\hline
 NGC 2011 &2.9 & 0.7 & 0.9 &  0.5 & 1.4 &  0.3 & 6.0 &  0.6 & 0.5 &  0.7 & 4.4 & 2.6 &-1.2 &  1.2 &-0.4 &  0.3 \\
 NGC 1805 & 3.9 & 2.0 & 0.3 & 1.7 & 2.0 &  1.2 & 5.5 &  1.1 & 1.0 &  1.2 & 7.7 & 9.3 & 3.2 &  3.2 &-2.1 &  2.2 \\
 NGC 1984 & 3.7 & 1.5 &-0.8 & 1.1 & 1.6 &  0.6 & 3.6 &  0.5 & 1.5 &  0.4 & -2.1 & 4.8 & 1.0 &  1.3 &-0.3 & 0.6 \\
 NGC 2100 & 2.1 & 1.9 & -2.5 & 1.5 & 0.9 &  0.9 & 6.9 &  0.9 & 2.4 &  1.1 & 0.3 & 8.1 & 1.1 &  3.0 &-0.0 & 1.4 \\   
 NGC 2004 & 0.8 & 0.8 &-1.3 &  0.8 & 1.3 &  0.6 & 6.2 &  0.6 & 2.0 &  0.6 &12.7 & 4.2 & 0.3 & 1.3 &-0.7 &  0.6 \\ 
 NGC 1818 & 1.5 & 1.3 & -0.7 &  1.1 & 1.2 &  0.7 & 4.9 &  0.6 & 1.6 &  0.7 &6.2 & 5.5 &  0.0 &  1.5 & 0.1 & 1.0 \\  
 NGC 1847 & 1.6 & 1.7 & -1.3 &  1.1 & 1.4 &  0.7 & 4.7 &  0.6 & 1.4 &  0.6 & 3.6 & 5.1 &-2.1 &  1.9 &-1.3 & 0.9 \\  
NGC 1711 & 2.2 & 1.5 & -1.2 &  1.4 & 1.6 &  0.9 & 6.1 &  0.8 & 1.8 &  0.9 & 4.1 & 6.8 & 2.7 &  2.2 &-1.1 &  1.3 \\  
NGC 1866 & 2.4 & 1.6 & -0.1 &  1.4 & 2.0 &  0.9 & 7.1 &  0.5 & 2.4 &  0.5 & 12 & 5.2 & 1.4 &  1.4 &-0.8 & 1.0 
\\ 
NGC 1850 & 0.7 & 1.8 & -2.7 &  1.5 & 1.2 &  0.9 & 6.1 &  0.9 & 2.8 &  0.9 & 8.6 & 7.1 & 0.1 &  2.5 &-1.4 & 1.5 \\  
 \hline
 \end{tabular}
 \caption{Same as B.1 for mid-$UV$ indices.}
  \label{tab:ews2}
\end{table*}

  \section{Age and metallicity estimates of LMC globular clusters based on model line-indices}
  \label{sec:age_met}
  \begin{table}[!ht]
    \centering
    \begin{tabular}{lrrrrrr}\hline
      Cluster & $\log (\mathrm{age})$ & $[\mathrm{Z}/\mathrm{H}]$ & $\log
      (\mathrm{age})$ & $[\mathrm{Z}/\mathrm{H}]$ &
       $\log (\mathrm{age})$ &  $[\mathrm{Z}/\mathrm{H}]$\\
                     & \multicolumn{2}{c}{(Literature)} &       
              \multicolumn{2}{c}{empirical models} &
              \multicolumn{2}{c}{Kurucz-based models}\\
      \hline\hline
NGC 1711 &  7.70 & -0.57 & 7.10 & -1.00 &   7.50 & -0.10 \\
NGC 1805 &  7.00 & -0.20 & 7.33 & -0.33 &   7.20 & -0.33 \\
NGC 1818 &  7.40 & -0.20 & 7.15 & -1.00 &   7.50 & -0.20  \\
NGC 1847 &  7.42 & -0.37 & 7.10 &  -0.60 &   6.85 & -0.40 \\
NGC 1850 &  7.40 & -0.12 & 7.10 &  -1.00 &   8.00 & -0.10  \\
NGC 1866 &  8.12 & -0.50 & 8.25 &  0.10 &   8.00 &  -0.50 \\
NGC 1984 &  7.06 & -0.90 & 6.95 &  -0.60 &   7.10 &  -0.10  \\
NGC 2004 &  7.30 & -0.56 & 6.95 & -0.50 &   7.20 &  0.00 \\
NGC 2011 &  6.99 & -0.47 & 6.80 &  -0.33 &   6.70 &  0.00 \\
NGC 2100 &  7.20 & -0.32 & 7.00 &  -0.60 &   7.35 &  0.20  \\
\hline
    \end{tabular}
    \caption{Ages and metallicities derived from 7 empirical model indices and from 11 Kurucz-based model indices as discussed in Section 6 (see Fig~\ref{fig:age_est}
		and \ref{fig:met_est}).}
    \label{tab:agemetest}
  \end{table}
 \end{appendix} 

\begin{thebibliography}{115}
\expandafter\ifx\csname natexlab\endcsname\relax\def\natexlab#1{#1}\fi

\bibitem[{{Alexander}(1967)}]{alexander67}
{Alexander}, J.~B. 1967, \mnras, 137, 41

\bibitem[{{Beasley} {et~al.}(2002){Beasley}, {Hoyle}, \&
  {Sharples}}]{beaetal02}
{Beasley}, M.~A., {Hoyle}, F., \& {Sharples}, R.~M. 2002, \mnras, 336, 168

\bibitem[{{Bell} \& {Rodgers}(1965)}]{bell65}
{Bell}, R.~A. \& {Rodgers}, A.~W. 1965, \mnras, 129, 127

\bibitem[{{Bertelli} {et~al.}(1994){Bertelli}, {Bressan}, {Chiosi}, {Fagotto},
  \& {Nasi}}]{bertelli94}
{Bertelli}, G., {Bressan}, A., {Chiosi}, C., {Fagotto}, F., \& {Nasi}, E. 1994,
  \aaps, 106, 275

\bibitem[{{Boesgaard}(1989)}]{boesgaard89}
{Boesgaard}, A.~M. 1989, \apj, 336, 798

\bibitem[{{Boesgaard} \& {Friel}(1990)}]{boesgaard90}
{Boesgaard}, A.~M. \& {Friel}, E.~D. 1990, \apj, 351, 467

\bibitem[{{Bonatto} {et~al.}(1995){Bonatto}, {Bica}, \& {Alloin}}]{bonatto}
{Bonatto}, C., {Bica}, E., \& {Alloin}, D. 1995, \aaps, 112, 71

\bibitem[{{Bruzual} \& {Charlot}(2003)}]{bc03}
{Bruzual}, G. \& {Charlot}, S. 2003, \mnras, 344, 1000

\bibitem[{{Bruzual A.} \& {Charlot}(1993)}]{bc93}
{Bruzual A.}, G. \& {Charlot}, S. 1993, \apj, 405, 538

\bibitem[{{Burstein} {et~al.}(1988){Burstein}, {Bertola}, {Buson}, {Faber}, \&
  {Lauer}}]{burstein}
{Burstein}, D., {Bertola}, F., {Buson}, L.~M., {Faber}, S.~M., \& {Lauer},
  T.~R. 1988, \apj, 328, 440

\bibitem[{{Burstein} {et~al.}(1984){Burstein}, {Faber}, {Gaskell}, \&
  {Krumm}}]{buretal84}
{Burstein}, D., {Faber}, S.~M., {Gaskell}, C.~M., \& {Krumm}, N. 1984, \apj,
  287, 586

\bibitem[{{Buzzoni}(1989)}]{buzzoni}
{Buzzoni}, A. 1989, \apjs, 71, 817

\bibitem[{{Buzzoni} {et~al.}(1992){Buzzoni}, {Gariboldi}, \&
  {Mantegazza}}]{buzetal92}
{Buzzoni}, A., {Gariboldi}, G., \& {Mantegazza}, L. 1992, \aj, 103, 1814

\bibitem[{{Cardiel} {et~al.}(1998){Cardiel}, {Gorgas}, {Cenarro}, \&
  {Gonzalez}}]{cardiel}
{Cardiel}, N., {Gorgas}, J., {Cenarro}, J., \& {Gonzalez}, J.~J. 1998, \aaps,
  127, 597

\bibitem[{{Cassatella} {et~al.}(1987){Cassatella}, {Barbero}, \&
  {Geyer}}]{cassatella}
{Cassatella}, A., {Barbero}, J., \& {Geyer}, E.~H. 1987, \apjs, 64, 83

\bibitem[{{Castellani} {et~al.}(1992){Castellani}, {Chieffi}, \&
  {Straniero}}]{castellani}
{Castellani}, V., {Chieffi}, A., \& {Straniero}, O. 1992, \apjs, 78, 517

\bibitem[{{Castelli} {et~al.}(1984){Castelli}, {Cornachin}, {Morossi}, \&
  {Hack}}]{casetal84}
{Castelli}, F., {Cornachin}, M., {Morossi}, C., \& {Hack}, M. 1984, \aap, 141,
  223

\bibitem[{{Cayrel} {et~al.}(1985){Cayrel}, {Cayrel de Strobel}, \&
  {Campbell}}]{cayrel85}
{Cayrel}, R., {Cayrel de Strobel}, G., \& {Campbell}, B. 1985, \aap, 146, 249

\bibitem[{{Cayrel de Strobel} {et~al.}(1970){Cayrel de Strobel},
  {Chauve-Godard}, {Hernandez}, \& {Vaziaga}}]{cayrel70}
{Cayrel de Strobel}, G., {Chauve-Godard}, J., {Hernandez}, G., \& {Vaziaga},
  M.~J. 1970, \aap, 7, 408

\bibitem[{{Cayrel de Strobel} {et~al.}(1997){Cayrel de Strobel}, {Soubiran},
  {Friel}, {Ralite}, \& {Francois}}]{cayrel97}
{Cayrel de Strobel}, G., {Soubiran}, C., {Friel}, E.~D., {Ralite}, N., \&
  {Francois}, P. 1997, \aaps, 124, 299

\bibitem[{{Cenarro} {et~al.}(2002){Cenarro}, {Gorgas}, {Cardiel}, {Vazdekis},
  \& {Peletier}}]{cenetal02}
{Cenarro}, A.~J., {Gorgas}, J., {Cardiel}, N., {Vazdekis}, A., \& {Peletier},
  R.~F. 2002, \mnras, 329, 863

\bibitem[{{Chaffee} {et~al.}(1971){Chaffee}, {Carbon}, \& {Strom}}]{chaffee71}
{Chaffee}, Jr., F.~H., {Carbon}, D.~F., \& {Strom}, S.~E. 1971, \apj, 166, 593

\bibitem[{{Chavez} {et~al.}(2007){Chavez}, {Bertone}, {Buzzoni}, {Franchini},
  {Malagnini}, {Morossi}, \& {Rodriguez-Merino}}]{chaetal07}
{Chavez}, M., {Bertone}, E., {Buzzoni}, A., {et~al.} 2007, \apj, 657, 1046

\bibitem[{{Cimatti} {et~al.}(2004){Cimatti}, {Daddi}, {Renzini}, {Cassata},
  {Vanzella}, {Pozzetti}, {Cristiani}, {Fontana}, {Rodighiero}, {Mignoli}, \&
  {Zamorani}}]{cimattietal04}
{Cimatti}, A., {Daddi}, E., {Renzini}, A., {et~al.} 2004, \nat, 430, 184

\bibitem[{{Coluzzi}(1993)}]{coluzzi}
{Coluzzi}, R. 1993, Bulletin d'Information du Centre de Donnees Stellaires, 43,
  7

\bibitem[{{Conti} {et~al.}(1965){Conti}, {Wallerstein}, \& {Wing}}]{conti65}
{Conti}, P.~S., {Wallerstein}, G., \& {Wing}, R.~F. 1965, \apj, 142, 999

\bibitem[{{Daddi} {et~al.}(2005){Daddi}, {Renzini}, {Pirzkal}, {Cimatti},
  {Malhotra}, {Stiavelli}, {Xu}, {Pasquali}, {Rhoads}, {Brusa}, {di Serego
  Alighieri}, {Ferguson}, {Koekemoer}, {Moustakas}, {Panagia}, \&
  {Windhorst}}]{daddietal05}
{Daddi}, E., {Renzini}, A., {Pirzkal}, N., {et~al.} 2005, \apj, 626, 680

\bibitem[{{de Boer}(1985)}]{deboer}
{de Boer}, K.~S. 1985, \aap, 142, 321

\bibitem[{{de Grijs} {et~al.}(2002){de Grijs}, {Gilmore}, {Johnson}, \&
  {Mackey}}]{degrijs}
{de Grijs}, R., {Gilmore}, G.~F., {Johnson}, R.~A., \& {Mackey}, A.~D. 2002,
  \mnras, 331, 245

\bibitem[{{de Jager} \& {Nieuwenhuijzen}(1987)}]{dejager}
{de Jager}, C. \& {Nieuwenhuijzen}, H. 1987, \aap, 177, 217

\bibitem[{{de Mello} {et~al.}(2004){de Mello}, {Daddi}, {Renzini}, {Cimatti},
  {di Serego Alighieri}, {Pozzetti}, \& {Zamorani}}]{demello04}
{de Mello}, D.~F., {Daddi}, E., {Renzini}, A., {et~al.} 2004, \apjl, 608, L29

\bibitem[{{de Mello} {et~al.}(2000){de Mello}, {Leitherer}, \&
  {Heckman}}]{demello}
{de Mello}, D.~F., {Leitherer}, C., \& {Heckman}, T.~M. 2000, \apj, 530, 251

\bibitem[{{Dean} \& {Bruhweiler}(1985)}]{dean}
{Dean}, C.~A. \& {Bruhweiler}, F.~C. 1985, \apjs, 57, 133

\bibitem[{{Dirsch} {et~al.}(2000){Dirsch}, {Richtler}, {Gieren}, \&
  {Hilker}}]{dirsch}
{Dirsch}, B., {Richtler}, T., {Gieren}, W.~P., \& {Hilker}, M. 2000, \aap, 360,
  133

\bibitem[{{Dorman} {et~al.}(1995){Dorman}, {O'Connell}, \& {Rood}}]{dorman}
{Dorman}, B., {O'Connell}, R.~W., \& {Rood}, R.~T. 1995, \apj, 442, 105

\bibitem[{{Dorman} {et~al.}(1993){Dorman}, {Rood}, \& {O'Connell}}]{doretal93}
{Dorman}, B., {Rood}, R.~T., \& {O'Connell}, R.~W. 1993, \apj, 419, 596

\bibitem[{{Elson} \& {Fall}(1988)}]{elson_fall}
{Elson}, R.~A. \& {Fall}, S.~M. 1988, \aj, 96, 1383

\bibitem[{{Elson}(1991)}]{elson}
{Elson}, R.~A.~W. 1991, \apjs, 76, 185

\bibitem[{{Faber} {et~al.}(1985){Faber}, {Friel}, {Burstein}, \&
  {Gaskell}}]{fabetal85}
{Faber}, S.~M., {Friel}, E.~D., {Burstein}, D., \& {Gaskell}, C.~M. 1985,
  \apjs, 57, 711

\bibitem[{{Fanelli} {et~al.}(1990){Fanelli}, {O'Connell}, {Burstein}, \&
  {Wu}}]{fan_iii}
{Fanelli}, M.~N., {O'Connell}, R.~W., {Burstein}, D., \& {Wu}, C. 1990, \apj,
  364, 272

\bibitem[{{Fanelli} {et~al.}(1992){Fanelli}, {O'Connell}, {Burstein}, \&
  {Wu}}]{fan_iv}
{Fanelli}, M.~N., {O'Connell}, R.~W., {Burstein}, D., \& {Wu}, C. 1992, \apjs,
  82, 197

\bibitem[{{Fanelli} {et~al.}(1987){Fanelli}, {O'Connell}, \& {Thuan}}]{fan_i}
{Fanelli}, M.~N., {O'Connell}, R.~W., \& {Thuan}, T.~X. 1987, \apj, 321, 768

\bibitem[{{Fanelli} {et~al.}(1988){Fanelli}, {O'Connell}, \& {Thuan}}]{fan_ii}
{Fanelli}, M.~N., {O'Connell}, R.~W., \& {Thuan}, T.~X. 1988, \apj, 334, 665

\bibitem[{{Fioc} \& {Rocca-Volmerange}(1997)}]{fioc}
{Fioc}, M. \& {Rocca-Volmerange}, B. 1997, \aap, 326, 950

\bibitem[{{Girardi} {et~al.}(2002){Girardi}, {Bertelli}, {Bressan}, {Chiosi},
  {Groenewegen}, {Marigo}, {Salasnich}, \& {Weiss}}]{girardi02}
{Girardi}, L., {Bertelli}, G., {Bressan}, A., {et~al.} 2002, \aap, 391, 195

\bibitem[{{Gorgas} {et~al.}(1993){Gorgas}, {Faber}, {Burstein}, {Gonzalez},
  {Courteau}, \& {Prosser}}]{goretal93}
{Gorgas}, J., {Faber}, S.~M., {Burstein}, D., {et~al.} 1993, \apjs, 86, 153

\bibitem[{{Graves} \& {Schiavon}(2008)}]{grasch08}
{Graves}, G.~J. \& {Schiavon}, R.~P. 2008, ArXiv e-prints, 803

\bibitem[{{Greggio} \& {Renzini}(1990)}]{gr_ren}
{Greggio}, L. \& {Renzini}, A. 1990, \apj, 364, 35

\bibitem[{{Heap} {et~al.}(1998){Heap}, {Brown}, {Hubeny}, {Landsman}, {Yi},
  {Fanelli}, {Gardner}, {Lanz}, {Maran}, {Sweigart}, {Kaiser}, {Linsky},
  {Timothy}, {Lindler}, {Beck}, {Bohlin}, {Clampin}, {Grady}, {Loiacono}, \&
  {Krebs}}]{heapetal98}
{Heap}, S.~R., {Brown}, T.~M., {Hubeny}, I., {et~al.} 1998, \apjl, 492, L131+

\bibitem[{{Hearnshaw}(1974)}]{hearnshaw74}
{Hearnshaw}, J.~B. 1974, \aap, 36, 191

\bibitem[{{Heckman} {et~al.}(1998){Heckman}, {Robert}, {Leitherer}, {Garnett},
  \& {van der Rydt}}]{heckman98}
{Heckman}, T.~M., {Robert}, C., {Leitherer}, C., {Garnett}, D.~R., \& {van der
  Rydt}, F. 1998, \apj, 503, 646

\bibitem[{{Helfer} {et~al.}(1960){Helfer}, {Wallerstein}, \&
  {Greenstein}}]{helfer60}
{Helfer}, H.~L., {Wallerstein}, G., \& {Greenstein}, J.~L. 1960, \apj, 132, 553

\bibitem[{{Hill} {et~al.}(2000){Hill}, {Fran{\c c}ois}, {Spite}, {Primas}, \&
  {Spite}}]{hill}
{Hill}, V., {Fran{\c c}ois}, P., {Spite}, M., {Primas}, F., \& {Spite}, F.
  2000, \aap, 364, L19

\bibitem[{{Howarth} \& {Prinja}(1989)}]{howarth89}
{Howarth}, I.~D. \& {Prinja}, R.~K. 1989, \apjs, 69, 527

\bibitem[{{Humphreys} \& {McElroy}(1984)}]{hump}
{Humphreys}, R.~M. \& {McElroy}, D.~B. 1984, \apj, 284, 565

\bibitem[{{Jasniewicz} \& {Thevenin}(1994)}]{jasn}
{Jasniewicz}, G. \& {Thevenin}, F. 1994, \aap, 282, 717

\bibitem[{{Johnson} {et~al.}(2001){Johnson}, {Beaulieu}, {Gilmore}, {Hurley},
  {Santiago}, {Tanvir}, \& {Elson}}]{johnson}
{Johnson}, R.~A., {Beaulieu}, S.~F., {Gilmore}, G.~F., {et~al.} 2001, \mnras,
  324, 367

\bibitem[{{Kinney} {et~al.}(1993){Kinney}, {Bohlin}, {Calzetti}, {Panagia}, \&
  {Wyse}}]{kinney}
{Kinney}, A.~L., {Bohlin}, R.~C., {Calzetti}, D., {Panagia}, N., \& {Wyse},
  R.~F.~G. 1993, \apjs, 86, 5

\bibitem[{{Koleva} {et~al.}(2008){Koleva}, {Prugniel}, {Ocvirk}, {Le Borgne},
  \& {Soubiran}}]{koletal08}
{Koleva}, M., {Prugniel}, P., {Ocvirk}, P., {Le Borgne}, D., \& {Soubiran}, C.
  2008, \mnras, 385, 1998

\bibitem[{{Kroupa}(2001)}]{Kroupa2001}
{Kroupa}, P. 2001, \mnras, 322, 231

\bibitem[{{Kudritzki} {et~al.}(1987){Kudritzki}, {Pauldrach}, \&
  {Puls}}]{kudritzki87}
{Kudritzki}, R.~P., {Pauldrach}, A., \& {Puls}, J. 1987, \aap, 173, 293

\bibitem[{{Lee} \& {Worthey}(2005)}]{leewor05}
{Lee}, H.-c. \& {Worthey}, G. 2005, \apjs, 160, 176

\bibitem[{{Leitherer} \& {Heckman}(1995)}]{leitherer95}
{Leitherer}, C. \& {Heckman}, T.~M. 1995, \apjs, 96, 9

\bibitem[{{Leitherer} \& {Lamers}(1991)}]{leitherer91}
{Leitherer}, C. \& {Lamers}, H.~J.~G.~L. 1991, \apj, 373, 89

\bibitem[{{Leitherer} {et~al.}(1999){Leitherer}, {Schaerer}, {Goldader},
  {Delgado}, {Robert}, {Kune}, {de Mello}, {Devost}, \&
  {Heckman}}]{leitherer99}
{Leitherer}, C., {Schaerer}, D., {Goldader}, J.~D., {et~al.} 1999, \apjs, 123,
  3

\bibitem[{{Lejeune} \& {Schaerer}(2001)}]{lej01}
{Lejeune}, T. \& {Schaerer}, D. 2001, \aap, 366, 538

\bibitem[{{Leonardi} \& {Rose}(2003)}]{leoros03}
{Leonardi}, A.~J. \& {Rose}, J.~A. 2003, \aj, 126, 1811

\bibitem[{{Lotz} {et~al.}(2000){Lotz}, {Ferguson}, \& {Bohlin}}]{lotz00}
{Lotz}, J.~M., {Ferguson}, H.~C., \& {Bohlin}, R.~C. 2000, \apj, 532, 830

\bibitem[{{Luck}(1979)}]{luck79}
{Luck}, R.~E. 1979, \apj, 232, 797

\bibitem[{{Luck} \& {Lambert}(1981)}]{luck81}
{Luck}, R.~E. \& {Lambert}, D.~L. 1981, \apj, 245, 1018

\bibitem[{{Mackey} \& {Gilmore}(2003)}]{mackey}
{Mackey}, A.~D. \& {Gilmore}, G.~F. 2003, \mnras, 338, 85

\bibitem[{{Maraston}(1998)}]{claudia_mod}
{Maraston}, C. 1998, \mnras, 300, 872

\bibitem[{{Maraston}(2005)}]{claudia2005}
{Maraston}, C. 2005, \mnras, 362, 799

\bibitem[{{Maraston} {et~al.}(2003){Maraston}, {Greggio}, {Renzini},
  {Ortolani}, {Saglia}, {Puzia}, \& {Kissler-Patig}}]{claudia_cl}
{Maraston}, C., {Greggio}, L., {Renzini}, A., {et~al.} 2003, \aap, 400, 823

\bibitem[{{Massa}(1989)}]{massa}
{Massa}, D. 1989, \aap, 224, 131

\bibitem[{{McCarthy} {et~al.}(2004){McCarthy}, {Le Borgne}, {Crampton}, {Chen},
  {Abraham}, {Glazebrook}, {Savaglio}, {Carlberg}, {Marzke}, {Roth},
  {J{\o}rgensen}, {Hook}, {Murowinski}, \& {Juneau}}]{mcaretal04}
{McCarthy}, P.~J., {Le Borgne}, D., {Crampton}, D., {et~al.} 2004, \apjl, 614,
  L9

\bibitem[{{Mehlert} {et~al.}(2001){Mehlert}, {Seitz}, {Saglia}, {Appenzeller},
  {Bender}, {Fricke}, {Hoffmann}, {Hopp}, {Kudritzki}, \&
  {Pauldrach}}]{mehlert01}
{Mehlert}, D., {Seitz}, S., {Saglia}, R.~P., {et~al.} 2001, \aap, 379, 96

\bibitem[{{Meynet} {et~al.}(1994){Meynet}, {Maeder}, {Schaller}, {Schaerer}, \&
  {Charbonnel}}]{meynet94}
{Meynet}, G., {Maeder}, A., {Schaller}, G., {Schaerer}, D., \& {Charbonnel}, C.
  1994, \aaps, 103, 97

\bibitem[{{Moore}(1952)}]{moore}
{Moore}, C.~E. 1952, {An ultraviolet multiplet table} (NBS Circular,
  Washington: US Government Printing Office (USGPO), |c1952)

\bibitem[{{Morton}(1975)}]{moretal75}
{Morton}, D.~C. 1975, \apj, 197, 85

\bibitem[{{O'Connell}(1976)}]{oconnell}
{O'Connell}, R.~W. 1976, \apj, 206, 370

\bibitem[{{Oliva} \& {Origlia}(1998)}]{oliva}
{Oliva}, E. \& {Origlia}, L. 1998, \aap, 332, 46

\bibitem[{{Parker} {et~al.}(1961){Parker}, {Greenstein}, {Helfer}, \&
  {Wallerstein}}]{parker61}
{Parker}, R., {Greenstein}, J.~L., {Helfer}, H.~L., \& {Wallerstein}, G. 1961,
  \apj, 133, 101

\bibitem[{{Pauldrach} {et~al.}(2001){Pauldrach}, {Hoffmann}, \& {Lennon}}]{adi}
{Pauldrach}, A.~W.~A., {Hoffmann}, T.~L., \& {Lennon}, M. 2001, \aap, 375, 161

\bibitem[{{Persson} {et~al.}(1983){Persson}, {Aaronson}, {Cohen}, {Frogel}, \&
  {Matthews}}]{persson}
{Persson}, S.~E., {Aaronson}, M., {Cohen}, J.~G., {Frogel}, J.~A., \&
  {Matthews}, K. 1983, \apj, 266, 105

\bibitem[{{Pettini} {et~al.}(2000){Pettini}, {Steidel}, {Adelberger},
  {Dickinson}, \& {Giavalisco}}]{pettini}
{Pettini}, M., {Steidel}, C.~C., {Adelberger}, K.~L., {Dickinson}, M., \&
  {Giavalisco}, M. 2000, \apj, 528, 96

\bibitem[{{Ponder} {et~al.}(1998){Ponder}, {Burstein}, {O'Connell}, {Rose},
  {Frogel}, {Wu}, {Crenshaw}, {Rieke}, \& {Tripicco}}]{ponder}
{Ponder}, J.~M., {Burstein}, D., {O'Connell}, R.~W., {et~al.} 1998, \aj, 116,
  2297

\bibitem[{{Popesso} {et~al.}(2008){Popesso}, {Dickinson}, {Nonino}, {Vanzella},
  {Daddi}, {Fosbury}, {Kuntschner}, {Mainieri}, {Cristiani}, {Cesarsky},
  {Giavalisco}, {Renzini}, \& {the GOODS Team}}]{popetal08}
{Popesso}, P., {Dickinson}, M., {Nonino}, M., {et~al.} 2008, ArXiv e-prints,
  802

\bibitem[{{Proctor} {et~al.}(2004){Proctor}, {Forbes}, \&
  {Beasley}}]{proetal04}
{Proctor}, R.~N., {Forbes}, D.~A., \& {Beasley}, M.~A. 2004, \mnras, 355, 1327

\bibitem[{{Renzini} \& {Fusi Pecci}(1988)}]{renfus88}
{Renzini}, A. \& {Fusi Pecci}, F. 1988, \araa, 26, 199

\bibitem[{{Rix} {et~al.}(2004){Rix}, {Pettini}, {Leitherer}, {Bresolin},
  {Kudritzki}, \& {Steidel}}]{rix2004}
{Rix}, S.~A., {Pettini}, M., {Leitherer}, C., {et~al.} 2004, \apj, 615, 98

\bibitem[{{Robert} {et~al.}(1993){Robert}, {Leitherer}, \&
  {Heckman}}]{robert1993}
{Robert}, C., {Leitherer}, C., \& {Heckman}, T.~M. 1993, \apj, 418, 749

\bibitem[{{Rodr{\'{\i}}guez-Merino} {et~al.}(2005){Rodr{\'{\i}}guez-Merino},
  {Chavez}, {Bertone}, \& {Buzzoni}}]{rodetal05}
{Rodr{\'{\i}}guez-Merino}, L.~H., {Chavez}, M., {Bertone}, E., \& {Buzzoni}, A.
  2005, \apj, 626, 411

\bibitem[{Salpeter(1955)}]{salp}
Salpeter, E. 1955, ApJ, 121, 161

\bibitem[{{Schaerer} {et~al.}(1993){Schaerer}, {Meynet}, {Maeder}, \&
  {Schaller}}]{schaerer93}
{Schaerer}, D., {Meynet}, G., {Maeder}, A., \& {Schaller}, G. 1993, \aaps, 98,
  523

\bibitem[{{Schaller} {et~al.}(1992){Schaller}, {Schaerer}, {Meynet}, \&
  {Maeder}}]{gmodelsun}
{Schaller}, G., {Schaerer}, D., {Meynet}, G., \& {Maeder}, A. 1992, \aaps, 96,
  269

\bibitem[{{Schiavon}(2007)}]{sch07}
{Schiavon}, R.~P. 2007, \apjs, 171, 146

\bibitem[{{Schmidt-Kaler}(1982)}]{sk}
{Schmidt-Kaler}, T. 1982, Intrinsic colors and visual absolute magnitudes
  (calibration of the MK system), ed. L.~H. {Aller}, I.~{Appenzeller},
  B.~{Baschek}, H.~W. {Duerbeck}, T.~{Herczeg}, E.~{Lamla},
  E.~{Meyer-Hofmeister}, T.~{Schmidt-Kaler}, M.~{Scholz}, W.~{Seggewiss}, W.~C.
  {Seitter}, \& V.~{Weidemann}, Vol.~2, 14

\bibitem[{{Steidel} {et~al.}(1996){Steidel}, {Giavalisco}, {Dickinson}, \&
  {Adelberger}}]{steetal96}
{Steidel}, C.~C., {Giavalisco}, M., {Dickinson}, M., \& {Adelberger}, K.~L.
  1996, \aj, 112, 352

\bibitem[{{Storchi-Bergmann} {et~al.}(1994){Storchi-Bergmann}, {Calzetti}, \&
  {Kinney}}]{storchi}
{Storchi-Bergmann}, T., {Calzetti}, D., \& {Kinney}, A.~L. 1994, \apj, 429, 572

\bibitem[{{Thomas} {et~al.}(2003){Thomas}, {Maraston}, \& {Bender}}]{tmb03}
{Thomas}, D., {Maraston}, C., \& {Bender}, R. 2003, \mnras, 339, 897

\bibitem[{{Tomkin} \& {Lambert}(1978)}]{tomkin78}
{Tomkin}, J. \& {Lambert}, D.~L. 1978, \apj, 223, 937

\bibitem[{{Underhill} {et~al.}(1972){Underhill}, {Leckrone}, \&
  {West}}]{underhill}
{Underhill}, A.~B., {Leckrone}, D.~S., \& {West}, D.~K. 1972, \apj, 171, 63

\bibitem[{{Vazdekis} {et~al.}(1996){Vazdekis}, {Casuso}, {Peletier}, \&
  {Beckman}}]{vazdekis96}
{Vazdekis}, A., {Casuso}, E., {Peletier}, R.~F., \& {Beckman}, J.~E. 1996,
  \apjs, 106, 307

\bibitem[{{Walborn} \& {Panek}(1984)}]{walborn_siiv}
{Walborn}, N.~R. \& {Panek}, R.~J. 1984, \apjl, 280, L27

\bibitem[{{Wallerstein}(1962)}]{wallerstein62}
{Wallerstein}, G. 1962, \apjs, 6, 407

\bibitem[{{Wallerstein} \& {Helfer}(1959)}]{wallerstein59}
{Wallerstein}, G. \& {Helfer}, H.~L. 1959, \apj, 129, 347

\bibitem[{{Welty} {et~al.}(1999){Welty}, {Frisch}, {Sonneborn}, \&
  {York}}]{weletal99}
{Welty}, D.~E., {Frisch}, P.~C., {Sonneborn}, G., \& {York}, D.~G. 1999, \apj,
  512, 636

\bibitem[{{Worthey}(1994{\natexlab{a}})}]{worthey_models}
{Worthey}, G. 1994{\natexlab{a}}, \apjs, 95, 107

\bibitem[{{Worthey}(1994{\natexlab{b}})}]{wor94}
{Worthey}, G. 1994{\natexlab{b}}, \apjs, 95, 107

\bibitem[{{Worthey} {et~al.}(1992){Worthey}, {Faber}, \&
  {Gonzalez}}]{woretal92}
{Worthey}, G., {Faber}, S.~M., \& {Gonzalez}, J.~J. 1992, \apj, 398, 69

\bibitem[{{Worthey} {et~al.}(1994){Worthey}, {Faber}, {Gonzalez}, \&
  {Burstein}}]{worthey_lick}
{Worthey}, G., {Faber}, S.~M., {Gonzalez}, J.~J., \& {Burstein}, D. 1994,
  \apjs, 94, 687

\bibitem[{{Wu} {et~al.}(1983){Wu}, {Ake}, {Boggess}, {Bohlin}, {Imhoff},
  {Holm}, {Levay}, {Panek}, {Schiffer}, \& {Tunrose}}]{wu}
{Wu}, C.-C., {Ake}, T.~B., {Boggess}, A., {et~al.} 1983, NASA IUE Newsletter,
  22, 1

\bibitem[{{Wu} {et~al.}(1991){Wu}, {Crenshaw}, {Blackwell}, {Wilson-Díaz},
  {Schiffer}, {Burnstein}, {Fanelli}, \& {O'Connell}}]{wu_i}
{Wu}, C.-C., {Crenshaw}, D.~M., {Blackwell}, J.~H., {et~al.} 1991, NASA IUE
  Newsletter, 43

\bibitem[{{Yee} {et~al.}(1996){Yee}, {Ellingson}, {Bechtold}, {Carlberg}, \&
  {Cuillandre}}]{yee96}
{Yee}, H.~K.~C., {Ellingson}, E., {Bechtold}, J., {Carlberg}, R.~G., \&
  {Cuillandre}, J.-C. 1996, \aj, 111, 1783

\end{thebibliography}
\end{document}